\documentclass[prb,aps,twocolumn,amsmath,amssymb]{revtex4}
\usepackage{graphicx,bm,dcolumn,txfonts,color}

\begin{document}

\title{A Renormalization-Group Study of Interacting Bose--Einstein Condensates:\\
 II. Anomalous Dimension $\eta$ for $d\lesssim 4$ at Finite Temperatures}

\author{Takafumi Kita}
\affiliation{Department of Physics, Hokkaido University, Sapporo 060-0810, Japan}

\begin{abstract}
We study the anomalous dimension $\eta$ of homogeneous interacting single-component Bose--Einstein condensates 
at finite temperatures for $d\lesssim 4$ dimensions.
This $\eta$ is defined in terms of the one-particle density matrix $\rho({\bf r})\equiv \langle \hat\psi^\dagger({\bf r}_1)\hat\psi({\bf r}_1+{\bf r})\rangle$ 
through its asymptotic behavior $\rho({\bf r})\rightarrow N_{\bf 0}/V+C r^{-d+2-\eta}$ for $r\rightarrow \infty$, where $N_{\bf 0}/V$ is the condensate density and
$C$ is a constant. It is shown that the anomalous dimension is given by $\eta=0.181\epsilon^2$ to the leading order in $\epsilon\equiv d-4$.
The change of the prefactor $0.181$ from the value $0.02$  at the transition point of the ${\rm O}(2)$ symmetric $\phi^4$ model
is attributed to the emergence of three-point vertices and the anomalous Green's function
when $N_{\bf 0}$ acquires a finite value.
\end{abstract}

\maketitle
\section{Introduction}

In a previous paper,\cite{Kita19} which is referred to as I hereafter, exact renormalization-group equations have been derived for interacting single-component 
Bose-Einstein condensates 
based on the functional renormalization-group formalism\cite{Wetterich93,Morris94,Salmhofer99,BTW02,SK06,KBS10,MSHMS12}
in such a way as to satisfy the Hugenholtz-Pines theorem\cite{HP59} and Goldstone's theorem I.\cite{GSW62,Weinberg96}
Using them, it has been shown that the interaction vertex $g_\Lambda$ vanishes  below $d_{\rm c}=4$ dimensions at finite temperatures
as the infrared cutoff $\Lambda$ reduces to $0$, thereby causing
disappearance of the Bogoliubov mode\cite{Bogoliubov47} with a linear dispersion relation at long wavelengths.
Specifically, $g_\Lambda$ approaches zero as
$g_\Lambda\propto \Lambda^\epsilon$
with the exponent $\epsilon\equiv 4-d$  for $d\lesssim 4$ dimensions at finite temperatures.
Moreover, it is predicted that this vanishing of $g_\Lambda$ is accompanied by the development of the
anomalous dimension $\eta>0$ in the single-particle density matrix 
$\rho({\bf r})\equiv \langle \hat\psi^\dagger({\bf r}_1)\hat\psi({\bf r}_1+{\bf r})\rangle$ 
as
\begin{align}
\rho({\bf r})\stackrel{r\rightarrow\infty}{\longrightarrow} \frac{N_{\bf 0}}{V}+\frac{C}{r^{\,d-2+\eta}},
\label{rho}
\end{align}
where ${N_{\bf 0}}$ is the number of condensed particles, $V$ is the volume, and $C$ is a constant.
The exponent $\eta$ is predicted to be expressible as
$
\eta\propto \epsilon^2
$
for $d\lesssim 4$ dimensions, which has the importance of distinguishing the interacting Bose-Einstein condensates 
from the ideal ones with $\eta=0$.
Since the phase rigidity and coherence are expected to emerge due to the interaction,\cite{Kita17,Kita18}
which is also responsible for a finite $\eta>0$,
we call $\eta$ alternatively as {\em coherence exponent} here.

The purpose of the present paper is to confirm $\eta\propto \epsilon^2$ and also derive the prefactor
through careful calculations of exhausting all the processes contributing to it.
It will be shown that $\eta$ for $d\lesssim 4$ dimensions is given by
\begin{align}
\eta=&\,\left(\frac{1}{4}-\frac{\sqrt{3}}{8\pi}\right)\epsilon^2
\approx 0.1811\epsilon^2 ,
\label{eta-total}
\end{align}
which is exact up to the order of $\epsilon^2$ and valid at any finite temperature with $N_{\bf 0}>0$.
The prefactor is distinct from $0.02$ of the ${\rm O}(2)$ symmetric $\phi^4$ model at the transition point;\cite{Wilson72,Wilson74,Fisher74,Amit,Justin96} 
the difference is caused by the emergence of three-point vertices and the anomalous Green's function upon Bose-Einstein condensation.
The emergence of $\eta>0$ is expected to cause nonanalytic behaviors in various thermodynamic quantities of Bose-Einstein condensates at low temperatures.\cite{Kita19-2} However, the methods of extracting the exact value of $\eta$ experimentally are yet to be clarified theoretically.

This paper is organized as follows.
Section \ref{Sec2} presents basic formulas for obtaining $\eta$. Sections \ref{Sec:3}-\ref{Sec:5} 
consider the contributions of Fig.\ \ref{Fig1} (2a)-(2c) to $\eta$ separately to obtain
Eqs.\ (\ref{eta^(2a)}), (\ref{eta^(2b)}), and (\ref{eta^(2c)}), respectively,
which add up to Eq.\ (\ref{eta-total})  with Eq.\ (\ref{epsilon-def}).
We set $\hbar=k_{\rm B}=2m=1$ throughout with $m$ and $k_{\rm B}$ denoting the 
mass and Boltzmann constant, respectively.

\section{Key Quantities for Calculating $\eta$}
\label{Sec2}

\begin{figure}[b]
\begin{center}
\includegraphics[width=0.95\linewidth]{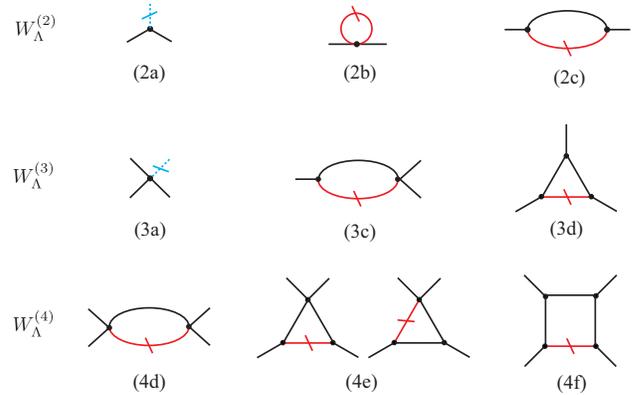}
\end{center}
\caption{Diagrammatic expressions of $W^{(n)}_{\Lambda,j_1\cdots j_n}$ for $n=2,3,4$.
A line with a dash (dotted line with a dash) denotes $\dot{G}_{\Lambda,j_1j_2}$ 
($\partial_\Lambda\Psi_\Lambda$).
\label{Fig1}}
\end{figure}

According to Eq.\ (81)  of I,\cite{Kita19} the exponent $\eta$ in Eq.\ (\ref{rho}) can be calculated by the formula
\begin{align}
\eta =\left. \frac{1}{2}\sum_{\alpha={\rm a},{\rm b},{\rm c}}
\frac{\partial^2 \delta\tilde{W}_{\infty}^{(2\alpha)}(\tilde{k})}{\partial \tilde{k}^2}\right|_{\tilde{k}=0}.
\label{eta-def2}
\end{align}
Functions $\delta\tilde{W}_{\infty}^{(2\alpha)}(\tilde{k})$ represent the momentum-dependent processes of
Fig.\ \ref{Fig1} (2a)-(2c). 
These diagrams and Eq.\ (\ref{eta-def2}) indicate that a finite $\eta$ originates mostly from the momentum dependences 
of the three- and four-point vertices, which in turn are caused by the loops in Fig.\ \ref{Fig1} (3c)-(4f).
A complete analysis of the loops will turn out laborious even for $d\lesssim 4$ owing to (i) the emergence of the three-point vertices and 
(ii) the internal degrees of freedom in the three- and four-point vertices.
Our goal is to derive Eqs.\ (\ref{eta^(2a)}), (\ref{eta^(2b)}), and (\ref{eta^(2c)}) for the three contributions in Eq.\ (\ref{eta-def2}), 
which add up to Eq.\ (\ref{eta-total}) as seen by using Eq.\ (\ref{epsilon-def}).

To start with, $\delta\tilde{W}_{\infty}^{(2\alpha)}(\tilde{k})$ are given analytically by Eq.\ (85) of I, 
which for $d\lesssim 4$ can be approximated by 
\begin{subequations}
\label{tW^(2)'s}
\begin{align}
\delta \tilde{W}_{\infty}^{(2{\rm a})}(\tilde{k})
=&\, -\frac{K_d}{4\beta\Psi}\int_0^1 \frac{d\lambda}{\lambda}\left[2\delta\tilde{W}^{(3)}_{\infty}(\lambda\tilde{\bf k},{\bf 0};-\lambda\tilde{\bf k})\right.
\notag \\
&\,
\left.-\delta\tilde{W}^{(3)}_{\infty}(\lambda\tilde{\bf k},-\lambda\tilde{\bf k};{\bf 0}) \right],
\label{tW^(2a)}
\end{align}
\begin{align}
&\,\delta\tilde{W}_{\infty}^{(2{\rm b})}(\tilde{k})
\notag \\
=&\,\frac{1}{4\beta}\int_0^1 \frac{d\lambda}{\lambda} \int\frac{d^d\tilde{q}}{(2\pi)^d}
\delta(\tilde{q}-1)
\Bigl[2\delta\tilde{W}^{(4)}_{\infty}(\lambda\tilde{\bf k},\lambda\tilde{\bf q};-\lambda\tilde{\bf k},-\lambda\tilde{\bf q})
\notag \\
&\,
+\delta \tilde{W}^{(4)}_{\infty}(\lambda\tilde{\bf k},-\lambda\tilde{\bf k};\lambda\tilde{\bf q},-\lambda\tilde{\bf q}) \Bigr] ,
\label{tW^(2b)}
\end{align}
\begin{align}
&\,\delta \tilde{W}^{(2{\rm c})}_\infty(\tilde{k})
\notag \\
=&\,-g_*\int\frac{d^d \tilde{q}}{(2\pi)^dK_d}\delta(\tilde{q}-1)\varTheta(|\tilde{\bf k}+\tilde{\bf q}|-1)
\left(\frac{1}{|\tilde{\bf k}+\tilde{\bf q}|^2}+1\right)
\notag \\
&\, -\frac{1}{\beta \Psi}\int\frac{d\lambda}{\lambda}\int\frac{d^d \tilde{q}}{(2\pi)^d}
\delta(\tilde{q}-1)\varTheta(|\tilde{\bf k}+\tilde{\bf q}|-1)
\notag \\
&\,\times
\Biggl\{ \delta \tilde{W}^{(3)}_{\infty}(\lambda\tilde{\bf q},-\lambda\tilde{\bf k}-\lambda\tilde{\bf q};\lambda\tilde{\bf k})
\left(\frac{1}{|\tilde{\bf k}+\tilde{\bf q}|^2}+1\right)
\notag \\
&\,
+ \left[\delta \tilde{W}^{(3)}_{\infty}(\lambda\tilde{\bf k},\lambda\tilde{\bf q};-\lambda\tilde{\bf k}-\lambda\tilde{\bf q})
-\delta \tilde{W}^{(3)}_{\infty}(\lambda\tilde{\bf k},-\lambda\tilde{\bf k}-\lambda\tilde{\bf q};\lambda\tilde{\bf q})
\right]
\notag \\
&\, \left.\times \left(\frac{1}{|\tilde{\bf k}+\tilde{\bf q}|^2}-1\right)\right\} .
\label{tW^(2c)}
\end{align}
\end{subequations}
Here $\Psi\equiv \lim_{\Lambda\rightarrow 0}\Psi_\Lambda$ is the condensate wave function, $\beta\equiv T^{-1}$ with $T$ denoting the temperature, 
$K_d$ and $g_*$ are given by
\begin{align}
K_d\equiv \frac{S_d}{(2\pi)^d}, \hspace{10mm}g_*\equiv 2\epsilon ,
\label{epsilon-def}
\end{align}
with $S_d\equiv 2\pi^{d/2}/\varGamma(d/2)$ the area of the unit sphere in $d$ dimensions,
and $\varTheta(x)$ and $\delta(x)$ are the Heaviside step function and Dirac delta function, respectively.
The integrals over $\lambda$ in Eq.\ (\ref{tW^(2)'s}) have the effect of producing the $n$-point vertices $\tilde{\Gamma}^{(n)}$ ($n=3,4$) from 
the source functions $\delta\tilde{W}^{(n)}$, as seen from Eq.\ (84) of I.

The key quantities  in Eq.\ (\ref{tW^(2)'s}) are $\delta\tilde{W}^{(3)}_x$ and $\delta\tilde{W}^{(4)}_x$ for $x\rightarrow \infty$, which are obtained from Eq.\ (76) of I
through the rescaling given by Eq.\ (79d) in I as
\begin{subequations}
\label{tW^(3,4)}
\begin{align}
&\, \delta\tilde{W}_x^{(3)}(\tilde{\bf k}_1,\tilde{\bf k}_2;\tilde{\bf k}_3)
\notag \\
\equiv&\, \frac{z_{\Lambda,-}^2}{\Lambda^{3-d}}\left[
W^{(3)}_{\Lambda,112}({\bf k}_1,{\bf k}_2,{\bf k}_3)-W^{(3)}_{\Lambda,111}({\bf k}_1,{\bf k}_2,{\bf k}_3)\right],
\label{tW^(3)}
\end{align}
\begin{align}
&\, \delta\tilde{W}_x^{(4)}(\tilde{\bf k}_1,\tilde{\bf k}_2;\tilde{\bf k}_3,\tilde{\bf k}_4)
\notag \\
\equiv &\, \frac{z_{\Lambda,-}^2}{3\Lambda^{3-d}}\left[
3W^{(4)}_{\Lambda,1122}({\bf k}_1,{\bf k}_2,{\bf k}_3,{\bf k}_4)-W^{(4)}_{\Lambda,1112}({\bf k}_1,{\bf k}_2,{\bf k}_3,{\bf k}_4)
\right.
\notag \\
&\, -W^{(4)}_{\Lambda,1121}({\bf k}_1,{\bf k}_2,{\bf k}_3,{\bf k}_4)-W^{(4)}_{\Lambda,1211}({\bf k}_1,{\bf k}_2,{\bf k}_3,{\bf k}_4)
\notag \\
&\,\left. -W^{(4)}_{\Lambda,2111}({\bf k}_1,{\bf k}_2,{\bf k}_3,{\bf k}_4)+W^{(4)}_{\Lambda,1111}({\bf k}_1,{\bf k}_2,{\bf k}_3,{\bf k}_4)\right],
\label{tW^(4)}
\end{align}
\end{subequations}
where $x$ and $\tilde{\bf k}$ denote
\begin{align}
x\equiv -\ln\frac{\Lambda}{\Lambda_0} ,\hspace{10mm}\tilde{\bf k}\equiv\frac{{\bf k}}{\Lambda} ,
\label{tk-k}
\end{align}
and $z_{\Lambda,-}$ is a renormalization factor defined by Eq.\ (59) of I.
Functions $W^{(3)}_{\Lambda,j_1j_2j_3}$ and $W^{(4)}_{\Lambda,j_1j_2j_3j_4}$ on the right-hand sides of 
Eqs.\ (\ref{tW^(3)}) and (\ref{tW^(4)}) are expressible diagrammatically as Fig.\ \ref{Fig1} (3a)-(3d) and (4d)-(4f), respectively, where we have 
omitted: (i) vertices with more than five legs as irrelevant; and (ii) $j=1,2$ degrees of freedom corresponding to the annihilation and creation operators
for simplicity. See Eqs.\ (47d) and (47e) of I for their analytic expressions.
Correspondingly, we can divide each of $\delta\tilde{W}^{(3)}_x$ and $\delta\tilde{W}^{(4)}_x$ into the three contributions.
Moreover, it follows from Eq.\ (77) of I that they both vanish when all the momenta are set equal to ${\bf 0}$.
Hence, we can express them as
\begin{subequations}
\label{tW^(3,4)-2}
\begin{align}
\delta\tilde{W}^{(3)}_x(\tilde{\bf k}_1,\tilde{\bf k}_2;\tilde{\bf k}_3)
=&\,\sum_{\alpha={\rm a},{\rm c},{\rm d}}
\Bigl[\delta\tilde{W}^{(3\alpha)}_x(\tilde{\bf k}_1,\tilde{\bf k}_2;\tilde{\bf k}_3) 
\notag \\
&\,-\lim_{\lambda\rightarrow 0} \delta\tilde{W}^{(3\alpha)}_x(\lambda\tilde{\bf k}_1,\lambda\tilde{\bf k}_2;\lambda\tilde{\bf k}_3) \Bigr],
\label{tW^(3)-2}
\end{align}
\begin{align}
\delta\tilde{W}^{(4)}_x(\tilde{\bf k}_1,\tilde{\bf k}_2;\tilde{\bf k}_3,\tilde{\bf k}_4)
=&\,\sum_{\alpha={\rm d},{\rm e},{\rm f}}\Bigl[
\delta\tilde{W}^{(4\alpha)}_x(\tilde{\bf k}_1,\tilde{\bf k}_2;\tilde{\bf k}_3,\tilde{\bf k}_4)
\notag \\
&\,
-\lim_{\lambda\rightarrow 0} \delta\tilde{W}^{(4\alpha)}_x(\lambda\tilde{\bf k}_1,\lambda\tilde{\bf k}_2;\lambda\tilde{\bf k}_3,\lambda\tilde{\bf k}_4) \Bigr] .
\label{tW^(4)-2}
\end{align}
\end{subequations}
This prescription is useful for considering each contribution separately in Eq.\ (\ref{tW^(2)'s}),
because its unphysical divergences, which cancel out eventually, are absent from the beginning.
We will adopt it throughout.

The expected behavior $\eta\propto \epsilon^2$ is one order of magnitude smaller 
in $\epsilon$ than the exponent of $g_\Lambda\propto \Lambda^\epsilon$.
This fact enables us to calculate Eq.\ (\ref{tW^(3,4)}) by using the 
leading-order expression of $g_\Lambda$ obtained as Eq.\ (67) of I, i.e.,
\begin{align}
g_\Lambda= g_*\frac{\beta }{K_dz_{\Lambda,-}^2}\Lambda^\epsilon .
\label{g_Lambda-asymp}
\end{align}
The vertices in Fig.\ \ref{Fig1} (3c)-(4f) are given in terms of $g_\Lambda$ by
\begin{subequations}
\label{Gamma^(3,4)}
\begin{align}
&\,\Gamma^{(3)}_{112}({\bf k}_1,{\bf k}_2,{\bf k}_3)=\Gamma^{(3)}_{221}({\bf k}_1,{\bf k}_2,{\bf k}_3)
=2g_\Lambda\Psi_\Lambda\delta_{{\bf k}_1+{\bf k}_2+{\bf k}_3,{\bf 0}},
\label{Gamma^(3)}
\end{align}
\begin{align}
\Gamma^{(4)}_{1122}({\bf k}_1,{\bf k}_2,{\bf k}_3,{\bf k}_4)=2g_\Lambda\delta_{{\bf k}_1+{\bf k}_2+{\bf k}_3+{\bf k}_4,{\bf 0}}.
\label{Gamma^(4)}
\end{align}
\end{subequations}
The other vertices such as $\Gamma^{(3)}_{111}$, $\Gamma^{(4)}_{1112}$, and $\Gamma^{(4)}_{1111}$ are omitted as irrelevant.
See Eq.\ (53) of I on this point.

For calculating $\eta$ at finite temperatures, 
the loops in Fig.\ \ref{Fig1} (3c)-(4f) are expressible in terms of the $2\times 2$ matrix Green's functions 
$\hat{G}_\Lambda(k)$ and $\hat{\dot{G}}_\Lambda(k)$
with zero Matsubara frequency. 
It follows from Eq.\ (30) of I that $\hat{G}_\Lambda(k)$ can be written as
\begin{align}
\hat{G}_\Lambda(k)=&\, \begin{bmatrix} -F_{\Lambda}(k) &  G_{\Lambda}(k) \\
G_{\Lambda}(k) & -F_{\Lambda}(k)
\end{bmatrix}.
\label{hatG}
\end{align}
Moreover, we can express the elements
$(G_\Lambda,F_\Lambda)$ for $\Lambda\rightarrow 0$ and $k\gtrsim \Lambda$ as
Eqs.\ (50) and (62) of I, i.e., 
\begin{subequations}
\label{GF}
\begin{align}
G_\Lambda(k)=&\,-\left(\frac{z_{\Lambda,-}}{2k^2}+\frac{1}{4g_\Lambda\Psi_\Lambda^2}\right)\varTheta(k-\Lambda),
\\
F_\Lambda(k)=&\,-\left(\frac{z_{\Lambda,-}}{2k^2}-\frac{1}{4g_\Lambda\Psi_\Lambda^2}\right)\varTheta(k-\Lambda).
\label{F}
\end{align}
\end{subequations}
Hence, $G_\Lambda(k)\approx F_\Lambda(k)$ holds within the leading order for $k\rightarrow 0$, and the difference
\begin{align}
\delta G_\Lambda(k)\equiv G_\Lambda(k)-F_\Lambda(k)\approx -\frac{\varTheta(k-\Lambda)}{2g_\Lambda\Psi_\Lambda^2}
\label{dG}
\end{align}
emerges in the next-to-leading order.

Another matrix Green's function $\hat{\dot{G}}$ is also expressible as Eq.\ (\ref{hatG}), where
the elements $(\dot{G}_\Lambda,\dot{F}_\Lambda)$ are obtained from Eq.\ (\ref{GF}) by the replacement
$\varTheta(k-\Lambda)\rightarrow -\delta(k-\Lambda)$.

\section{Calculation of $\eta^{(2{\bf a})}$}
\label{Sec:3}

We first calculate the 2a contribution of Eq.\ (\ref{eta-def2}) given by Eqs.\ (\ref{tW^(2a)}) and (\ref{tW^(3)}).
There are three kinds of topologically distinct diagrams for $\delta \tilde{W}^{(3)}_x$, i.e., those in the second row of Fig.\ \ref{Fig1}. 
Among them, the diagram of Fig.\ \ref{Fig1} (3a) gives rise to no ${\bf k}$ dependence within the leading-order approximation of 
Eq.\ (\ref{Gamma^(4)}). 
Hence, we can set 
\begin{align}
\delta\tilde{W}^{(3{\rm a})}_x(\tilde{\bf k}_1,\tilde{\bf k}_2;\tilde{\bf k}_3)=0 .
\label{W^(3a)}
\end{align}
We focus on the other two contributions below.

\subsection{The 2a-3c contribution}

\begin{figure}[t]
\begin{center}
\includegraphics[width=0.95\linewidth]{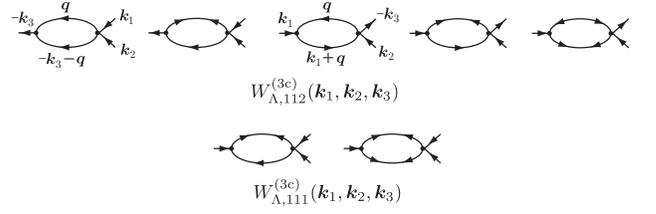}
\end{center}
\caption{Diagrammatic expressions of $W^{(3{\rm c})}_{\Lambda,112}$ and
$W^{(3{\rm c})}_{\Lambda,111}$. A single dash distinguishing $\dot{G}_{\Lambda,j_1j_2}$ from ${G}_{\Lambda,j_1j_2}$ 
should be incorporated into an internal line in all possible ways.
\label{Fig2}}
\end{figure}

First, we consider $\delta\tilde{W}^{(3{\rm c})}_\infty$.
Terms contributing to $\delta\tilde{W}^{(3{\rm c})}_x$ are expressible diagrammatically as Fig.\ \ref{Fig2}
by approximating the vertices in Fig.\ \ref{Fig1} (3c) as Eq.\ (\ref{Gamma^(3,4)}) 
and adding an incoming (outgoing) arrow around each vertex for $j=1$ ($j=2$).
The corresponding analytic expressions of $W^{(3{\rm c})}_{\Lambda,112}$ and $W^{(3{\rm c})}_{\Lambda,111}$ can be
obtained from Eq.\ (47d) of I as
\begin{subequations}
\label{W^(3c)}
\begin{align}
&\,W^{(3{\rm c})}_{\Lambda,112}({\bf k}_1,{\bf k}_2,{\bf k}_3)
\notag \\
=&\, (2g_\Lambda)^2\Psi_\Lambda
\left[\chi_{\Lambda,G\dot{G}}(k_3)-\chi_{\Lambda,G\dot{F}}(k_3)-\chi_{\Lambda,F\dot{G}}(k_3)
\right.
\notag \\
&\, +2\chi_{\Lambda,G\dot{G}}(k_1)-\chi_{\Lambda,G\dot{F}}(k_1)-\chi_{\Lambda,F\dot{G}}(k_1)+2\chi_{\Lambda,F\dot{F}}(k_1)
\notag \\
&\, \left.+2\chi_{\Lambda,G\dot{G}}(k_2)-\chi_{\Lambda,G\dot{F}}(k_2)-\chi_{\Lambda,F\dot{G}}(k_2)+2\chi_{\Lambda,F\dot{F}}(k_2)\right]
\notag \\
&\,\times 
\delta_{{\bf k}_1+{\bf k}_2+{\bf k}_3,{\bf 0}},
\label{W^(3c)_112}
\end{align}
\begin{align}
&\,W^{(3{\rm c})}_{\Lambda,111}({\bf k}_1,{\bf k}_2,{\bf k}_3)
\notag \\
=&\, (2g_\Lambda)^2\Psi_\Lambda
\left[-\chi_{\Lambda,G\dot{F}}(k_1)-\chi_{\Lambda,F\dot{G}}(k_1)+\chi_{\Lambda,F\dot{F}}(k_1) 
\right.
\notag \\
&\,-\chi_{\Lambda,G\dot{F}}(k_2)-\chi_{\Lambda,F\dot{G}}(k_2)+\chi_{\Lambda,F\dot{F}}(k_2)
\notag \\
&\, \left.-\chi_{\Lambda,G\dot{F}}(k_3)-\chi_{\Lambda,F\dot{G}}(k_3)+\chi_{\Lambda,F\dot{F}}(k_3)\right]
\delta_{{\bf k}_1+{\bf k}_2+{\bf k}_3,{\bf 0}}  ,
\label{W^(3c)_111}
\end{align}
\end{subequations}
where $\chi_{\Lambda,F\dot{F}}(k)$, for example, is defined by
\begin{align}
\chi_{\Lambda,F\dot{F}}(k)\equiv \frac{1}{\beta}\int\frac{d^d q}{(2\pi)^d}F_\Lambda(|{\bf k}+{\bf q}|)\dot{F}_\Lambda(q)  .
\label{chi_FF}
\end{align}
The minus signs in the square brackets of Eq.\ (\ref{W^(3c)}) originate from our definitions in Eq.\ (\ref{hatG}), i.e., $G_{\Lambda,12}\equiv G_\Lambda$ and $G_{\Lambda,11}\equiv -F_\Lambda$. 
The first (second) line in Eq.\ (\ref{W^(3c)_112}) corresponds to the first two (third through fifth) diagrams in Fig.\ \ref{Fig2}; the third line in Eq.\ (\ref{W^(3c)_112}) has been obtained from the second line by exchanging ${\bf k}_1$ and ${\bf k}_2$.

It suffices for calculating $\delta\tilde{W}^{(3{\rm c})}_x$ to use the leading-order expressions of $(G_\Lambda,F_\Lambda)$ in Eq.\ (\ref{GF}),
because they already give a finite value to $\delta\tilde{W}^{(3{\rm c})}_x$.
Hence, we set $G_\Lambda(k)\approx F_\Lambda(k)$ in Eq.\ (\ref{W^(3c)}),
substitute the resulting $(W^{(3)}_{\Lambda,112},W^{(3)}_{\Lambda,111})$ into Eq.\ (\ref{tW^(3)}), 
perform the transformation of Eq.\ (\ref{tk-k}), use Eq.\ (\ref{g_Lambda-asymp}), and take the limit $x\rightarrow\infty$ ($\Lambda\rightarrow 0$). 
We thereby obtain $\delta \tilde{W}^{(3{\rm c})}_\infty$ in a form without the renormalization factors $(\Lambda,z_{\Lambda,-})$ as
\begin{align}
\delta \tilde{W}^{(3{\rm c})}_\infty(\tilde{\bf k}_1,\tilde{\bf k}_2;\tilde{\bf k}_3)=&\,\frac{12g_*^2\beta\Psi}{K_d}
\left[\tilde{\chi}_{F\dot{F}}(\tilde{k}_1)+\tilde{\chi}_{F\dot{F}}(\tilde{k}_2)\right] \delta_{\tilde{\bf k}_1+\tilde{\bf k}_2+\tilde{\bf k}_3,{\bf 0}},
\label{tW^(3c)}
\end{align}
where $\tilde{\chi}_{F\dot{F}}(\tilde{k})$ is defined by
\begin{subequations}
\label{chi}
\begin{align}
\tilde{\chi}_{F\dot{F}}(\tilde{k})\equiv-\frac{1}{4}\int \frac{d^d\tilde{q}_1}{(2\pi)^d K_d}\delta(\tilde{q}_1-1)\frac{\varTheta(|\tilde{\bf k}+\tilde{\bf q}_1|-1)}{
|\tilde{\bf k}+\tilde{\bf q}_1|^2} .
\label{chi-def}
\end{align}
Let us express $|\tilde{\bf k}+\tilde{\bf q}_1|=(1+\tilde{k}^2+2\tilde{k}\cos\theta_1)^{1/2}$ for $\tilde{q}_1=1$
in terms of the angle $\theta_1$ between ${\bf q}_1$ and ${\bf k}$,
transform $\varTheta(|\tilde{\bf k}+\tilde{\bf q}_1|-1)=\varTheta(|\tilde{\bf k}+\tilde{\bf q}_1|^2-1)=\varTheta(\cos\theta_1+\tilde{k}/2)$,
and set $d= 4$ in the integrand as justified for $d\lesssim 4$. We thereby obtain
\begin{align}
\tilde{\chi}_{F\dot{F}}(\tilde{k})\approx &\, -\frac{1}{4}\int_0^{\xi_{\tilde{k}}}d\theta_1f_{\tilde{k}}(\theta_1),
\label{chi-approx}
\end{align}
\end{subequations}
where $f_x(\theta)$ and $\xi_x$ are defined by
\begin{align}
f_x(\theta)\equiv \frac{2}{\pi}\frac{\sin^2\theta}{1+2x\cos\theta+x^2} ,
\label{f_k}
\end{align}
\begin{align}
\xi_x\equiv \arccos\left(-\frac{x}{2}\right) =\frac{\pi}{2}+\frac{x}{2}+{\rm O}(x^3).
\label{xi_k-def}
\end{align}
The normalization constant $2/\pi$ in Eq.\ (\ref{f_k}) originates from integrating the Jacobian
$\sin^2\theta_1$ of the four-dimensional spherical coordinates over $\theta_1\in[0,\pi]$.

We subtract the momentum-independent contribution from Eq.\ (\ref{tW^(3c)}) based on Eq.\ (\ref{tW^(3)-2}) 
and substitute the resulting expression into Eq.\ (\ref{tW^(2a)}).
We then find immediately that the 3c contribution to $\delta \tilde{W}^{(2{\rm a})}_\infty$ 
vanishes.
Hence, it gives null to the exponent of Eq.\ (\ref{eta-def2}), i.e., 
\begin{align}
\eta^{(2{\rm a}3{\rm c})}=0 .
\label{eta^(2a3c)}
\end{align}

\subsection{The 2a-3d contribution}

\begin{figure}[t]
\begin{center}
\includegraphics[width=0.95\linewidth]{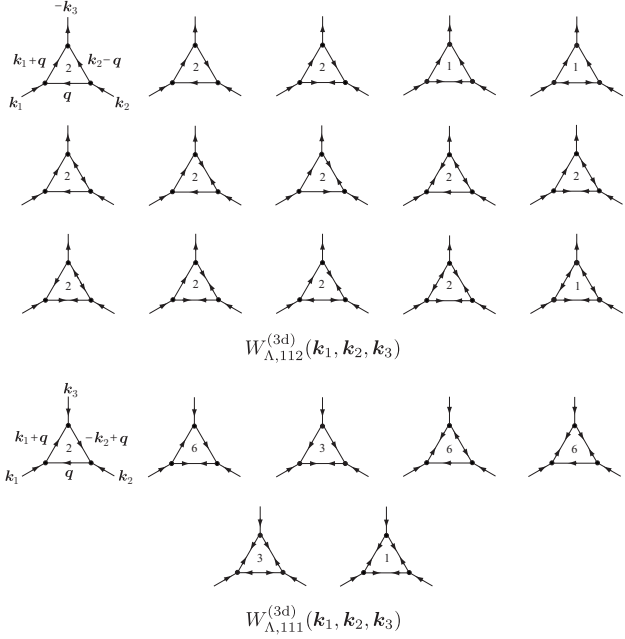}
\end{center}
\caption{Diagrammatic expressions of $W^{(3{\rm d})}_{\Lambda,112}$ and
$W^{(3{\rm d})}_{\Lambda,111}$. 
The number inside each diagram indicates its weight.
The total number of diagrams is $3^3$ for each of $W^{(3{\rm d})}_{\Lambda,112}$ and
$W^{(3{\rm d})}_{\Lambda,111}$, because for a $\Gamma_{\Lambda,jj_2j_3}^{(3)}$
with a given external $j=1$ or $2$, there are three choices of $(j_2,j_3)$ with Eq.\ (\ref{Gamma^(3)}).
 A single dash distinguishing $\dot{G}_{\Lambda,j_1j_2}$ from ${G}_{\Lambda,j_1j_2}$ 
should be incorporated into an internal line in all possible ways.
\label{Fig3}}
\end{figure}

Next, we focus on $\delta\tilde{W}^{(3{\rm d})}_x$.
Terms contributing to $\delta\tilde{W}^{(3{\rm d})}_x$ are expressible diagrammatically as Fig.\ \ref{Fig3}
by approximating the vertices in Fig.\ \ref{Fig1} (3d) as Eq.\ (\ref{Gamma^(3,4)})
and adding an incoming (outgoing) arrow for $j=1$ ($j=2$) around each vertex.
The corresponding analytic expressions of $W^{(3{\rm d})}_{\Lambda,112}$ and $W^{(3{\rm d})}_{\Lambda,111}$ are 
obtained from Eq.\ (47d) of I as
\begin{subequations}
\label{W^(3d)}
\begin{align}
&\,W^{(3{\rm d})}_{\Lambda,112}({\bf k}_1,{\bf k}_2,{\bf k}_3)
\notag \\
=&\,(2g_\Lambda\Psi_\Lambda)^3
\left(4\chi_{GGG}^{1,2}-4\chi_{GFG}^{1,2}-3\chi_{GGF}^{1,2}-3\chi_{FGG}^{1,2}+2\chi_{FGF}^{1,2}
\right.
\notag \\
&\,\left. +4\chi_{GFF}^{1,2}+4\chi_{FFG}^{1,2}-3\chi_{FFF}^{1,2}\right) \delta_{{\bf k}_1+{\bf k}_2+{\bf k}_3,{\bf 0}},
\label{W^(3d)_112}
\end{align}
\begin{align}
&\,W^{(3{\rm d})}_{\Lambda,111}({\bf k}_1,{\bf k}_2,{\bf k}_3)
\notag \\
=&\, (2g_\Lambda\Psi_\Lambda)^3
\left(2\chi_{GGG}^{1,2}-3\chi_{GFG}^{1,2}-3\chi_{GGF}^{1,2}-3\chi_{FGG}^{1,2}+4\chi_{FGF}^{1,2}
\right.
\notag \\
&\,\left. +4\chi_{GFF}^{1,2}+4\chi_{FFG}^{1,2}-4\chi_{FFF}^{1,2}\right)\delta_{{\bf k}_1+{\bf k}_2+{\bf k}_3,{\bf 0}},
\label{W^(3d)_111}
\end{align}
\end{subequations}
where $\chi_{ABC}^{1,2}$ is defined by
\begin{align}
\chi_{ABC}^{1,2}\equiv&\, \frac{1}{\beta}\int\frac{d^d q}{(2\pi)^d}\left[\dot{A}(|{\bf k}_1+{\bf q}|)B(q)C(|{\bf k}_2-{\bf q}|)
\right.
\notag \\
&\,+A(|{\bf k}_1+{\bf q}|)\dot{B}(q)C(|{\bf k}_2-{\bf q}|)
\notag \\
&\,\left.+A(|{\bf k}_1+{\bf q}|)B(q)\dot{C}(|{\bf k}_2-{\bf q}|)\right].
\label{chi_ABC}
\end{align}
For example, $4\chi_{GGG}^{1,2}$ ($-3\chi_{GGF}^{1,2}-3\chi_{FGG}^{1,2}$) in the round brackets of Eq.\ (\ref{W^(3d)_112})
corresponds to the first two diagrams in the first line of Fig.\ \ref{Fig3} (the first three diagrams in the second line of Fig.\ \ref{Fig3}).
Function $\chi_{ABC}^{1,2}$ satisfies
\begin{align}
\chi_{ABC}^{1,2}=\chi_{CBA}^{2,1},\hspace{10mm}\chi_{ABC}^{1,2}=\chi_{ABC}^{-1,-2},
\label{chi_ABC-symm}
\end{align}
where $\chi_{ABC}^{1,-2}$ denotes the function obtained from $\chi_{ABC}^{1,2}$ by ${\bf k}_2\rightarrow -{\bf k}_2$.

Let us substitute Eq.\ (\ref{W^(3d)}) into Eq.\ (\ref{tW^(3)}), express $G=F+\delta G$, and expand the resulting expression
in terms of $\delta G$. We then find that terms of ${\rm O}((\delta G)^0)$ cancel out, and the next-to-leading order terms yield
\begin{align}
\delta \tilde{W}^{(3{\rm d})}_x(\tilde{\bf k}_1,\tilde{\bf k}_2;\tilde{\bf k}_3)
=&\, z_{\Lambda,-}^2\Lambda^{d-3}
(2g_\Lambda\Psi_\Lambda)^3\left(\chi_{\delta G FF}^{1,2}+\chi_{FF\delta G}^{1,2}\right)
\notag \\
&\,\times\delta_{\tilde{\bf k}_1+\tilde{\bf k}_2+\tilde{\bf k}_3,{\bf 0}} .
\label{tW^(3d)-0}
\end{align}
The key function $\chi_{\delta GFF}^{1,2}$ defined by Eq.\ (\ref{chi_ABC}) can be simplified substantially 
by using Eqs.\ (\ref{GF}) and (\ref{dG}), performing the transformation of Eq.\ (\ref{tk-k}),
and making changes of integration variables so that the arguments of $\dot{F}$ and $\delta\dot{G}$ are given in terms of a common vector ${\bf q}_1$.
We thereby obtain
\begin{align}
\chi_{\delta GFF}^{1,2}
=&\, \frac{z_{\Lambda,-}^2\Lambda^{d-5}K_d}{8g_\Lambda\Psi_\Lambda^2\beta}\Bigl[\phi_{22}(\tilde{\bf k}_1,\tilde{\bf k}_1+\tilde{\bf k}_2)
+\phi_{02}(\tilde{\bf k}_1,-\tilde{\bf k}_2)
\notag \\
&\, +\phi_{20}(\tilde{\bf k}_2,\tilde{\bf k}_1+\tilde{\bf k}_2)\Bigr],
\label{chi_dGFF}
\end{align}
where $\phi_{n_1n_2}(\tilde{\bf k}_1,\tilde{\bf k}_2)$ is defined by
\begin{align}
\phi_{n_1n_2}(\tilde{\bf k}_1,\tilde{\bf k}_2)
\equiv &\,
\int\frac{d^d \tilde{q}_1}{(2\pi)^d K_d}
\delta(\tilde{q}_1-1)
\frac{\varTheta(|\tilde{\bf k}_1+\tilde{\bf q}_1|-1)}{
|\tilde{\bf k}_1+\tilde{\bf q}_1|^{n_1}}
\notag \\
&\,\times \frac{\varTheta(|\tilde{\bf k}_2+\tilde{\bf q}_1|-1)}{|\tilde{\bf k}_2+\tilde{\bf q}_1|^{n_2}},
\label{phi^(3d)}
\end{align}
satisfying $\phi_{n_1n_2}(-\tilde{\bf k}_1,-\tilde{\bf k}_2)=\phi_{n_1n_2}(\tilde{\bf k}_1,\tilde{\bf k}_2)$
as shown by setting $\tilde{\bf q}_1\rightarrow -\tilde{\bf q}_1$ in the integrand.
Let us substitute Eq.\ (\ref{chi_dGFF}) into Eq.\ (\ref{tW^(3d)-0}), 
rewrite $g_\Lambda$ as Eq.\ (\ref{g_Lambda-asymp}), 
and set $x\rightarrow\infty$ ($\Lambda\rightarrow 0$). 
We thereby obtain $\delta \tilde{W}^{(3{\rm d})}_\infty$ in a form without the renormalization factors $(\Lambda,z_{\Lambda,-})$ as
\begin{align}
&\,\delta \tilde{W}^{(3{\rm d})}_\infty(\tilde{\bf k}_1,\tilde{\bf k}_2;\tilde{\bf k}_3)
\notag \\
=&\, \frac{g_*^2 \beta\Psi}{K_d}\Bigl[ 
\phi_{22}(\tilde{\bf k}_1,\tilde{\bf k}_1+\tilde{\bf k}_2)
+\phi_{22}(\tilde{\bf k}_2,\tilde{\bf k}_1+\tilde{\bf k}_2)+\phi_{02}(\tilde{\bf k}_1,-\tilde{\bf k}_2) 
\notag \\
&\, +\phi_{20}(\tilde{\bf k}_1,-\tilde{\bf k}_2)
+\phi_{20}(\tilde{\bf k}_1,\tilde{\bf k}_1+\tilde{\bf k}_2)
+\phi_{20}(\tilde{\bf k}_2,\tilde{\bf k}_1+\tilde{\bf k}_2)\Bigr]
\notag \\
&\,\times \delta_{\tilde{\bf k}_1+\tilde{\bf k}_2+\tilde{\bf k}_3,{\bf 0}}.
\label{tW^(3d)}
\end{align}

We substitute Eq.\ (\ref{tW^(3d)}) into Eq.\ (\ref{tW^(2a)}),
subtract the momentum-independent contribution from Eq.\ (\ref{tW^(3d)}) based on Eq.\ (\ref{tW^(3)-2}),
 and simplify the resulting expression
by using $\delta(x)\varTheta(x)=\frac{1}{2}\delta(x)$ and $[\varTheta(x)]^2=\varTheta(x)$.
The procedure yields the 3d contribution to $\delta \tilde{W}^{(2{\rm a})}_\infty$ as
\begin{align}
&\,\delta \tilde{W}_{\infty}^{(2{\rm a}3{\rm d})}(\tilde{k})
\notag \\
=&\, -\frac{g_*^2}{2}\int_0^1 \frac{d\lambda}{\lambda}\Bigl[\phi_{22}(\lambda\tilde{\bf k},\lambda\tilde{\bf k})+2
\phi_{02}(\lambda\tilde{\bf k},0)-(\lambda\rightarrow 0)\Bigr]
\notag \\
\approx &\, -\frac{g_*^2}{2}\int_0^1 \frac{d\lambda}{ \lambda}\Biggl\{\int_0^{\xi_{\lambda\tilde{k}}}\!d\theta_q f_0(\theta_q)
\Biggl[\frac{1}{(1+2\lambda\tilde{k}\cos\theta_q+\lambda^2\tilde{k}^2)^2}+1\Biggr]
\notag \\
&\, -(\lambda\rightarrow 0) \Biggr\},
\label{tW^(2a3d)}
\end{align}
where we have made a transformation similar to Eq.\ (\ref{chi}) to derive the second expression.
Let us differentiate Eq.\  (\ref{tW^(2a3d)}) twice with respect to $\tilde{k}$, set $\tilde{k}=0$ subsequently, substitute the resulting expression into 
Eq.\ (\ref{eta-def2}), and evaluate the integrals. We thereby obtain the 2a-3d contribution to the coherence exponent as
\begin{align}
\eta^{(2{\rm a}3{\rm d})}=-\frac{1}{8} g_*^2.
\label{eta^(2a3d)}
\end{align}

\subsection{Sum of various 2a contributions}

The net 2a contribution is obtained by adding Eqs.\ (\ref{eta^(2a3c)}) and (\ref{eta^(2a3d)}) as
\begin{align}
\eta^{(2{\rm a})}=-\frac{1}{8}g_*^2 .
\label{eta^(2a)}
\end{align}
It is worth noting that this finite result could not have been obtained without exhausting all the processes,
as has been done in Fig.\ \ref{Fig3} for Fig.\ \ref{Fig1} (3d),
where a complete cancellation of the leading-order terms exists as mentioned above Eq.\ (\ref{tW^(3d)-0}),
which removes the divergence in each of them.
We will encounter this kind of cancellation two times below, in one of which it even extends up to the next-to-leading order terms.

\section{Calculation of $\eta^{(2{\bf b})}$}
\label{Sec:4}

We proceed to calculate the 2b contribution
in Eq.\ (\ref{eta-def2}) given by Eqs.\ (\ref{tW^(2b)}) and (\ref{tW^(4)}).  
There are three kinds of topologically distinct diagrams for $\delta \tilde{W}^{(4)}_x$, i.e., those in the third row 
of Fig.\ \ref{Fig1}. Among them, the 4d contribution has already been studied to yield Eq.\ (93) of I, i.e., 
\begin{align}
\eta^{(2{\rm b}4{\rm d})}=\frac{3}{8}g_*^2 .
\label{eta^(2b4d)}
\end{align}
Hence, we here focus on the other two diagrams.

\subsection{The 2b-4e contribution}

\begin{figure}[t]
\begin{center}
\includegraphics[width=0.95\linewidth]{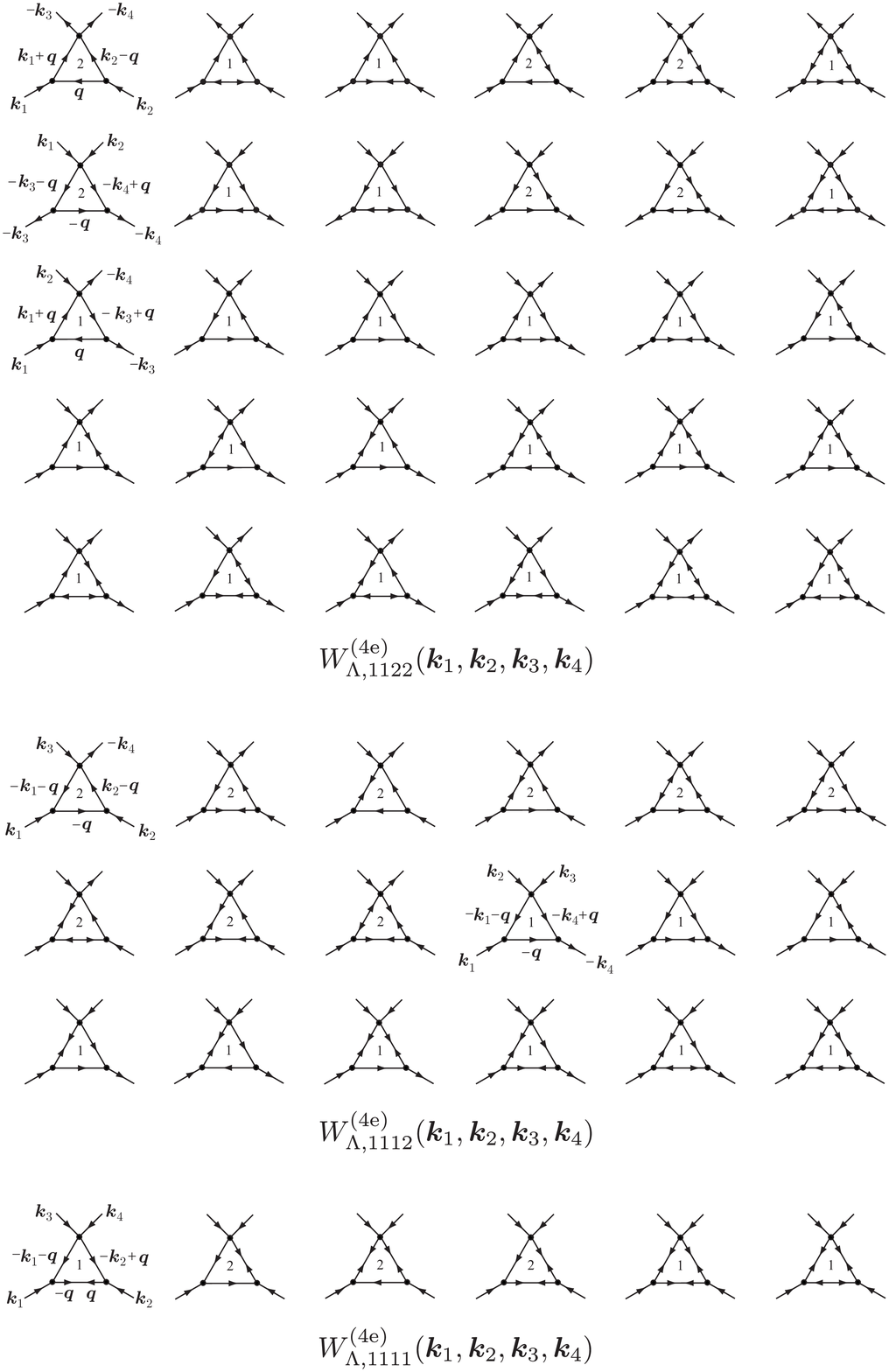}
\end{center}
\caption{Diagrammatic expressions of $W^{(4{\rm e})}_{\Lambda,1122}$,  $W^{(4{\rm e})}_{\Lambda,1112}$, and $W^{(4{\rm e})}_{\Lambda,1111}$.
The number inside each closed particle line indicates its weight.
The total numbers of diagrams for the three functions are $3^2+3^2+2\cdot 3^2=36$,  $2\cdot 3^2+3^2=27$, and $3^2=9$, respectively,
in agreement with the following combinatorial consideration in terms of Eq.\ (\ref{Gamma^(3,4)}):
For $\Gamma_{\Lambda,jj_2j_3}^{(3)}$ with a given external $j=1$ or $2$, there are three choices of $(j_2,j_3)$.
For $\Gamma^{(4)}_{\Lambda,jjj_3j_4}$ with a given $j=1$ or $2$, only the case $(j_3,j_4)=(3-j,3-j)$ is relevant, 
whereas for $\Gamma^{(4)}_{\Lambda,12j_3j_4}$, there are two choices of  $(j_3,j_4)=(1,2),(2,1)$.
\label{Fig4}}
\end{figure}

First, we consider the diagram of Fig.\ \ref{Fig1} (4e).
Its contribution to Eq.\ (\ref{tW^(4)}) is  expressible in more detail as Fig.\ \ref{Fig4}
by approximating the vertices in Fig.\ \ref{Fig1} (4e) as Eq.\ (\ref{Gamma^(3,4)})
and adding an incoming (outgoing) arrow for $j=1$ ($j=2$) around each vertex.
The corresponding analytic expressions of $W^{(4{\rm e})}_{\Lambda,1122}$, $W^{(4{\rm e})}_{\Lambda,1112}$, and $W^{(4{\rm e})}_{\Lambda,1111}$ can be 
obtained from Eq.\ (47e) of I. 
They are written in terms of Eq.\ (\ref{chi_ABC}) as
\begin{subequations}
\label{W^(4e)}
\begin{align}
&\,W^{(4{\rm e})}_{\Lambda,1122}({\bf k}_1,{\bf k}_2,{\bf k}_3,{\bf k}_4)
\notag \\
=&\, (2g_\Lambda)^3\Psi_\Lambda^2
\biggl[\left(2\chi_{GGG}^{1,2}-2\chi_{GFG}^{1,2}-\chi_{GGF}^{1,2}-\chi_{FGG}^{1,2}+\chi_{GFF}^{1,2}
\right.
\notag \\
&\,\left. +\chi_{FFG}^{1,2}-\chi_{FFF}^{1,2}\right)+\left(\chi_{}^{1,2}\rightarrow \chi_{}^{3,4}\right)
\notag \\
&\,
+\left(3\chi_{GGG}^{1,3}-2\chi_{GFG}^{1,3}-2\chi_{GGF}^{1,3}-2\chi_{FGG}^{1,3}+3\chi_{FGF}^{1,3}+2\chi_{GFF}^{1,3}
\right.
\notag \\
&\,
\left. +2\chi_{FFG}^{1,3}-2\chi_{FFF}^{1,3}\right)+\left(\chi^{1,3}\rightarrow \chi_{}^{1,4}\right)+\left(\chi^{1,3}\rightarrow \chi_{}^{2,3}\right)
\notag \\
&\, +\left(\chi^{1,3}\rightarrow \chi_{}^{2,4}\right)
\biggr]\,\delta_{{\bf k}_1+{\bf k}_2+{\bf k}_3+{\bf k}_4,{\bf 0}},
\label{W^(4e)_1122}
\end{align}
\begin{align}
&\,W^{(4{\rm e})}_{\Lambda,1112}({\bf k}_1,{\bf k}_2,{\bf k}_3,{\bf k}_4)
\notag \\
=&\, (2g_\Lambda)^3\Psi_\Lambda^2\biggl[
\left(2\chi_{GGG}^{1,2}-2\chi_{GFG}^{1,2}-2\chi_{GGF}^{1,2}-2\chi_{FGG}^{1,2}+2\chi_{FGF}^{1,2}
\right.
\notag \\
&\,\left. +3\chi_{GFF}^{1,2}+3\chi_{FFG}^{1,2}-2\chi_{FFF}^{1,2}\right)+\left(\chi^{1,2}\rightarrow \chi_{}^{1,3}\right)+ \left(\chi^{1,2}\rightarrow \chi_{}^{2,3}\right)
\notag \\
&\, +\left(\chi_{GGG}^{1,4}-\chi_{GFG}^{1,4}-\chi_{GGF}^{1,4}-2\chi_{FGG}^{1,4}+\chi_{FGF}^{1,4}+2\chi_{FFG}^{1,4}
\right.
\notag \\
&\,\left.-\chi_{FFF}^{1,4}\right)+\left(\chi^{1,4}\rightarrow \chi_{}^{2,4}\right)+ \left(\chi^{1,4}\rightarrow \chi_{}^{3,4}\right)
\biggr]\,\delta_{{\bf k}_1+{\bf k}_2+{\bf k}_3+{\bf k}_4,{\bf 0}} ,
\label{W^(4e)_1112}
\end{align}
\begin{align}
&\,W^{(4{\rm e})}_{\Lambda,1111}({\bf k}_1,{\bf k}_2,{\bf k}_3,{\bf k}_4)
\notag \\
=&\, (2g_\Lambda)^3\Psi_\Lambda^2\biggl[
\left(-\chi_{GFG}^{1,2}-\chi_{GGF}^{1,2}-\chi_{FGG}^{1,2}+2\chi_{FGF}^{1,2}+\chi_{GFF}^{1,2}
\right.
\notag \\
&\,\left. +\chi_{FFG}^{1,2}-2\chi_{FFF}^{1,2}\right)+\left(\chi^{1,2}\rightarrow \chi_{}^{1,3}\right)+ \left(\chi^{1,2}\rightarrow \chi_{}^{1,4}\right)
\notag \\
&\,+\left(\chi^{1,2}\rightarrow \chi_{}^{2,3}\right)
+\left(\chi^{1,2}\rightarrow \chi_{}^{2,4}\right)+\left(\chi^{1,2}\rightarrow \chi_{}^{3,4}\right)
\biggr]
\notag \\
&\,\times \delta_{{\bf k}_1+{\bf k}_2+{\bf k}_3+{\bf k}_4,{\bf 0}} ,
\label{W^(4e)_1111}
\end{align}
\end{subequations}
where $\left(\chi^{1,2}\rightarrow \chi^{3,4}\right)$ in Eq.\ (\ref{W^(4e)_1122}), for example, denotes the term obtained from the first term
in the square brackets by the replacement $({\bf k}_1,{\bf k}_2)\rightarrow({\bf k}_3,{\bf k}_4)$.

Using Eqs.\ (\ref{tW^(4)}) and (\ref{W^(4e)}) and noting the symmetries of Eq.\ (\ref{chi_ABC-symm}), 
we obtain the key function in Eq.\ (\ref{tW^(2b)}) as
\begin{align*}
&\, 2\delta \tilde{W}^{(4{\rm e})}_x(\tilde{\bf k}_1,\tilde{\bf k}_2;-\tilde{\bf k}_1,-\tilde{\bf k}_2)
+\delta \tilde{W}^{(4{\rm e})}_x(\tilde{\bf k}_1,-\tilde{\bf k}_1;\tilde{\bf k}_2,-\tilde{\bf k}_2)
\notag \\
=&\, 
(2g_\Lambda)^3\Psi_\Lambda^2 z_{\Lambda,-}^2\Lambda^{d-3}\delta_{\tilde{\bf k}_1+\tilde{\bf k}_2+\tilde{\bf k}_3+\tilde{\bf k}_4,{\bf 0}} 
\biggl[\left(2\chi_{GGG}^{1,2}-2\chi_{GFG}^{1,2}\right.
\notag \\
&\,\left.+4\chi_{GGF}^{1,2}+4\chi_{FGG}^{1,2}-2\chi_{FGF}^{1,2}-6\chi_{FFG}^{1,2}-6\chi_{GFF}^{1,2}\right)
\notag \\
&\,+\left(6\chi_{GGG}^{1,-2}-2\chi_{GFG}^{1,-2}+10\chi_{FGF}^{1,-2}-2\chi_{FFG}^{1,-2}-2\chi_{GFF}^{1,-2}\right.
\notag \\
&\,\left.-4\chi_{FFF}^{1,-2}\right)+\left(2\chi_{GGG}^{1,-1}-\chi_{GFG}^{1,-1}+\chi_{GGF}^{1,-1}+\chi_{FGG}^{1,-1}+2\chi_{FGF}^{1,-1}\right.
\notag \\
&\,\left.-2\chi_{FFG}^{1,-1}-2\chi_{GFF}^{1,-1}-\chi_{FFF}^{1,-1}\right)+\left(2\chi_{GGG}^{2,-2}-\chi_{GFG}^{2,-2}+\chi_{GGF}^{2,-2}
\right.
\notag \\
&\, \left.+\chi_{FGG}^{2,-2}+2\chi_{FGF}^{2,-2}-2\chi_{FFG}^{2,-2}-2\chi_{GFF}^{2,-2}-\chi_{FFF}^{2,-2}\right)\biggr].
\end{align*}
Subsequently, we write $G=F+\delta G$ and expand the resulting expression in terms of $\delta G$ given by Eq.\ (\ref{dG}).
We then find that terms of ${\rm O}((\delta G)^0)$ cancel out, and the next-to-leading order terms yield
\begin{align}
&\, 2\delta \tilde{W}^{(4{\rm e})}_x(\tilde{\bf k}_1,\tilde{\bf k}_2;-\tilde{\bf k}_1,-\tilde{\bf k}_2)
+\delta \tilde{W}^{(4{\rm e})}_x(\tilde{\bf k}_1,-\tilde{\bf k}_1;\tilde{\bf k}_2,-\tilde{\bf k}_2)
\notag \\
=&\, 
(2g_\Lambda)^3\Psi_\Lambda^2 z_{\Lambda,-}^2\Lambda^{d-3}\delta_{\tilde{\bf k}_1+\tilde{\bf k}_2+\tilde{\bf k}_3+\tilde{\bf k}_4,{\bf 0}} 
\biggl[6\left(\chi_{FFF}^{1,2}-\chi_{FFF}^{1,-2}\right)
\notag \\
&\,\left(-2\chi_{\delta GFF}^{1,2}+8\chi_{F\delta G F}^{1,2}-2\chi_{FF\delta G}^{1,2}\right)
+\left(2\chi_{\delta GFF}^{1,-2}+16\chi_{F\delta G F}^{1,-2}\right.
\notag \\
&\,\left.+2\chi_{FF\delta G}^{1,-2}\right)+6\chi_{F\delta GF}^{1,-1}+6\chi_{F\delta GF}^{2,-2}+{\rm O}\Bigl((\delta G)^2\Bigr) \biggr] .
\label{tW^(4e)}
\end{align}
Function $\chi_{\delta GFF}^{1,2}$ is given by Eq.\ (\ref{chi_dGFF}).
Similarly, $\chi_{F\delta GF}^{1,2}$ is expressible as
\begin{align}
\chi_{F\delta GF}^{1,2}
=&\, \frac{z_{\Lambda,-}^2\Lambda^{d-5}K_d}{8g_\Lambda\Psi_\Lambda^2\beta}\Bigl[\phi_{20}(\tilde{\bf k}_1+\tilde{\bf k}_2,\tilde{\bf k}_1)
+\phi_{22}(\tilde{\bf k}_1,-\tilde{\bf k}_2)
\notag \\
&\,+\phi_{20}(\tilde{\bf k}_1+\tilde{\bf k}_2,\tilde{\bf k}_2)\Bigr] ,
\label{chi_FdGF}
\end{align}
where $\phi_{n_1n_2}$ is defined by Eq.\ (\ref{phi^(3d)}).

Let us substitute Eq.\ (\ref{tW^(4e)}) into Eq.\ (\ref{tW^(2b)})
and make a change of variables $\tilde{\bf q}\rightarrow-\tilde{\bf q}$ for the $\chi_{ABC}^{1,-2}$ contribution in Eq.\ (\ref{tW^(4e)})
to combine it with that of $\chi_{ABC}^{1,2}$.
We then find that terms of ${\rm O}\bigl((\delta G)^0\bigr)$ cancel out, and only the contribution of Eq.\ (\ref{chi_FdGF}) survives
in the next-to-leading order.
Subsequently, we substitute Eq.\ (\ref{g_Lambda-asymp}) and approximate $d\approx 4$ in the integrand as justified
for $\epsilon\equiv 4-d\ll 1$.
We thereby obtain $\delta\tilde{W}_{\infty}^{(2{\rm b}4{\rm e})}$ as
\begin{align}
&\,\delta\tilde{W}_{\infty}^{(2{\rm b}4{\rm e})}(\tilde{k}) 
\notag \\
=&\,\frac{3}{2}g_*^2\int_0^1\frac{d\lambda}{\lambda} \!\int\! \frac{d^4\tilde{q}}{(2\pi)^4K_4}
\delta(\tilde{q}-1)
\Bigl\{\bigl[ \phi_{22}(\lambda\tilde{\bf k},\lambda\tilde{\bf k})+2\phi_{02}(\lambda\tilde{\bf k},{\bf 0})\bigr]
\notag \\
&\,+4\phi_{22}(\lambda\tilde{\bf q},\lambda\tilde{\bf k})+4 \phi_{20}(\lambda\tilde{\bf q}+\lambda\tilde{\bf k},\lambda\tilde{\bf q})
+4 \phi_{20}(\lambda\tilde{\bf q}+\lambda\tilde{\bf k},\lambda\tilde{\bf k})
\notag \\
&\,+\bigl[\phi_{22}(\lambda\tilde{\bf q},\lambda\tilde{\bf q})+2\phi_{02}(\lambda\tilde{\bf q},{\bf 0})\bigr]-(\lambda\rightarrow 0)\Bigr\}.
\label{tW^(2b4e)}
\end{align}

We consider each term in the curly brackets separately. 
First, we focus on the first term and
substitute Eq.\ (\ref{phi^(3d)}).
Transforming the resulting expression  in a way similar to Eqs.\ (\ref{chi}) and (\ref{tW^(2a3d)}), 
we obtain
\begin{align}
&\,\delta W^{(2{\rm b}4{\rm e}1)}_\infty(\tilde{k})
\notag \\
= &\, \frac{3}{2}g_*^2\int_0^1\frac{d\lambda}{\lambda}\Biggl\{\int_0^{\xi_{\lambda\tilde{k}}} d\theta_1 f_0(\theta_1)\Biggl[\frac{1}{(1+2\lambda\tilde{k}\cos\theta_1+\lambda^2\tilde{k}^2)^2}+1\Biggr]
\notag \\
&\,-(\lambda\rightarrow 0)\Biggr\} ,
\label{tW^(2b4e1)}
\end{align}
which is identical with Eq.\ (\ref{tW^(2a3d)}) except for the numerical factor.
Hence, its contribution to Eq.\ (\ref{eta-def2}) is immediately found to be
\begin{align}
\eta^{(2{\rm b}4{\rm e}1)}= \frac{3}{8}g_*^2 .
\label{eta^(2b4e1)}
\end{align}

Second, we focus on the second term in the curly brackets of Eq.\ (\ref{tW^(2b4e)}).
Let us substitute Eq.\ (\ref{phi^(3d)}) into it, 
subsequently exchange the order of integrations between $\tilde{\bf q}$ and $\tilde{\bf q}_1$,
express the $\tilde{\bf q}$ integral in the four-dimensional spherical coordinates where $\tilde{\bf q}_1$ lies along the first axis,
and transform the resulting expression in a way similar to Eq.\ (\ref{chi}).
We thereby obtain $\delta\tilde{W}_{\infty}^{(2{\rm b}4{\rm e}1)}(\tilde{k})$ in terms of the functions in Eqs.\ (\ref{f_k}) and (\ref{xi_k-def}) as
\begin{align}
&\,\delta\tilde{W}_{\infty}^{(2{\rm b}4{\rm e}2)}(\tilde{k}) 
\notag \\
= &\,6g_*^2 \int_0^1 \frac{d\lambda}{\lambda}\Biggl[\int_0^{\xi_{\lambda\tilde{k}}}d\theta_1 f_{\lambda\tilde{k}}(\theta_1)\int_0^{\xi_\lambda}d\theta_q f_\lambda(\theta_q) 
-(\lambda\rightarrow 0)\Biggr] ,
\label{tW^(2b4e2)}
\end{align}
where $\theta_1$ ($\theta_q$) is the angle between $\tilde{\bf k}$ and $\tilde{\bf q}_1$ ($\tilde{\bf q}$ and $\tilde{\bf q}_1$).
Let us differentiate Eq.\ (\ref{tW^(2b4e2)}) twice with respect to $\tilde{k}$,  set $\tilde{k}=0$ subsequently, 
substitute the resulting expression into Eq.\ (\ref{eta-def2}),
and perform the integrations.
The procedure yields
\begin{align}
\eta^{(2{\rm b}4{\rm e}2)}= 0 .
\label{eta^(2b4e2)}
\end{align}

Third, we focus on the third term in the curly brackets of Eq.\ (\ref{tW^(2b4e)}), which is given explicitly by
\begin{align}
&\,\delta\tilde{W}_{\infty}^{(2{\rm b}4{\rm e}3)}(\tilde{k}) 
\notag \\
=&\, 6g_*^2 \int_0^1 \frac{d\lambda}{\lambda}\Biggl[\int\frac{d^4\tilde{q}}{(2\pi)^4K_4}\delta(\tilde{q}-1)
\int\frac{d^4\tilde{q}_1}{(2\pi)^4K_4}
\delta(\tilde{q}_1-1)
\notag \\
&\,
\times \frac{\varTheta(|\tilde{\bf q}_1+\lambda\tilde{\bf q}'|-1)\varTheta(|\tilde{\bf q}_1+\lambda\tilde{\bf q}|-1)}
{|\tilde{\bf q}_1+\lambda\tilde{\bf q}'|^2} 
-(\lambda\rightarrow 0)\Biggr],
\label{tW^(2b4e3)-0}
\end{align}
with $\tilde{\bf q}'\equiv \tilde{\bf q}+\tilde{\bf k}$. 
The calculation of this term requires a new and lengthy treatment.
However, we will eventually arrive at a simple analytic expression of Eq.\ (\ref{eta^(2b4e3)}) below
for its contribution to $\eta$.

To start with, we note that the integral over $\tilde{\bf q}_1$ for $\tilde{q}=1$ depends only on two variables, i.e., 
the magnitude $\tilde{k}$
and angle $\theta_q$ between $(\tilde{\bf q},\tilde{\bf k})$. 
This fact enables us to write the $\tilde{\bf q}_1$ integral in the coordinate system 
where $\tilde{\bf q}$ lies along the first axis
and $\tilde{\bf k}$ lies in the $12$ plane. 
The key vectors are expressible in the four dimensional spherical coordinates 
with $ \theta_{1}\in [0,\pi]$, $\varphi_{1}\in [0,\pi]$, and $\phi_{1}\in[0, 2\pi]$ as
\begin{align}
&\,\tilde{\bf q}=\begin{bmatrix}1 \\ 0 \\ 0 \\ 0 \end{bmatrix},\hspace{5mm}
\tilde{\bf k}=\begin{bmatrix}\tilde{k}\cos\theta_q \\ \tilde{k}\sin\theta_q \\ 0 \\ 0 \end{bmatrix},\hspace{5mm}
\tilde{\bf q}_1=\begin{bmatrix}\cos\theta_1 \\ \sin\theta_1 \cos\varphi_1 \\ \sin\theta_1 \sin\varphi_1\cos\phi_1   \\ \sin\theta_1 \sin\varphi_1\sin\phi_1 \end{bmatrix},
\notag \\
&\,\tilde{\bf q}'\equiv \tilde{\bf q}+\tilde{\bf k}=\begin{bmatrix}1+\tilde{k}\cos\theta_q \\ \tilde{k}\sin\theta_q \\ 0 \\ 0 \end{bmatrix}
=\begin{bmatrix}\tilde{q}'\cos\theta_{q'q} \\ \tilde{q}'\sin\theta_{q'q} \\ 0 \\ 0 \end{bmatrix},
\label{SpheC}
\end{align}
where $\theta_{q'q}$ is the angle between $(\tilde{\bf q}',\tilde{\bf q})$,
and $\tilde{q}'$ is given in terms of $(\tilde{k},\theta_q)$ as
\begin{align}
\tilde{q}'=(1+2\tilde{k}\cos\theta_q+\tilde{k}^2)^{1/2} .
\label{q'}
\end{align}
Using Eq.\ (\ref{SpheC}) and the corresponding Jacobian $\sin^2\theta_{1}\sin\varphi_{1}$ for the $\tilde{\bf q}_1$ integral, 
we can transform Eq.\ (\ref{tW^(2b4e3)-0}) into
\begin{align}
&\,\delta\tilde{W}_{\infty}^{(2{\rm b}4{\rm e}3)}(\tilde{k}) 
\notag \\
=&\, 6g_*^2 \int_0^1 \frac{d\lambda}{\lambda}\Biggl[\int_0^{\pi} d\theta_q f_0(\theta_q)\int_0^{\xi_\lambda} d\theta_1 f_{0}(\theta_1)
\notag \\
&\,\times 
\frac{1}{2}\int_{-1}^1ds_1 \frac{\varTheta(\cos\theta_{q'1}-\cos\xi_{\lambda\tilde{q}'})}{1+\lambda^2\tilde{q}^{\prime 2}+2\lambda\tilde{q}'\cos\theta_{q'1}}
-(\lambda\rightarrow 0)\Biggr] .
\label{tW^(2b4e3)-01}
\end{align}
Here $f_0$ and $\xi_\lambda$ are given by Eqs.\ (\ref{f_k}) and (\ref{xi_k-def}), respectively, 
$s_1$ denotes $s_1 \equiv \cos\varphi_1$,
and $\theta_{q'1}$ is the angle between $(\tilde{\bf q}',\tilde{\bf q}_1)$ that satisfies
\begin{align}
\cos\theta_{q'1}=\cos\theta_{q'q}\cos\theta_1+s_1 \sin\theta_{q'q}\sin\theta_1,
\label{theta_q'1}
\end{align}
as seen from Eq.\ (\ref{SpheC}).

\begin{figure}[b]
\begin{center}
\includegraphics[width=0.9\linewidth]{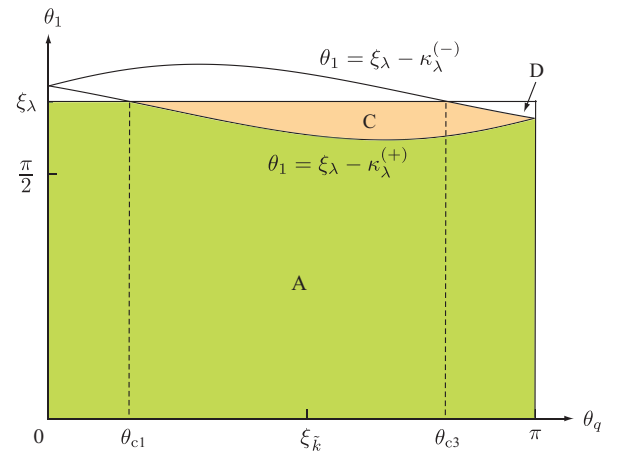}
\end{center}
\caption{Distinct regions of the double integral over $(\theta_q,\theta_1)$; we set
$(\lambda,\tilde{k})=(0.9,0.2)$ to see the basic features clearly. 
The ranges of integration over $s_1$ for regions A and C are $s_1\in [-1,1]$ and $s_1\in [s_{{\rm c}2},1]$, respectively. 
Region C disappears as $\tilde{k}\rightarrow 0$.
\label{Fig5}}
\end{figure}
We can draw Fig.\ \ref{Fig5} that divides the $(\theta_q,\theta_1)$ plane into two regions
according to the range of integration over $s_1$: region A with $s_1\in [-1,1]$ and region C with $s_1\in [s_{{\rm c}2},1]$, 
where $s_{{\rm c}2}$ is defined as the solution of the equation $(\cos\theta_{q'1}-\cos\xi_{\lambda\tilde{q}'})\bigr|_{s_1=s_{{\rm c}2}}=0$
given explicitly by
\begin{align}
s_{{\rm c}2}= -\frac{\cos\theta_{q'q}\cos\theta_1-\cos\xi_{\lambda\tilde{q}'}}{\sin\theta_{q'q}\sin\theta_1} .
\label{s_c2}
\end{align}
The boundary of C is determined partly by $s_{{\rm c}2}=\pm  1$, which can be solved as $\theta_1=\xi_{\lambda\tilde{q}'}\pm \theta_{q'q}$.
They yield the two curves in Fig.\ \ref{Fig5}, which are expressed alternatively in terms of the function $\kappa_\lambda^{(\pm)}=\kappa_\lambda^{(\pm)}(\theta_q,\tilde{k})$ defined by
\begin{align}
\kappa_\lambda^{(\pm)}\equiv \xi_\lambda-\xi_{\lambda\tilde{q}'}\pm\theta_{q'q} 
\label{kappa^(pm)-1}
\end{align}
for convenience.
The quantities  $\theta_{{\rm c}1}\!=\!\theta_{{\rm c}1}(\lambda,\tilde{k})$ and $\theta_{{\rm c}3}\!=\!\theta_{{\rm c}3}(\lambda,\tilde{k})$ in Fig.\ \ref{Fig5} are solutions to the equations
$\kappa_\lambda^{(+)}(\theta_{{\rm c}1},\tilde{k})=0$ and  $\kappa_\lambda^{(-)}(\theta_{{\rm c}3},\tilde{k})=0$,
which can be solved analytically as
\begin{align}
\theta_{{\rm c}1}=\xi_\lambda-\pi+\xi_{\lambda\tilde{k}},\hspace{5mm}
\theta_{{\rm c}3}=\pi+\xi_{\lambda\tilde{k}}-\xi_\lambda .
\label{theta_cj}
\end{align}
The expression of $\theta_{{\rm c}1}$, for example, has been obtained by: (i) expressing $\kappa_\lambda^{(+)}(\theta_{{\rm c}1},\tilde{k})=0$
as $\tilde{q}'\cos(\xi_\lambda+\theta_{q'q})=\tilde{q}'\cos\xi_{\lambda\tilde{q}'}$;
(ii) writing $\tilde{q}'\cos\xi_{\lambda\tilde{q}'}=\tilde{q}^{\prime 2}\cos\xi_\lambda$,
$\tilde{q}'\cos\theta_{q'q}=1+\tilde{k}\cos\theta_q$, $\tilde{q}'\sin\theta_{q'q}=\tilde{k}\sin\theta_q$, and $\tilde{k}\cos\xi_\lambda=\cos\xi_{\lambda\tilde{k}}$
based on Eqs.\ (\ref{xi_k-def}) and (\ref{SpheC}); and (iii) noting $\theta_{{\rm c}1}\in[0,\frac{\pi}{2}]$.

On the basis of these considerations, 
we can perform the integration over $s_1$ in Eq.\ (\ref{tW^(2b4e3)-01}) elementarily.
To express the result concisely, it is convenient to introduce three local functions for considering
2{\rm b}-4{\rm e} contributions by
\begin{subequations}
\label{tJ_x^(pm)-J_x-2b4e}
\begin{align}
&\,\tilde{J}_{\lambda\tilde{k}}^{(\pm)}(\lambda,\theta_q,\theta_1)\equiv  \frac{1}{2} \int_{s_{{\rm c}2}}^{\mp1}\frac{ds_1}{
1+\lambda^2\tilde{q}^{\prime 2}+2\lambda\tilde{q}'\cos\theta_{q'1}}
\notag \\
=&\, \frac{\ln \Bigl\{1\!+\!\lambda^2 \!+\!2\lambda\cos\theta_1\!+\!2\lambda\tilde{k}\bigl[\lambda\cos\theta_q
\!+\!\cos(\theta_q\!\pm\!\theta_1) \bigr]\!+\!\lambda^2\tilde{k}^2\Bigr\}}{4\lambda\tilde{k}\sin\theta_{q}\sin\theta_1},
\label{tJ_x^(pm)-2b4e}
\end{align}
\begin{align}
J_{\lambda\tilde{k}}(\lambda,\theta_q,\theta_1)\equiv &\, 
\frac{1}{2} \int_{-1}^{1}\frac{ds_1}{
1+\lambda^2\tilde{q}^{\prime 2}+2\lambda\tilde{q}'\cos\theta_{q'1}}
\notag \\
=&\, \tilde{J}_{\lambda\tilde{k}}^{(-)}(\lambda,\theta_q,\theta_1)
-\tilde{J}_{\lambda\tilde{k}}^{(+)}(\lambda,\theta_q,\theta_1)  ,
\label{J_x-2b4e}
\end{align}
\end{subequations}
where the second expression of Eq.\ (\ref{tJ_x^(pm)-2b4e}) has been obtained by substituting Eq.\ (\ref{s_c2}), 
performing the integration over $s_1$,
and using Eqs.\ (\ref{s_c2}), (\ref{xi_k-def}), (\ref{SpheC}) and (\ref{q'}) successively.
Now, we can write Eq.\ (\ref{tW^(2b4e3)-01}) as
\begin{align}
&\, \delta\tilde{W}_{\infty}^{(2{\rm b}4{\rm e}3)}(\tilde{k}) 
\notag \\
=&\, 6g_*^2 \int_0^1 \frac{d\lambda}{\lambda}\Biggl[ \int\!\!\!\int_{\rm A} d\theta_q d\theta_1 f_0(\theta_q) f_{0}(\theta_1)J_{\lambda\tilde{k}}(\lambda,\theta_q,\theta_1)
\notag \\
&\,
+\int\!\!\!\int_{\rm C} d\theta_q d\theta_1 f_0(\theta_q) f_{0}(\theta_1)\tilde{J}_{\lambda\tilde{k}}^{(-)}(\lambda,\theta_q,\theta_1)
-(\lambda\rightarrow 0)\Biggr] .
\label{tW^(2b4e3)-02}
\end{align}
This expression needs a further improvement before differentiating it with respect to $\tilde{k}$.
Specifically, we express regions A and C in Fig.\ \ref{Fig5} as $A=({\rm A}+{\rm C}+{\rm D})-({\rm C}+{\rm D})$
and $C=({\rm C}+{\rm D})-{\rm D}$, subsequently
combine the contributions of $({\rm C}+{\rm D})$, and use Eq.\ (\ref{J_x-2b4e}).
We can thereby transform Eq.\ (\ref{tW^(2b4e3)-02}) into
\begin{align}
&\,\delta\tilde{W}_{\infty}^{(2{\rm b}4{\rm e}3)}(\tilde{k}) 
\notag \\
=&\, 6g_*^2 \int_0^1 \frac{d\lambda}{\lambda}\Biggl[\int_0^{\pi} d\theta_q\int_0^{\xi_{\lambda}}d\theta_1 f_0(\theta_q) f_0(\theta_1)
J_{\lambda\tilde{k}}(\lambda,\theta_q,\theta_1)
\notag \\
&\,
+\int_{\theta_{{\rm c}1}}^\pi d\theta_q  \bar{f}_0(\theta_q){\cal J}_{\tilde{k}}^{(+)}(\lambda,\theta_q)
-\int_{\theta_{{\rm c}3}}^\pi d\theta_q \bar{f}_0(\theta_q) 
{\cal J}_{\tilde{k}}^{(-)}(\lambda,\theta_q)
\notag \\
&\,-(\lambda\rightarrow 0)\Biggr],
\label{tW^(2b4e3)}
\end{align}
where  ${\cal J}_{\tilde{k}}^{(\pm)}(\lambda,\theta_q)$ is defined by
\begin{align}
{\cal J}_{\tilde{k}}^{(\pm)}(\lambda,\theta_q)\equiv \sin\theta_q \int_{\xi_\lambda-\kappa_\lambda^{(\pm)}}^{\xi_\lambda}d\theta_1f_0(\theta_1)\tilde{J}_{\lambda\tilde{k}}^{(\pm)}(\lambda,\theta_q,\theta_1).
\label{calJ^(pm)}
\end{align}
The contribution of Eq.\ (\ref{tW^(2b4e3)}) to Eq.\ (\ref{eta-def2}) is obtained 
by differentiating Eq.\ (\ref{tW^(2b4e3)}) twice with respect to $\tilde{k}$ and setting $\tilde{k}=0$ subsequently.
Terms with derivatives of $\theta_{{\rm c}j}$ ($j=1,3)$ all vanish due to ${\cal J}_{\tilde{k}}^{(+)}(\lambda,\theta_{{\rm c}1})=
{\cal J}_{\tilde{k}}^{(-)}(\lambda,\theta_{{\rm c}3})=0$ for Eq.\ (\ref{calJ^(pm)}), which result from
$\kappa_\lambda^{(+)}(\theta_{{\rm c}1},\tilde{k})=\kappa_\lambda^{(-)}(\theta_{{\rm c}3},\tilde{k})=0$.
Also using Eqs.\ (\ref{xi_k-def}) and (\ref{theta_cj}), we obtain
\begin{align}
&\,\eta^{(2{\rm b}4{\rm e}3)}
\notag \\
=&\, 3g_*^2 \int_0^1 \frac{d\lambda}{\lambda}\Biggl[\int_0^{\xi_{\lambda}}d\theta_1 f_0(\theta_1) \int_0^{\pi} d\theta_q f_0(\theta_q)
 \lambda^2 J_{0}^{(2)}(\lambda,\theta_q,\theta_1)
\notag \\
&\,
+\int_{\xi_\lambda-\frac{\pi}{2}}^\pi d\theta_q  \bar{f}_0(\theta_q){\cal J}_0^{(+2)}(\lambda,\theta_q)
\notag \\
&\,
-\int_{\frac{3\pi}{2}-\xi_\lambda}^\pi d\theta_q \bar{f}_0(\theta_q) 
{\cal J}_0^{(- 2)}(\lambda,\theta_q)\Biggr],
\label{eta^(2b4e3)-0}
\end{align}
where $J_{0}^{(2)}(\lambda,\theta_q,\theta_1)\!\equiv\! \partial^2 J_x(\lambda,\theta_q,\theta_1)/\partial x^2\bigl|_{x=0}$,
${\cal J}_0^{(\pm2)}(\lambda,\theta_q)\!\equiv\! \partial^2 {\cal J}_{\tilde{k}}^{(\pm)}(\lambda,\theta_q)/\partial\tilde{k}^2\bigl|_{\tilde{k}=0}$,
and $\bar{f}_0(\theta)$ is defined in terms of Eq.\ (\ref{f_k}) more generally by
\begin{align}
\bar{f}_x(\theta)\equiv \frac{f_x(\theta)}{\sin\theta}=\frac{2}{\pi}\frac{\sin\theta}{1+x^2+2x\cos\theta}.
\label{barf_k}
\end{align}
Since the integrand turns out to vanish in the limit, we have removed $(\lambda\rightarrow 0)$ from Eq.\ (\ref{eta^(2b4e3)-0}).
The coefficient $J_{0}^{(2)}$ in Eq.\ (\ref{eta^(2b4e3)-0}) can be calculated straightforwardly from Eq.\ (\ref{tJ_x^(pm)-J_x-2b4e}) as
\begin{align}
&\, J_0^{(2)}(\lambda,\theta_q,\theta_1)
\notag \\
=&\,
\frac{2}{(1+\lambda^2+2\lambda\cos\theta_1)^2}\left[-1+\frac{4(\lambda+\cos\theta_1)^2}{1+\lambda^2+2\lambda\cos\theta_1}\cos^2\theta_q
\right.
\notag \\
&\,\left.
+ \frac{4\sin^2\theta_1}{3(1+\lambda^2+2\lambda\cos\theta_1)}\sin^2\theta_q\right] ,
\label{J_0^(2)}
\end{align}
while ${\cal J}_0^{(\pm2)}$ is obtained in Appendix \ref{App: A1}
as
\begin{align}
&\,{\cal J}^{(\pm 2)}_0(\lambda,\theta_q)
\notag \\
=&\, -\frac{1}{\pi}\Biggl[\lambda\cos(\theta_q\mp\xi_\lambda) \cos2(\theta_q\mp\xi_\lambda)
+\frac{\cos^2(\theta_q\mp\xi_\lambda)}{\sin\xi_\lambda} 
\notag \\
&\,
\times \sin(3\xi_\lambda\mp\theta_q)-\frac{4}{3}\cos\xi_\lambda\cos^3(\theta_q\mp\xi_\lambda)\Biggr] .
\label{calJ^(pm2)-2b4e}
\end{align}
Substituting them into Eq.\ (\ref{eta^(2b4e3)-0}), we find that the contribution of $J_0^{(2)}$ vanishes 
upon the integration over $\theta_q$.
Moreover, the two integrals of ${\cal J}^{(\pm 2)}_0$ in Eq.\ (\ref{eta^(2b4e3)-0}) can be combined 
by using the symmetry ${\cal J}^{(- 2)}_0(\lambda,\theta_q)=-{\cal J}^{(+2)}_0(\lambda,\pi-\theta_q)$
into a single integral over $\theta_q\in[0,\pi]$. 
Equation (\ref{eta^(2b4e3)-0}) is thereby transformed into
\begin{align}
\eta^{(2{\rm b}4{\rm e}3)}
=&\, 3g_*^2 \int_0^1 \frac{d\lambda}{\lambda}\int_0^\pi d\theta_q  \bar{f}_0(\theta_q){\cal J}_0^{(+2)}(\lambda,\theta_q) .
\label{eta^(2b4e3)-1}
\end{align}
The double integral can be evaluated both numerically and analytically. 
We obtain
\begin{align}
\eta^{(2{\rm b}4{\rm e}3)}= -\frac{1}{8}g_*^2 .
\label{eta^(2b4e3)}
\end{align}

Fourth, we focus on the fourth term in the curly brackets of Eq.\ (\ref{tW^(2b4e)}).
The calculation of this term also requires a new and lengthy treatment, 
but we will eventually arrive at the simple analytic result of Eq.\ (\ref{eta^(2b4e4)}) below.
We start by expressing the contribution in the coordinate system of Eq.\ (\ref{SpheC})
as
\begin{align}
&\,\delta\tilde{W}_{\infty}^{(2{\rm b}4{\rm e}4)}(\tilde{k}) 
\notag \\
=&\, 6g_*^2 \int_0^1 \frac{d\lambda}{\lambda}\int_0^{\pi} d\theta_q f_0(\theta_q)\int_0^{\pi} d\theta_1f_0(\theta_1)
\notag \\
&\,\times
\frac{1}{2}\int_{-1}^1ds_1 \frac{\varTheta(\cos\theta_{k1}-\cos\xi_{\lambda\tilde{k}})\varTheta(\cos\theta_{q'1}-\cos\xi_{\lambda\tilde{q}'})}{1+\lambda^2\tilde{q}^{\prime 2}+2\lambda\tilde{q}'\cos\theta_{q'1}},
\label{tW^(2b4e4)-01}
\end{align}
where $f_0$ and $\xi_\lambda$ are given by Eqs.\ (\ref{f_k}) and (\ref{xi_k-def}), respectively, $s_1$ denotes $s_1 \equiv \cos\varphi_1$,
$\theta_{q'1}$ is given by Eq.\ (\ref{theta_q'1}), and $\theta_{k1}$ is the angle between $(\tilde{\bf k},\tilde{\bf q}_1)$ that satisfies
\begin{align}
\cos\theta_{k1}=\cos\theta_{q}\cos\theta_1+s_1 \sin\theta_{q}\sin\theta_1,
\label{theta_k1}
\end{align}
as seen from Eq.\ (\ref{SpheC}).

\begin{figure}[t]
\begin{center}
\includegraphics[width=0.9\linewidth]{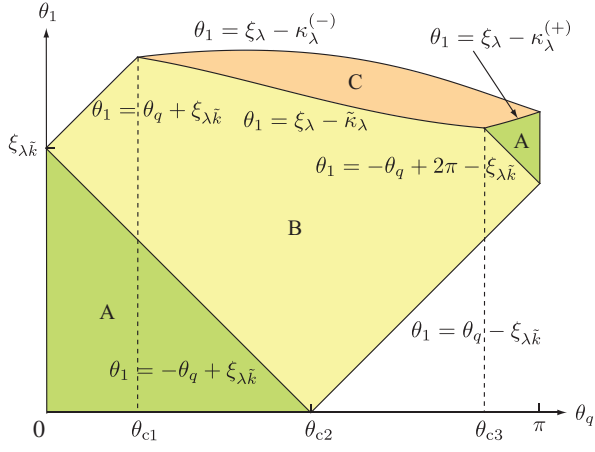}
\end{center}
\caption{Distinct regions of the double integral over $(\theta_q,\theta_1)$ for $(\lambda,\tilde{k})=(0.9,0.25)$.
The ranges of integration over $s_1$ for regions A, B, and C are
$s_1\in [-1,1]$, $s_1\in [s_{{\rm c}1},1]$, and $s_1\in [s_{{\rm c}2},1]$, respectively. 
Region C disappears as $\tilde{k}\rightarrow 0$.
\label{Fig6}}
\end{figure}
We can draw Fig.\ \ref{Fig6} that divides the $(\theta_q,\theta_1)$ plane into three regions
according to the range of integration over $s_1$ in Eq.\ (\ref{tW^(2b4e4)-01}):  region A with $s_1\in [-1,1]$, region B with $s_1\in [s_{{\rm c}1},1]$,
and region C with $s_1\in [s_{{\rm c}2},1]$, where $s_{{\rm c}2}$ is given by Eq.\ (\ref{s_c2}), 
and $s_{{\rm c}1}$ is defined by $(\cos\theta_{k1}-\cos\xi_{\lambda\tilde{k}})\bigr|_{s_1=s_{{\rm c}1}}=0$, i.e., 
\begin{align}
s_{{\rm c}1}\equiv -\frac{\cos\theta_q\cos\theta_1-\cos\xi_{\lambda\tilde{k}}}{\sin\theta_q\sin\theta_1} .
\label{s_c1}
\end{align}
The boundary of B is determined partly by  $s_{{\rm c}1}=\pm  1$,  which is solved as $\theta_1=\theta_{q}\pm \xi_{\lambda\tilde{k}}$,
$\xi_{\lambda\tilde{k}}-\theta_{q}$, $2\pi-\xi_{\lambda\tilde{k}}-\theta_{q}$.
The boundary of C is determined partly by $s_{{\rm c}2}=\pm  1$ in terms of Eq.\ (\ref{s_c2}), 
which have yielded $\theta_1=\xi_{\lambda\tilde{q}'}\pm \theta_{q'q}$;
these curves in Fig.\ \ref{Fig6} are expressed conveniently in terms of $\kappa_\lambda^{(\pm)}$ defined by 
Eq.\ (\ref{kappa^(pm)-1}) similarly as in Fig.\ \ref{Fig5}.
There is another nontrivial one, i.e., the B-C boundary determined by $s_{{\rm c}1}=s_{{\rm c}2}$,
which can be transformed by using Eq.\ (\ref{xi_k-def}) and the last equality of Eq.\ (\ref{SpheC}) into 
$\cos\theta_1=\tilde{q}'\cos\xi_{\lambda\tilde{q}'}-\tilde{k}\cos\xi_{\lambda\tilde{k}}
=-\frac{\lambda}{2}(1+2\tilde{k}\cos\theta_q)$, i.e., $\theta_1=\arccos\bigl[(1+2\tilde{k}\cos\theta_q)\cos\xi_\lambda\bigr]$.
Introducing the function $\tilde{\kappa}_\lambda\equiv \tilde{\kappa}_\lambda(\theta_q,\tilde{k})$ by
\begin{align}
\tilde{\kappa}_\lambda\equiv &\, \xi_\lambda-\arccos\bigl[(1+2\tilde{k}\cos\theta_q)\cos\xi_\lambda\bigr] ,
\label{tkappa}
\end{align}
we can express the B-C boundary as $\theta_1=\xi_\lambda-\tilde{\kappa}_\lambda$.
On the other hand,  $\theta_{{\rm c}1}$ and $\theta_{{\rm c}3}$ in Fig.\ \ref{Fig6} are solutions to the equations
\begin{align*}
\xi_\lambda-\tilde{\kappa}_\lambda(\theta_{{\rm c}1},\tilde{k})=&\,\theta_{{\rm c}1}+\xi_{\lambda\tilde{k}},
\\
\xi_\lambda-\tilde{\kappa}_\lambda(\theta_{{\rm c}3},\tilde{k})=&\,2\pi-\theta_{{\rm c}3}-\xi_{\lambda\tilde{k}} .
\end{align*}
They can be solved analytically to yield Eq.\ (\ref{theta_cj}) once again; for example, the equation
for $\theta_{{\rm c}3}\in [\frac{\pi}{2},\pi]$ is expressible by using Eq.\ (\ref{tkappa}) 
as $(1+2\tilde{k}\cos\theta_{{\rm c}3})\cos\xi_\lambda=\cos(2\pi-\theta_{{\rm c}3}-\xi_{\lambda\tilde{k}})$,
from which we easily obtain $\theta_{{\rm c}3}=\pi+\xi_{\lambda\tilde{k}}-\xi_\lambda$. Another angle $\theta_{{\rm c}2}$ in Fig.\ \ref{Fig6} is given simply by
\begin{align}
\theta_{{\rm c}2}=\xi_{\lambda\tilde{k}} .
\label{theta_c2}
\end{align}

On the basis of these considerations and using Eq.\ (\ref{theta_q'1}), 
we can perform the integration over $s_1$ in Eq.\ (\ref{tW^(2b4e4)-01}) elementarily.
To express the result concisely, it is convenient to introduce additional local functions for considering
 2b-4e contributions by
\begin{align}
&\,J_{\lambda\tilde{k}}^{(\pm)}(\lambda,\theta_q,\theta_1)
\equiv  \frac{1}{2} \int_{s_{{\rm c}1}}^{\mp1}
\frac{ds_1}{1+\lambda^2\tilde{q}^{\prime 2}+2\lambda\tilde{q}'\cos\theta_{q'1}}
\notag \\
=&\,\tilde{J}_{\lambda\tilde{k}}^{(\pm)}(\lambda,\theta_q,\theta_1)
- \frac{\ln \Bigl(
1\!+\!\lambda^2\!+\!2\lambda\cos\theta_1\!+\!2\lambda^2\tilde{k}\cos\theta_q\Bigr)}{4\lambda\tilde{k}\sin\theta_{q}\sin\theta_1},
\label{J_x^(pm)-2b4e}
\end{align}
where $\tilde{J}_{\lambda\tilde{k}}^{(\pm)}$ are defined by Eq.\ (\ref{tJ_x^(pm)-2b4e}),
and the second term originates from the lower bound of the $s_1$ integral, which we have transformed by using Eqs.\ (\ref{SpheC}), 
(\ref{q'}), (\ref{s_c1}), and (\ref{xi_k-def}) successively.
Now, Eq.\ (\ref{tW^(2b4e4)-01}) can be written in terms of the functions in Eqs.\ (\ref{tJ_x^(pm)-J_x-2b4e}) and (\ref{J_x^(pm)-2b4e}) as
\begin{align}
&\, \delta\tilde{W}_{\infty}^{(2{\rm b}4{\rm e}4)}(\tilde{k}) 
\notag \\
=&\, 6g_*^2 \int_0^1 \frac{d\lambda}{\lambda}\Biggl[ \int\!\!\!\int_{\rm A} d\theta_q d\theta_1 f_0(\theta_q) f_{0}(\theta_1)J_{\lambda\tilde{k}}(\lambda,\theta_q,\theta_1)
\notag \\
&\,
+\int\!\!\!\int_{\rm B} d\theta_q d\theta_1 f_0(\theta_q) f_{0}(\theta_1)J_{\lambda\tilde{k}}^{(-)}(\lambda,\theta_q,\theta_1)\notag \\
&\,
+\int\!\!\!\int_{\rm C} d\theta_q d\theta_1 f_0(\theta_q) f_{0}(\theta_1)\tilde{J}_{\lambda\tilde{k}}^{(-)}(\lambda,\theta_q,\theta_1)
-(\lambda\rightarrow 0)\Biggr] .
\label{tW^(2b4e4)-02}
\end{align}
Let us express the integral over region C as
\begin{align*}
\int\!\!\!\int_{\rm C} d\theta_q d\theta_1
=&\,\int_{\theta_{{\rm c}1}}^{\theta_{{\rm c}3}}d\theta_q\int_{\xi_\lambda-\tilde{\kappa}_\lambda}^{\xi_\lambda}d\theta_1
-\int_{\theta_{{\rm c}1}}^{\pi}d\theta_q\int_{\xi_\lambda-\kappa_\lambda^{(-)}}^{\xi_\lambda}d\theta_1
\notag \\
&\,
+\int_{\theta_{{\rm c}3}}^{\pi}d\theta_q\int_{\xi_\lambda-\kappa_\lambda^{(+)}}^{\xi_\lambda}d\theta_1.
\end{align*}
Subsequently, we write the integrand of the third term as $\tilde{J}_{\lambda\tilde{k}}^{(-)}=
\tilde{J}_{\lambda\tilde{k}}^{(+)}+J_{\lambda\tilde{k}}$
based on Eq.\ (\ref{J_x-2b4e}), and combine its $J_{\lambda\tilde{k}}$ contribution with that of region A.
We can thereby express Eq.\ (\ref{tW^(2b4e4)-02}) as
\begin{align}
&\,\delta\tilde{W}_{\infty}^{(2{\rm b}4{\rm e}4)}(\tilde{k}) 
\notag \\
=&\, 6g_*^2 \int_0^1 \frac{d\lambda}{\lambda}\Biggl[\Biggl(\int_0^{\theta_{{\rm c}2}} d\theta_q \int_0^{\xi_{\lambda\tilde{k}}-\theta_q}d\theta_1
+\int_{\theta_{{\rm c}3}}^\pi d\theta_q \int_{2\pi-\xi_{\lambda\tilde{k}}-\theta_q}^{\xi_\lambda}d\theta_1 \Biggr)  
\notag \\
&\,\times f_0(\theta_q) f_0(\theta_1) J_{\lambda\tilde{k}}(\lambda,\theta_q,\theta_1)
\notag \\
&\, 
+\Biggl(\int_0^{\theta_{{\rm c}1}} d\theta_q \int_{\xi_{\lambda\tilde{k}}-\theta_q}^{\xi_{\lambda\tilde{k}}+\theta_q} d\theta_1 
+\int_{\theta_{{\rm c}1}}^{\theta_{{\rm c}2}} d\theta_q \int_{\xi_{\lambda\tilde{k}}-\theta_q}^{\xi_\lambda-\tilde{\kappa}_\lambda} d\theta_1 
\notag \\
&\,+\int_{\theta_{{\rm c}2}}^{\theta_{{\rm c}3}} d\theta_q \int_{\theta_q-\xi_{\lambda\tilde{k}}}^{\xi_\lambda-\tilde{\kappa}_\lambda} d\theta_1
+\int_{\theta_{{\rm c}3}}^\pi d\theta_q \int_{\theta_q-\xi_{\lambda\tilde{k}}}^{2\pi-\xi_{\lambda\tilde{k}}-\theta_q} d\theta_1 \Biggr) 
\notag \\
&\,\times f_0(\theta_q) f_0(\theta_1) J_{\lambda\tilde{k}}^{(-)}(\lambda,\theta_q,\theta_1)
\notag \\
&\, +\int_{\theta_{{\rm c}1}}^{\theta_{{\rm c}3}} d\theta_q  \bar{f}_0(\theta_q)\tilde{\cal J}_{\tilde{k}}(\lambda,\theta_q)
-\int_{\theta_{{\rm c}1}}^\pi d\theta_q  \bar{f}_0(\theta_q){\cal J}_{\tilde{k}}^{(-)}(\lambda,\theta_q)
\notag \\
&\,
+\int_{\theta_{{\rm c}3}}^\pi d\theta_q \bar{f}_0(\theta_q) {\cal J}_{\tilde{k}}^{(+)}(\lambda,\theta_q)\Biggr],
\label{tW^(2b4e4)}
\end{align}
where ${\cal J}_{\tilde{k}}^{(\pm)}$ are given in Eq.\ (\ref{calJ^(pm)}), and 
$\tilde{\cal J}_{\tilde{k}}$ is defined by
\begin{align}
\tilde{\cal J}_{\tilde{k}}(\lambda,\theta_q)\equiv \sin\theta_q \int_{\xi_\lambda-\tilde{\kappa}_\lambda}^{\xi_\lambda}d\theta_1f_0(\theta_1)\tilde{J}_{\lambda\tilde{k}}^{(-)}(\lambda,\theta_q,\theta_1).
\label{tcalJ}
\end{align}

Let us differentiate Eq.\ (\ref{tW^(2b4e4)}) twice with respect to $\tilde{k}$, set $\tilde{k}=0$, 
substitute the resulting expression into Eq.\ (\ref{eta-def2}),
and evaluate the integrals to obtain $\eta^{(2{\rm b}4{\rm e}4)}$. 
The process of this tedious calculation is outlined as follows:
(i) The contributions originating from $J_{0}^{(2)}(\lambda,\theta_q,\theta_1)\!\equiv\! 
\partial^2 J_x(\lambda,\theta_q,\theta_1)/\partial x^2\bigl|_{x=0}$ in region A
and $J_{0}^{(-2)}(\lambda,\theta_q,\theta_1)\!\equiv\! \partial^2 J_x^{(-)}(\lambda,\theta_q,\theta_1)/\partial x^2\bigl|_{x=0}$ in region B
can be combined, by using $J_0^{(2)}=J_0^{(-2)}-J_0^{(+2)}$
and $J_{0}^{(+2)}(\lambda,\theta_q,\theta_1)\!\equiv\! \partial^2 J_x^{(+)}(\lambda,\theta_q,\theta_1)/\partial x^2\bigl|_{x=0}
=-J_{0}^{(-2)}(\lambda,\pi-\theta_q,\theta_1)$, 
into a single integral of $J_{0}^{(-2)}$ over $\theta_q\in [0,\pi]$ and $\theta_1\in [0,\xi_\lambda]$,
where $J_{0}^{(\pm 2)}$ are obtained from Eq.\ (\ref{J_x^(pm)-2b4e}) as
\begin{align}
&\, J_{0}^{(\pm2)}(\lambda,\theta_q,\theta_1)
\notag \\
=&\,\frac{1}{\sin\theta_{q}\sin\theta_1}\Biggl\{-\frac{\lambda\cos\theta_q+\cos(\theta_q\pm \theta_1)}{(1+\lambda^2+2\lambda\cos\theta_1)^2}
\notag \\
&\,+\frac{4\lambda\cos\theta_q\cos(\theta_q\pm\theta_1)\bigl[
\lambda\cos\theta_q+\cos(\theta_q\pm\theta_1)\bigr]}{(1+\lambda^2+2\lambda\cos\theta_1)^3}
\notag \\
&\, +\frac{4\cos^3(\theta_q\pm \theta_1)}{3(1+\lambda^2+2\lambda\cos\theta_1)^3}\Biggr\} .
\label{J_0^-2}
\end{align}
Performing integration over $\theta_q\in[0,\pi]$ first,
one can show that this contribution of $J_{0}^{(-2)}$ vanishes.
(ii) Terms with derivatives of $\theta_{{\rm c}j}$ ($j=1,2,3)$ vanish.
Specifically, they disappear from the first derivative of $\delta\tilde{W}_{\infty}^{(2{\rm b}4{\rm e}4)}(\tilde{k})$, 
as shown based on Eq.\ (\ref{tW^(2b4e4)-02}).
They also vanish from the second derivative of $\delta\tilde{W}_{\infty}^{(2{\rm b}4{\rm e}4)}$ 
after setting $\tilde{k}=0$, as shown by using 
$J_0^{(-)}(\lambda,\theta_q,\theta_q\pm\frac{\pi}{2})=0$, 
$J_0^{(+)}(\lambda,\theta_q,\frac{\pi}{2}- \theta_q)=J_0^{(+)}(\lambda,\theta_q,\frac{3\pi}{2}- \theta_q)=0$,
$\kappa_\lambda^{(-)}\!=\!\tilde{\kappa}_\lambda$ at $\theta_q\!=\!\theta_{{\rm c}1}$, 
$\kappa_\lambda^{(+)}\!=\!\tilde{\kappa}_\lambda$ at $\theta_q\!=\!\theta_{{\rm c}3}$, 
and $J_x=\tilde{J}_x^{(-)}-\tilde{J}_x^{(+)}=
J_x^{(-)}-J_x^{(+)}$.
(iii) The four boundary lines of region B depend on $\tilde{k}$ through $\xi_{\lambda\tilde{k}}$ given by Eq.\ (\ref{xi_k-def}).
However, the contributions with $\partial\xi_{\lambda\tilde{k}}/\partial\tilde{k}\bigr|_{\tilde{k}=0}=\lambda/2$ 
can be shown to cancel out exactly by using $J_x-J_x^{(-)}=-J_x^{(+)}$ and making a change of variables $\theta_q\rightarrow \pi-\theta_q$ 
in the integrals over $\theta_q\geq \frac{\pi}{2}$.
(iv) The contribution with $\tilde{\kappa}_\lambda^{(n)}\equiv \partial^n\tilde{\kappa}_\lambda/\partial\tilde{k}^n\bigr|_{\tilde{k}=0}$ $(n=1,2)$ from the B-C boundary exactly cancels the contribution of $\tilde{\cal J}_0^{(2)}(\lambda,\theta_q)\!\equiv\! \partial^2 \tilde{\cal J}_{\tilde{k}}(\lambda,\theta_q)/\partial\tilde{k}^2\bigl|_{\tilde{k}=0}$; see Eqs.\ (\ref{tkappa^(12)}) and (\ref{tcalJ^(2)-2b4e}) for the analytic expressions of $\tilde{\kappa}_\lambda^{(n)}$ and
$\tilde{\cal J}_0^{(2)}$ to confirm the statement.
(v) The remaining contributions are those with ${\cal J}_0^{(\pm 2)}(\lambda,\theta_q)\!\equiv\! \partial^2 {\cal J}_{\tilde{k}}^{(\pm)}(\lambda,\theta_q)/\partial\tilde{k}^2\bigl|_{\tilde{k}=0}$, which can be combined by using ${\cal J}_0^{(- 2)}(\lambda,\theta_q)=-{\cal J}_0^{(+ 2)}(\lambda,\pi-\theta_q)$
into a single integral of ${\cal J}_0^{(+2)}(\lambda,\theta_q)$ over $\theta_q\in[0,\pi]$,
where ${\cal J}_0^{(\pm2)}(\lambda,\theta_q)$ is given by Eq.\ (\ref{calJ^(pm2)-2b4e}).

Summarizing (i)-(v) above, we obtain
\begin{align}
\eta^{(2{\rm b}4{\rm e}4)}
=&\, 3g_*^2 \int_0^1 \frac{d\lambda}{\lambda}\int_0^\pi d\theta_q  \bar{f}_0(\theta_q){\cal J}_0^{(+2)}(\lambda,\theta_q) .
\label{eta^(2b4e4)-4}
\end{align}
This integral is the same as Eq.\ (\ref{eta^(2b4e3)-1}) yielding Eq.\ (\ref{eta^(2b4e3)}). Hence, we have
\begin{align}
\eta^{(2{\rm b}4{\rm e}4)}= -\frac{1}{8}g_*^2 
\label{eta^(2b4e4)}
\end{align}
once again.

Adding Eqs.\ (\ref{eta^(2b4e1)}), (\ref{eta^(2b4e2)}), (\ref{eta^(2b4e3)}), and (\ref{eta^(2b4e4)}),
we obtain the 2b-4e contribution to the coherence exponent as
\begin{align}
\eta^{(2{\rm b}4{\rm e})}= \frac{1}{8}g_*^2 .
\label{eta^(2b4e)}
\end{align}

\subsection{The 2b-4f contribution}

Next, we focus on the diagram of Fig.\ \ref{Fig1} (4f).
It is shown in Appendix\ref{App-2b4f} that its contribution to Eq.\ (\ref{tW^(2b)})
is expressible as
\begin{align}
&\,\delta\tilde{W}_{\infty}^{(2{\rm b}4{\rm f})}(\tilde{k}) 
\notag \\
=&\, -g_*^2
\int_0^1 \frac{d\lambda}{\lambda} \int\frac{d^d\tilde{q}}{(2\pi)^4K_4}
\delta(\tilde{q}-1)\Bigl[\phi_{22}(\lambda\tilde{\bf q},\lambda\tilde{\bf k})
+\phi_{00}(\lambda\tilde{\bf q},\lambda\tilde{\bf k})
\notag \\
&\,
+\phi_{20}(\lambda\tilde{\bf q}+\lambda\tilde{\bf k},\lambda\tilde{\bf q})
+\phi_{20}(\lambda\tilde{\bf q}+\lambda\tilde{\bf k},\lambda\tilde{\bf k})
+\phi_{04}(\lambda\tilde{\bf q}+\lambda\tilde{\bf k},\lambda\tilde{\bf q})
\notag \\
&\, 
+\phi_{04}(\lambda\tilde{\bf q}+\lambda\tilde{\bf k},\lambda\tilde{\bf k})
+2\phi_{022}(\lambda\tilde{\bf q}+\lambda\tilde{\bf k},\lambda\tilde{\bf k},\lambda\tilde{\bf q})
\notag \\
&\, +2\phi_{200}(\lambda\tilde{\bf q}+\lambda\tilde{\bf k},\lambda\tilde{\bf k},\lambda\tilde{\bf q})-(\lambda\rightarrow 0)\Bigr] ,
\label{tW^(2b4f)}
\end{align}
where $\phi_{n_1n_2}(\tilde{\bf k}_1,\tilde{\bf k}_2)$ is given by Eq.\ (\ref{phi^(3d)}), 
and $\phi_{n_1n_2n_3}(\tilde{\bf k}_1,\tilde{\bf k}_2,\tilde{\bf k}_3)$ is defined similarly by
\begin{align}
\phi_{n_1n_2n_3}(\tilde{\bf k}_1,\tilde{\bf k}_2,\tilde{\bf k}_3)
\equiv &\,
\int\frac{d^4 \tilde{q}_1}{(2\pi)^4K_4}
\delta(\tilde{q}_1-1)
\frac{\varTheta(|\tilde{\bf k}_1+\tilde{\bf q}_1|-1)}{
|\tilde{\bf k}_1+\tilde{\bf q}_1|^{n_1}}
\notag \\
&\,\times \frac{\varTheta(|\tilde{\bf k}_2+\tilde{\bf q}_1|-1)\varTheta(|\tilde{\bf k}_3+\tilde{\bf q}_1|-1)}{
|\tilde{\bf k}_2+\tilde{\bf q}_1|^{n_2}|\tilde{\bf k}_3+\tilde{\bf q}_1|^{n_3}} .
\label{phi^(4f)}
\end{align}

Let us consider each term in the square brackets of Eq.\ (\ref{tW^(2b4f)}) separately. 
The first term is the same as the second term in the curly brackets of Eq.\ (\ref{tW^(2b4e)}) except for the numerical factor, 
whose contribution to $\eta$ has already been studied to yield Eq.\ (\ref{eta^(2b4e2)}). 
The contribution of the second term in the square brackets of Eq.\ (\ref{tW^(2b4f)}) can be calculated similarly.
We obtain
\begin{align}
\eta^{(2{\rm b}4{\rm f}1)}=\eta^{(2{\rm b}4{\rm f}2)}=0.
\label{eta^(2b4f1,2)}
\end{align}
The third and fourth terms in the square brackets of Eq.\ (\ref{tW^(2b4f)}) are also
the same as the third and fourth terms in the curly brackets of Eq.\ (\ref{tW^(2b4e)}), respectively, except for the numerical factor,
whose contributions to $\eta$ are given by Eqs.\ (\ref{eta^(2b4e3)}) and (\ref{eta^(2b4e4)}). 
Hence, we can conclude immediately that their contributions  to $\eta$  are given by
\begin{align}
\eta^{(2{\rm b}4{\rm f}3)}=\eta^{(2{\rm b}4{\rm f}4)}=\frac{1}{48}g_*^2 .
\label{eta^(2b4f3,4)}
\end{align}

Fifth, we focus on the fifth term in the square brackets of Eq.\ (\ref{tW^(2b4f)}),
which can be treated in the same way as the 2b-4e3 contribution described 
from Eq.\ (\ref{tW^(2b4e3)-0}) through Eq.\ (\ref{eta^(2b4e3)}). 
We need two modifications due to the change $|\tilde{\bf q}_1+\lambda\tilde{\bf q}'|^2 \rightarrow |\tilde{\bf q}_1+\lambda\tilde{\bf q}|^4$
in the denominator of the integrand. 
The first is to use
\begin{align}
\tilde{f}_\lambda(\theta_1)\equiv \frac{\pi}{2}\frac{\sin^2\theta_1}{(1+\lambda^2+2\lambda\cos\theta_1)^2}
\label{h_lambda}
\end{align}
instead of $f_0(\theta_1)$ in Eq.\ (\ref{tW^(2b4e3)}).
The second is to replace Eqs.\ (\ref{tJ_x^(pm)-2b4e}) and (\ref{J_x-2b4e}) by the local functions
\begin{align}
\tilde{J}_{\lambda\tilde{k}}^{(\pm)}(\lambda,\theta_q,\theta_1)\equiv  \frac{1}{2} \int_{s_{{\rm c}2}}^{\mp1}ds_1
\label{tJ_k-J_k1-2b4f5}
\end{align}
and $J_{\lambda\tilde{k}}\equiv\tilde{J}_{\lambda\tilde{k}}^{(-)}-\tilde{J}_{\lambda\tilde{k}}^{(+)}=1$, respectively.
We thereby obtain $\delta\tilde{W}_{\infty}^{(2{\rm b}4{\rm f}5)}$ in place of Eq.\ (\ref{tW^(2b4e3)}) as
\begin{align}
\delta\tilde{W}_{\infty}^{(2{\rm b}4{\rm f}5)}(\tilde{k}) 
=&\, -g_*^2 \int_0^1 \frac{d\lambda}{\lambda}\Biggl[\int_0^{\pi} d\theta_q\int_0^{\xi_{\lambda}}d\theta_1 f_0(\theta_q) \tilde{f}_\lambda(\theta_1)
\notag \\
&\,
+\int_{\theta_{{\rm c}1}}^\pi d\theta_q  \bar{f}_0(\theta_q){\cal J}_{\tilde{k}}^{(+)}(\lambda,\theta_q)
\notag \\
&\,-\int_{\theta_{{\rm c}3}}^\pi d\theta_q \bar{f}_0(\theta_q) 
{\cal J}_{\tilde{k}}^{(-)}(\lambda,\theta_q)
-(\lambda\rightarrow 0)\Biggr],
\label{tW^(2b4f5)-01}
\end{align}
where  ${\cal J}_{\tilde{k}}^{(\pm)}(\lambda,\theta_q)$ are now given in terms of Eq.\ (\ref{h_lambda}) by
\begin{align}
{\cal J}_{\tilde{k}}^{(\pm)}(\lambda,\theta_q)\equiv \sin\theta_q \int_{\xi_\lambda-\kappa_\lambda^{(\pm)}}^{\xi_\lambda}d\theta_1\tilde{f}_\lambda(\theta_1)\tilde{J}_{\lambda\tilde{k}}^{(\pm)}(\lambda,\theta_q,\theta_1).
\label{calJ^(pm)-2b4f5}
\end{align}
Let us differentiate Eq.\ (\ref{tW^(2b4f5)-01}) twice with respect to $\tilde{k}$, set $\tilde{k}=0$ subsequently, and
substitute the resulting expression into Eq.\ (\ref{eta-def2}).
Terms with derivatives of $\theta_{{\rm c}j}$ ($j=1,3)$ vanish once again owing to ${\cal J}_{\tilde{k}}^{(+)}(\lambda,\theta_{{\rm c}1})=
{\cal J}_{\tilde{k}}^{(-)}(\lambda,\theta_{{\rm c}3})=0$, and we obtain
\begin{align}
\eta^{(2{\rm b}4{\rm f}5)}
=&\, -\frac{g_*^2}{2} \int_0^1 \frac{d\lambda}{\lambda}\Biggl[\int_{\xi_\lambda-\frac{\pi}{2}}^\pi d\theta_q  \bar{f}_0(\theta_q){\cal J}_0^{(+2)}(\lambda,\theta_q)
\notag \\
&\,
-\int_{\frac{3\pi}{2}-\xi_\lambda}^\pi d\theta_q \bar{f}_0(\theta_q) 
{\cal J}_0^{(- 2)}(\lambda,\theta_q)\Biggr] .
\label{eta^(2b4f5)-0}
\end{align}
The coefficients ${\cal J}_0^{(\pm 2)}(\lambda,\theta_q)\equiv \partial^2 {\cal J}_{\tilde{k}}^{(\pm)}(\lambda,\theta_q)/\partial\tilde{k}^2\bigr|_{\tilde{k}=0}$
are obtained as Eq.\ (\ref{calJ_0^(pm2)-2b4f5}) in Appendix\ref{App-calJ-exp}. 
Using the symmetry ${\cal J}^{(-2 )}_0(\lambda,\theta_q)=-{\cal J}^{(+2 )}_0(\lambda,\pi-\theta_q)$, we can
combine the two integrals of Eq.\ (\ref{eta^(2b4f5)-0}) into a single integral with ${\cal J}^{(+2)}_0(\lambda,\theta_q)$
over $\theta_q\in[0,\pi]$, which can be evaluated both numerically and analytically.
We thereby obtain
\begin{align}
\eta^{(2{\rm b}4{\rm f}5)}=-\left(\frac{5}{48}+\frac{3\sqrt{3}}{16\pi}\right)g_*^2=-0.2075g_*^2.
\label{eta^(2b4f5)}
\end{align}

Sixth, we focus on the sixth term in the square brackets of Eq.\ (\ref{tW^(2b4f)}).
It can be treated in the same way as the fourth term (i.e., 2b-4e4 contribution) in the curly brackets of Eq.\  (\ref{tW^(2b4e)}) 
described from Eq.\ (\ref{tW^(2b4e4)-01}) through Eq.\ (\ref{eta^(2b4e4)}). 
Due to the change $|\tilde{\bf q}_1+\lambda\tilde{\bf q}'|^{2} \rightarrow |\tilde{\bf q}_1+\lambda\tilde{\bf k}|^{4}$
in the denominator of the integrand, 
we need to replace Eqs.\ (\ref{tJ_x^(pm)-2b4e}) and (\ref{J_x^(pm)-2b4e}) by the local functions
\begin{subequations}
\begin{align}
\tilde{J}_{\lambda\tilde{k}}^{(\pm)}(\lambda,\theta_q,\theta_1)
\equiv  \frac{1}{2} \int_{s_{{\rm c}2}}^{\mp1}
\frac{ds_1}{\bigl(1+\lambda^2\tilde{k}^2+2\lambda\tilde{k}\cos\theta_{k1}\bigr)^2},
\label{tJ_x^(pm)-2b4f6}
\end{align}
\begin{align}
{J}_{\lambda\tilde{k}}^{(\pm)}(\theta_q,\theta_1)\equiv &\, \frac{1}{2} \int_{s_{{\rm c}1}}^{\mp 1}
\frac{ds_1}{\bigl(1+\lambda^2\tilde{k}^2+2\lambda\tilde{k}\cos\theta_{k1}\bigr)^2}
\notag \\
=&\,\frac{\cos(\theta_q\pm\theta_1)+\frac{1}{2}\lambda\tilde{k}}{2\sin\theta_q\sin\theta_1\bigl[1+\lambda^2\tilde{k}^2+2\lambda\tilde{k}\cos(\theta_q\pm\theta_1)\bigr]},
\label{J_x^(pm)-2b4f6}
\end{align}
\end{subequations}
respectively,
where we have used Eqs.\ (\ref{s_c1}) and (\ref{xi_k-def}) to derive the second expression of Eq.\ (\ref{J_x^(pm)-2b4f6}).
Our $\delta \tilde{W}^{(2{\rm b}4{\rm f}6)}$ is expressible in terms of these functions in the same way as
$\delta \tilde{W}^{(2{\rm b}4{\rm e}4)}$ in Eq.\ (\ref{tW^(2b4e4)}).
The differences are summarized as follows : (i) the prefactor is now  $-g_*^2$ instead of $6g_*^2$; (ii)
functions ${J}_{\lambda\tilde{k}}^{(\pm)}(\lambda,\theta_q,\theta_1)$ and ${J}_{\lambda\tilde{k}}(\lambda,\theta_q,\theta_1)$ in Eq.\ (\ref{tW^(2b4e4)}) are replaced by Eq.\ (\ref{J_x^(pm)-2b4f6}) and 
$J_{\lambda\tilde{k}}(\theta_q,\theta_1)\equiv J_{\lambda\tilde{k}}^{(-)}(\theta_q,\theta_1)-J_{\lambda\tilde{k}}^{(+)}(\theta_q,\theta_1)$, respectively;
(iii) Eqs.\ (\ref{calJ^(pm)}) and (\ref{tcalJ}) are defined in terms of Eq.\ (\ref{tJ_x^(pm)-2b4f6}).

Let us substitute the resulting $\delta \tilde{W}^{(2{\rm b}4{\rm f}6)}$ into Eq.\ (\ref{eta-def2}), set $\tilde{k}=0$, 
and evaluate the integrals to obtain $\eta^{(2{\rm b}4{\rm f}6)}$.
This process can also be outlined in the same way as that described in the paragraph below Eq.\ (\ref{tcalJ}).
Indeed, items (i)-(v) given there also apply to this case except that our $J_{0}^{(-2)}(\theta_q,\theta_1)\!\equiv\! \partial^2 J_x^{(-)}(\theta_q,\theta_1)/\partial x^2\bigl|_{x=0}$ from Eq.\ (\ref{J_x^(pm)-2b4f6}),
\begin{align}
J_0^{(- 2)}(\theta_q,\theta_1)=&\,\frac{\cos3(\theta_q-\theta_1)+\cos(\theta_q-\theta_1)}{\sin\theta_{q}\sin\theta_1} ,
\end{align}
now yields a finite contribution to $\eta^{(2{\rm b}4{\rm f}6)}$; see also Eqs.\ (\ref{tkappa^(12)}) and (\ref{tcalJ_0^(2)-2b4f6}) 
for confirming (iv).
Thus, Eq.\ (\ref{eta^(2b4e4)-4}) is replaced by
\begin{align}
&\,\eta^{(2{\rm b}4{\rm f}6)}
\notag \\
=&\, -\frac{g_*^2}{2} \int_0^1 \frac{d\lambda}{\lambda}\Biggl[ \int_0^{\xi_\lambda}d\theta_1f_0(\theta_1)
\int_0^\pi d\theta_q f_0(\theta_q) 
\lambda^2 J_{0}^{(-2)}(\theta_q,\theta_1)
\notag \\
&\,  +\int_0^\pi d\theta_q  \bar{f}_0(\theta_q){\cal J}_0^{(+2)}(\lambda,\theta_q) \Biggr],
\label{eta^(2b4f6)-4}
\end{align}
where ${\cal J}_0^{(+2)}(\lambda,\theta_q)$ is obtained as Eq.\ (\ref{calJ_0^(pm2)-2b4f6}) in Appendix\ref{App-calJ-exp}.
We can evaluate the integrals of $\eta^{(2{\rm b}4{\rm f}6)}$ both numerically and analytically
to obtain
\begin{align}
\eta^{(2{\rm b}4{\rm f}6)}=\left(-\frac{1}{16}+\frac{5\sqrt{3}}{32\pi}\right)g_*^2=0.0236g_*^2.
\label{eta^(2b4f6)}
\end{align}

Seventh, we focus on the seventh term in the square brackets of Eq.\ (\ref{tW^(2b4f)}).
Its contribution is expressible in the coordinate system of Eq.\ (\ref{SpheC})
as
\begin{align}
&\,\delta\tilde{W}_{\infty}^{(2{\rm b}4{\rm f}7)}(\tilde{k}) 
\notag \\
=&\, -2g_*^2 \int_0^1 \frac{d\lambda}{\lambda}\Biggl[\int_0^{\pi} d\theta_q f_0(\theta_q)\int_0^{\xi_\lambda} d\theta_1f_\lambda(\theta_1)
\notag \\
&\,\times 
\frac{1}{2}\int_{-1}^1ds_1 \frac{\varTheta(\cos\theta_{k1}-\cos\xi_{\lambda\tilde{k}})\varTheta(\cos\theta_{q'1}-\cos\xi_{\lambda\tilde{q}'})}{1+\lambda^2\tilde{k}^2+2\lambda\tilde{k}\cos\theta_{k1}}
\notag \\
&\,-(\lambda\rightarrow 0)\Biggr],
\label{tW^(2b4f7)-01}
\end{align}
where $f_\lambda$, $\xi_\lambda$, $\theta_{q'1}$, and $\theta_{k1}$ are defined by Eqs.\ (\ref{f_k}), 
(\ref{xi_k-def}), (\ref{theta_q'1}), and (\ref{theta_k1}), respectively. 
\begin{figure}[t]
\begin{center}
\includegraphics[width=0.9\linewidth]{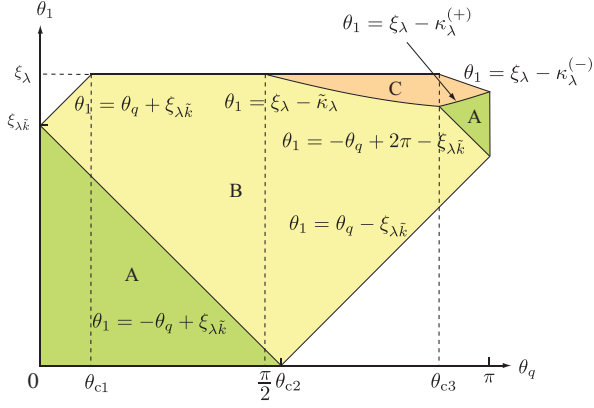}
\end{center}
\caption{Distinct regions of the double integral over $(\theta_q,\theta_1)$ for $(\lambda,\tilde{k})=(0.9,0.25)$.
The ranges of integration over $s_1$ for regions A, B, and C are
$s_1\in [-1,1]$, $s_1\in [s_{{\rm c}1},1]$, and $s_1\in [s_{{\rm c}2},1]$, respectively. 
Region C disappears as $\tilde{k}\rightarrow 0$.
\label{Fig7}}
\end{figure}
We can draw Fig.\ \ref{Fig7} that divides the $(\theta_q,\theta_1)$ plane 
into three regions
according to the range of integration over $s_1$ in Eq.\ (\ref{tW^(2b4f7)-01}):  region A with $s_1\in [-1,1]$, region B with $s_1\in [s_{{\rm c}1},1]$,
and region C with $s_1\in [s_{{\rm c}2},1]$, where $s_{{\rm c}1}$ and $s_{{\rm c}2}$
are given by Eqs.\ (\ref{s_c1}) and (\ref{s_c2}), respectively. 
The boundaries are the same as those in Fig.\ \ref{Fig6} except that the upper limit of the $\theta_1$
integral is now bounded by $\xi_\lambda$, and $\theta_{{\rm c}j}$ ($j=1,2,3$) are also given by 
Eqs.\ (\ref{theta_cj}) and (\ref{theta_c2}). One can also show that the curve $\theta_1=\xi_\lambda-\tilde{\kappa}_\lambda$ and
line $\theta_1=\xi_\lambda$ intersect at $\theta_q=\frac{\pi}{2}$.

On the basis of these considerations, 
we can perform the integration over $s_1$ in Eq.\ (\ref{tW^(2b4f7)-01}) elementarily.
To express the result concisely, it is convenient to introduce local functions for considering
2b-4f7 contributions by
\begin{subequations}
\begin{align}
{J}_{\lambda\tilde{k}}^{(\pm)}(\theta_q,\theta_1)\equiv &\, \frac{1}{2} \int_{s_{{\rm c}1}}^{\mp 1}ds_1
\frac{1}{1+\lambda^2\tilde{k}^2+2\lambda\tilde{k}\cos\theta_{k1}}
\notag \\
=&\, \frac{\ln \Bigl[1+\lambda^2\tilde{k}^2+2\lambda\tilde{k}\cos(\theta_q\pm \theta_1)\Bigr]}{4\lambda\tilde{k}\sin\theta_q\sin\theta_1},
\label{J_x^(pm)-2b4f7}
\\
{J}_{\lambda\tilde{k}}^{}(\theta_q,\theta_1)\equiv &\, {J}_{\lambda\tilde{k}}^{(-)}(\theta_q,\theta_1)-{J}_{\lambda\tilde{k}}^{(+)}(\theta_q,\theta_1),
\label{J_x-2b4f7}
\end{align}
\begin{align}
&\,\tilde{J}_{\lambda\tilde{k}}^{(\pm)}(\lambda,\theta_q,\theta_1)\equiv  \frac{1}{2} \int_{s_{{\rm c}2}}^{\mp1}ds_1
\frac{1}{1+\lambda^2\tilde{k}^2+2\lambda\tilde{k}\cos\theta_{k1}} .
\label{tJ_x^(pm)-2b4f7}
\end{align}
\end{subequations}
Using them, we can transform Eq.\ (\ref{tW^(2b4f7)-01}) in the same way as from Eq.\ (\ref{tW^(2b4e4)-02})
through Eq.\ (\ref{tW^(2b4e4)}). We thereby obtain
\begin{align}
&\,\delta\tilde{W}_{\infty}^{(2{\rm b}4{\rm f}7)}(\tilde{k}) 
\notag \\
=&\, -2g_*^2 \int_0^1 \frac{d\lambda}{\lambda}\Biggl[\Biggl(\int_0^{\theta_{{\rm c}2}} d\theta_q \int_0^{\xi_{\lambda\tilde{k}}-\theta_q}d\theta_1
\notag \\
&\,+\int_{\theta_{{\rm c}3}}^\pi d\theta_q \int_{2\pi-\xi_{\lambda\tilde{k}}-\theta_q}^{\xi_\lambda}d\theta_1 \Biggr)  
 f_0(\theta_q) f_\lambda(\theta_1) J_{\lambda\tilde{k}}(\theta_q,\theta_1)
\notag \\
&\, 
+\Biggl(\int_0^{\theta_{{\rm c}1}} d\theta_q \int_{\xi_{\lambda\tilde{k}}-\theta_q}^{\xi_{\lambda\tilde{k}}+\theta_q} d\theta_1 
+\int_{\theta_{{\rm c}1}}^{\frac{\pi}{2}} d\theta_q \int_{\xi_{\lambda\tilde{k}}-\theta_q}^{\xi_\lambda} d\theta_1 
\notag \\
&\,+\int_{\frac{\pi}{2}}^{\theta_{{\rm c}2}} d\theta_q  \int_{\xi_{\lambda\tilde{k}}-\theta_q}^{\xi_\lambda-\tilde{\kappa}_\lambda} d\theta_1
+\int_{\theta_{{\rm c}2}}^{\theta_{{\rm c}3}} d\theta_q \int_{\theta_q-\xi_{\lambda\tilde{k}}}^{\xi_\lambda-\tilde{\kappa}_\lambda} d\theta_1
\notag \\
&\,+\int_{\theta_{{\rm c}3}}^\pi d\theta_q \int_{\theta_q-\xi_{\lambda\tilde{k}}}^{2\pi-\xi_{\lambda\tilde{k}}-\theta_q} d\theta_1 \Biggr) 
 f_0(\theta_q) f_\lambda(\theta_1) J_{\lambda\tilde{k}}^{(-)}(\theta_q,\theta_1)
\notag \\
&\, +\int_{\frac{\pi}{2}}^{\theta_{{\rm c}3}} d\theta_q  \bar{f}_0(\theta_q)\tilde{\cal J}_{\tilde{k}}(\lambda,\theta_q)
-\int_{\theta_{{\rm c}3}}^\pi d\theta_q  \bar{f}_0(\theta_q){\cal J}_{\tilde{k}}^{(-)}(\lambda,\theta_q)
\notag \\
&\,
+\int_{\theta_{{\rm c}3}}^\pi d\theta_q \bar{f}_0(\theta_q) {\cal J}_{\tilde{k}}^{(+)}(\lambda,\theta_q)\Biggr],
\label{tW^(2b4f7)}
\end{align}
where ${\cal J}_{\tilde{k}}^{(\pm)}$ and $\tilde{\cal J}_{\tilde{k}}$ are now defined in terms of Eq.\ (\ref{tJ_x^(pm)-2b4f7}) by
\begin{subequations}
\label{calJ^(pm)-tcalJ-2b4f7}
\begin{align}
{\cal J}_{\tilde{k}}^{(\pm)}(\lambda,\theta_q)\equiv &\,\sin\theta_q \int_{\xi_\lambda-\kappa_\lambda^{(\pm)}}^{\xi_\lambda}d\theta_1f_\lambda(\theta_1)\tilde{J}_{\lambda\tilde{k}}^{(\pm)}(\lambda,\theta_q,\theta_1) ,
\label{calJ^(pm)-2b4f7}
\\
\tilde{\cal J}_{\tilde{k}}(\lambda,\theta_q)\equiv &\,\sin\theta_q \int_{\xi_\lambda-\tilde{\kappa}_\lambda}^{\xi_\lambda}d\theta_1f_\lambda(\theta_1)\tilde{J}_{\lambda\tilde{k}}^{(-)}(\lambda,\theta_q,\theta_1) .
\label{tcalJ-2b4f7}
\end{align}
\end{subequations}
Let us substitute Eq.\ (\ref{tW^(2b4f7)}) into Eq.\ (\ref{eta-def2}), set $\tilde{k}=0$, 
and evaluate the integrals to obtain $\eta^{(2{\rm b}4{\rm f}7)}$.
This process can be outlined in the same way as that described in the paragraph below Eq.\ (\ref{tcalJ}),
except that the two contributions described in (iv) do not cancel out owing to the fact that the lower bound of the $\theta_q$ integral is
$\frac{\pi}{2}$ instead of $\xi_\lambda-\frac{\pi}{2}$.
Thus, Eq.\ (\ref{eta^(2b4e4)-4}) is now replaced by
\begin{align}
&\,\eta^{(2{\rm b}4{\rm f}7)}
\notag \\
=&\, -g_*^2 \int_0^1 \frac{d\lambda}{\lambda}\Biggl\{
\int_{\frac{3\pi}{2}-\xi_\lambda}^\pi d\theta_q  \bar{f}_0(\theta_q)\Bigl[{\cal J}_0^{(+2)}(\lambda,\theta_q)-{\cal J}_0^{(-2)}(\lambda,\theta_q)\Bigr]
\notag \\
&\, +\int_{\frac{\pi}{2}}^{\frac{3\pi}{2}-\xi_\lambda} d\theta_q  \bar{f}_0(\theta_q)\tilde{\cal J}_0^{(2)}(\lambda,\theta_q)
-\int_{\frac{\pi}{2}}^{\frac{3\pi}{2}-\xi_\lambda} d\theta_q f_0(\theta_q) 
 f_\lambda(\theta_1)
\notag \\
&\,\times \tilde{\kappa}_\lambda^{(1)} \Biggl[ 
2\lambda {J}_0^{(-1)}(\theta_q,\theta_1)
-\tilde{\kappa}_\lambda^{(1)}
\frac{\partial {J}^{(-)}_0(\theta_q,\theta_1)}{\partial\theta_1}
+\frac{\tilde{\kappa}_\lambda^{(2)}}{\tilde{\kappa}_\lambda^{(1)}} {J}^{(-)}_0(\theta_q,\theta_1)
\notag \\
&\,-\tilde{\kappa}_\lambda^{(1)}\,\frac{d \ln f_\lambda(\theta_1)}{d\theta_1}{J}^{(-)}_0(\theta_q,\theta_1)
\Biggr]_{\theta_1=\xi_\lambda}\,\Biggr\}.
\label{eta^(2b4f7)-0}
\end{align}
Functions $J_0^{(-)}(\theta_q,\theta_1)$ and $J_0^{(-1)}(\theta_q,\theta_1)\equiv\partial J_x^{(-)}(\theta_q,\theta_1)/\partial x\bigr|_{x=0}$ are obtained easily from Eq.\  (\ref{J_x^(pm)-2b4f7}), $\tilde{\kappa}_\lambda^{(n)}$ for $n=1,2$ are given in Eq.\ (\ref{tkappa^(12)}),
and ${\cal J}_0^{(\pm 2)}(\lambda,\theta_q)$ and $\tilde{\cal J}_0^{(2)}(\lambda,\theta_q)$ can be calculated as Eq.\ (\ref{calJ_0^(pm2)-2b4f7})
and (\ref{tcalJ_0^(2)-2b4f7}) in Appendix\ref{App-calJ-exp}.
Using them in Eq.\ (\ref{eta^(2b4f7)-0}), we can perform the integrations of $\eta^{(2{\rm b}4{\rm f}7)}$ 
both numerically and analytically
to obtain
\begin{align}
\eta^{(2{\rm b}4{\rm f}7)}=-\left(\frac{1}{144}-\frac{\sqrt{3}}{48\pi} +\frac{37}{192\pi^2}\right)g_*^2=-0.0150g_*^2.
\label{eta^(2b4f7)}
\end{align}

Eighth, we focus on the eighth term in the square brackets of Eq.\ (\ref{tW^(2b4f)}).
Its contribution is expressible in the coordinate system of Eq.\ (\ref{SpheC}) as
\begin{align}
&\,\delta\tilde{W}_{\infty}^{(2{\rm b}4{\rm f}8)}(\tilde{k}) 
\notag \\
=&\,-2g_*^2 \int_0^1 \frac{d\lambda}{\lambda}\Biggl[\int_0^{\pi} d\theta_q f_0(\theta_q)\int_0^{\xi_\lambda} d\theta_1f_0(\theta_1)
\notag\\
&\,\times
\frac{1}{2}\int_{-1}^1ds_1 \frac{\varTheta(\cos\theta_{k1}-\cos\xi_{\lambda\tilde{k}})\varTheta(\cos\theta_{q'1}-\cos\xi_{\lambda\tilde{q}'})}{1+\lambda^2\tilde{q}^{\prime 2}+2\lambda\tilde{q}'\cos\theta_{q'1}}
\notag \\
&\,-(\lambda\rightarrow 0)\Biggr].
\label{tW^(2b4f8)-01}
\end{align}
We can transform this integral in exactly the same way as from Eq.\ (\ref{tW^(2b4f7)-01}) through Eq.\ (\ref{tW^(2b4f7)})
with two replacements, i.e., $f_\lambda(\theta_1)\rightarrow f_0(\theta_1)$ and $|\tilde{\bf q}_1+\lambda\tilde{\bf k}|^{-2}\rightarrow 
|\tilde{\bf q}_1+\lambda\tilde{\bf q}'|^{-2}$.
The corresponding contribution to the exponent, which we denote $\eta^{(2{\rm b}4{\rm f}8)}$, can also be written as 
Eq.\ (\ref{eta^(2b4f7)-0}), 
where $f_\lambda(\theta_1)$ should be replaced by $f_0(\theta_1)$, $({\cal J}_0^{(\pm 2)},\tilde{\cal J}_0^{(2)},\tilde{\kappa}^{(n)})$ ($n=1,2$)
are now given by
Eqs.\ (\ref{calJ^(pm2)-2b4e}), (\ref{tcalJ^(2)-2b4e}), and (\ref{tkappa^(12)}), respectively, and $J_0^{(-)}(\theta_q,\theta_1)$ and 
$J_0^{(-1)}(\theta_q,\theta_1)$
are replaced by  $J_0^{(-)}(\lambda,\theta_q,\theta_1)$ and $J_0^{(-1)}(\lambda,\theta_q,\theta_1)\equiv \partial J_x^{(-)}(\lambda,\theta_q,\theta_1)/\partial x\bigr|_{x=0}$
obtained from Eq.\ (\ref{J_x^(pm)-2b4e}), respectively. 
Using them, we can evaluate $\eta^{(2{\rm b}4{\rm f}8)}$ both numerically and analytically
to obtain
\begin{align}
\eta^{(2{\rm b}4{\rm f}8)}=\left(\frac{1}{144}+\frac{1}{192\pi^2}\right)=0.0075g_*^2.
\label{eta^(2b4f8)}
\end{align}

Adding Eqs.\ (\ref{eta^(2b4f1,2)}), (\ref{eta^(2b4f3,4)}), (\ref{eta^(2b4f5)}), (\ref{eta^(2b4f6)}), (\ref{eta^(2b4f7)}), and (\ref{eta^(2b4f8)})
yields the 2b-4f contribution to $\eta$ as
\begin{align}
\eta^{(2{\rm b}4{\rm f})}=\left(-\frac{1}{8}-\frac{\sqrt{3}}{96\pi}-\frac{3}{16\pi^2}\right)g_*^2=-0.1497g_*^2.
\label{eta^(2b4f)}
\end{align}

\subsection{Sum of various 2b contributions}

The net 2b contribution is obtained by adding Eqs.\  (\ref{eta^(2b4d)}), (\ref{eta^(2b4e)}) and (\ref{eta^(2b4f)}) as
\begin{align}
\eta^{(2{\rm b})}=\left(\frac{3}{8}-\frac{\sqrt{3}}{96\pi}-\frac{3}{16\pi^2}\right)g_*^2=0.3503g_*^2.
\label{eta^(2b)}
\end{align}

\section{Calculation of $\eta^{(2{\rm c})}$}
\label{Sec:5}

We here calculate the 2c contribution to Eq.\ (\ref{eta-def2}) given by Eq.\ (\ref{tW^(2c)}). 
The first term on the right-hand side of Eq.\ (\ref{tW^(2c)}) has already been studied to yield Eq.\ (95) of I, i.e., 
\begin{align}
\eta^{(2{\rm c}0)}=-\frac{1}{16} g_*^2 .
\label{eta^(2c0)}
\end{align}
Hence, we here focus on the second term on the right-hand side of Eq.\ (\ref{tW^(2c)}),
which is expressible diagrammatically as Fig.\ \ref{Fig1} (3a)-(3d). Among them,
we can exclude Fig.\ \ref{Fig1} (3a) owing to Eq.\ (\ref{W^(3a)}). 
Hence, we consider the other two contributions.

\subsection{2c-3c contribution}

First, we focus on $\delta\tilde{W}^{(3{\rm c})}_\infty$ given diagrammatically by Fig.\ \ref{Fig1} (3c).
Its analytic expression has already been derived as Eq.\ (\ref{tW^(3c)}).
We regularize it as Eq.\ (\ref{tW^(3)-2}) and substitute the resulting 
$\delta\tilde{W}^{(3{\rm c})}_\infty$ into the second term of Eq.\ (\ref{tW^(2c)}).
We thereby obtain the 3c contribution to $\delta \tilde{W}^{(2{\rm c})}_\infty$ as
\begin{align}
\delta \tilde{W}^{(2{\rm c}3{\rm c})}_\infty(\tilde{k})
=&\, -24g_*^2\int_0^1 \frac{d\lambda}{\lambda} \Biggl\{ \int_0^{\xi_{\tilde{k}}}  d\theta_qf_0(\theta_q)
\notag \\
&\,\times \left[
\tilde{\chi}_{F\dot{F}}\bigl(\lambda(1+\tilde{k}^2+2\tilde{k}\cos\theta_q)^{1/2}\bigr)-\tilde{\chi}_{F\dot{F}}(0)\right]
\notag \\
&\,+\left[\tilde{\chi}_{F\dot{F}}(\lambda)-\tilde{\chi}_{F\dot{F}}(0)\right]\int_0^{\xi_{\tilde{k}}}  d\theta_q f_{\tilde{k}}(\theta_q)\Biggr\},
\label{tW^(2c3c)}
\end{align}
where we have also made a transformation similar to Eq.\ (\ref{chi}).
Let us substitute Eq.\ (\ref{tW^(2c3c)}) into Eq.\ (\ref{eta-def2}), calculate the second derivative at $\tilde{k}=0$, and evaluate the 
integrals. The process is the same as Eqs.\ (91)-(93) of I except that
we have to take care of the additional $\tilde{k}$ dependences of (i) the upper limit $\xi_{\tilde{k}}$ of the $\theta_q$ integral
and (ii) $f_{\tilde{k}}(\theta_q)$.
However, one can show that these extra dependences give null contribution to $\eta$.
Also noting that the upper limit $\xi_{\tilde{k}}$ approaches $\pi/2$ instead of $\pi$ as $\tilde{k}\rightarrow 0$, we obtain 
\begin{align}
\eta^{(2{\rm c}3{\rm c})}=-\frac{1}{8}g_*^2  .
\label{eta^(2c3c)}
\end{align}

\subsection{2c-3d contribution}

Next, we consider the contribution of $\delta\tilde{W}^{(3{\rm d})}_\infty$  given diagrammatically by Fig.\ \ref{Fig1} (3d).
Its analytic expression has already been derived as Eq.\ (\ref{tW^(3d)}).
We regularize it as Eq.\ (\ref{tW^(3)-2}), substitute the resulting 
$\delta\tilde{W}^{(3{\rm d})}_\infty$ into the second term of Eq.\ (\ref{tW^(2c)}), 
and approximate $d\approx 4$ in the integrand as justified for $\epsilon\ll 1$.
The procedure yields the 3d contribution to $\delta \tilde{W}^{(2{\rm c})}_\infty$ as
\begin{align}
&\,\delta\tilde{W}_\infty^{(2{\rm c}3{\rm d})}(\tilde{k})
\notag \\
=&\,  -2g_*^2  \int_0^1 \frac{d\lambda}{\lambda}\int\frac{d^4 \tilde{q}}{(2\pi)^4K_4}
\delta(\tilde{q}-1)\varTheta(|\tilde{\bf k}+\tilde{\bf q}|-1)
\notag \\
&\,\times 
\Biggl[  \phi_{20}(\lambda\tilde{\bf q}+\lambda\tilde{\bf k},\lambda\tilde{\bf q})
+\frac{\phi_{02}(\lambda\tilde{\bf q}+\lambda\tilde{\bf k},\lambda\tilde{\bf q})}{|\tilde{\bf k}+\tilde{\bf q}|^2} 
+\phi_{22}(-\lambda\tilde{\bf q},\lambda\tilde{\bf k})
\notag \\
&\, 
+\frac{\phi_{20}(-\lambda\tilde{\bf q},\lambda\tilde{\bf k})}{|\tilde{\bf k}+\tilde{\bf q}|^2}
+ \phi_{20}(\lambda\tilde{\bf k}+\lambda\tilde{\bf q},\lambda\tilde{\bf k})
+\frac{\phi_{22}(\lambda\tilde{\bf k}+\lambda\tilde{\bf q},\lambda\tilde{\bf k})}{|\tilde{\bf k}+\tilde{\bf q}|^2} 
\notag \\
&\,-(\lambda\rightarrow 0)\Biggr] .
\label{W^(2c3d)}
\end{align}

We consider each term in the square brackets of Eq.\ (\ref{W^(2c3d)}) separately.
First, we focus on the first term, which can be transformed in the same way as the third term in the curly brackets of Eq.\ (\ref{tW^(2b4e)}), i.e., 
from Eq.\ (\ref{tW^(2b4e3)-0}) through Eq.\ (\ref{tW^(2b4e3)}).
The key difference lies in the additional factor $\varTheta(|\tilde{\bf k}+\tilde{\bf q}|-1)$,
which introduces $\xi_{\tilde{k}}$ defined by Eq.\ (\ref{xi_k-def}) as the upper limit of the $\theta_q$ integral.
Also noting $\theta_{{\rm c}1}<\xi_{\tilde{k}}<\theta_{{\rm c}3}$ as seen from Eqs.\ (\ref{xi_k-def}) and (\ref{theta_cj}), 
we obtain 
\begin{align}
&\,\delta\tilde{W}_{\infty}^{(2{\rm c}3{\rm d}1)}(\tilde{k}) 
\notag \\
=&\, -2g_*^2 \int_0^1 \frac{d\lambda}{\lambda}\Biggl[\int_0^{\xi_{\tilde{k}}} d\theta_q\int_0^{\xi_{\lambda}}d\theta_1 f_0(\theta_q) f_0(\theta_1)
J_{\lambda\tilde{k}}(\lambda,\theta_q,\theta_1)
\notag \\
&\,
+\int_{\theta_{{\rm c}1}}^{\xi_{\tilde{k}}} d\theta_q  \bar{f}_0(\theta_q){\cal J}_{\tilde{k}}^{(+)}(\lambda,\theta_q)
-(\lambda\rightarrow 0)\Biggr] ,
\label{tW^(2c3d1)-2}
\end{align}
where $J_{\lambda\tilde{k}}$ and ${\cal J}_{\tilde{k}}^{(+)}$ are given by Eqs.\  (\ref{J_x-2b4e}) and (\ref{calJ^(pm)}), respectively.

Let us substitute Eq.\ (\ref{tW^(2c3d1)-2}) into Eq.\ (\ref{eta-def2}).
Terms with derivatives of $\theta_{{\rm c}1}$ all vanish once again, and we obtain
\begin{align}
&\, \eta^{(2{\rm c}3{\rm d}1)}
\notag \\
=&\, -g_*^2 \int_0^1 \frac{d\lambda}{\lambda}\Biggl\{\int_0^{\xi_{\lambda}}d\theta_1 f_0(\theta_1)\int_0^{\frac{\pi}{2}} d\theta_q  f_0(\theta_q)
\lambda^2 J_{0}^{(2)}(\lambda,\theta_q,\theta_1)
\notag \\
&\,
+\int_{\xi_\lambda-\frac{\pi}{2}}^{\frac{\pi}{2}} d\theta_q  \bar{f}_0(\theta_q){\cal J}_0^{(+2)}(\lambda,\theta_q)
\notag \\
&\,
+ f_0\bigl({\textstyle\frac{\pi}{2}}\bigr)\!\int_0^{\xi_\lambda}\! d\theta_1  f_0(\theta_1)\!
\left[\lambda J_0^{(1)}(\lambda,\theta_q,\theta_1)+\!\frac{1}{4}\frac{\partial J_0(\lambda,\theta_q,\theta_1)}{\partial\theta_q}\right]_{\theta_q=\frac{\pi}{2}}
\notag \\
&\,
+\bar{f}_0\bigl({\textstyle\frac{\pi}{2}}\bigr){\cal J}_0^{(+1)}\bigl(\lambda,{\textstyle\frac{\pi}{2}}\bigr)-(\lambda\rightarrow 0)\Biggr\} ,
\label{eta^(2c3d1)-0}
\end{align}
where $J_{0}^{(2)}$, ${\cal J}_0^{(+2)}$, and ${\cal J}_0^{(+1)}$ are given by Eqs.\ (\ref{J_0^(2)}), (\ref{calJ^(pm2)-2b4e}), and (\ref{calJ^(pm1)-2b4e}), respectively, and we have used $\partial f_0(\theta_q)/\partial\theta_q\bigr|_{\theta_q=\pi/2}= {\cal J}_0^{(+)}(\lambda,\theta_q)=0$ as seen from 
Eqs.\ (\ref{f_k}) and  (\ref{calJ^(pm)-exp}), respectively.
It follows from Eq.\ (\ref{J_x-2b4e}) that $J_0^{(1)}(\lambda,\theta_q,\theta_1)\bigr|_{\theta_q=\frac{\pi}{2}}=\partial J_0(\lambda,\theta_q,\theta_1)/\partial\theta_q\bigr|_{\theta_q=\frac{\pi}{2}}=0$ holds in the integrand of Eq.\ (\ref{eta^(2c3d1)-0}).
Moreover, the first term in the curly brackets vanishes upon the integration over $\theta_q$.
The remaining integrals can be evaluated both numerically and analytically, and we obtain
\begin{align}
\eta^{(2{\rm c}3{\rm d}1)}= \left( \frac{5}{288} -\frac{13}{64\pi^2}\right)g_*^2=  -0.0032g_*^2 .
\label{eta^(2c3d1)}
\end{align}

Second, we consider the second term in the square brackets of Eq.\ (\ref{W^(2c3d)}).
We can transform the contribution
in the same way as the first term above to obtain
\begin{align}
\delta\tilde{W}_{\infty}^{(2{\rm c}3{\rm d}2)}(\tilde{k}) 
=&\, -2g_*^2 \int_0^1 \frac{d\lambda}{\lambda}\Biggl[\int_0^{\xi_{\tilde{k}}} d\theta_q\int_0^{\xi_{\lambda}}d\theta_1 f_{\tilde{k}}(\theta_q) f_\lambda(\theta_1)
\notag \\
&\,
+\int_{\theta_{{\rm c}1}}^{\xi_{\tilde{k}}} d\theta_q  \bar{f}_{\tilde{k}}(\theta_q){\cal J}_{\tilde{k}}^{(+)}(\lambda,\theta_q)
-(\lambda\rightarrow 0)\Biggr] ,
\label{tW^(2c3d2)-2}
\end{align}
where ${\cal J}_{\tilde{k}}^{(+)}$ is now defined by
\begin{align}
{\cal J}_{\tilde{k}}^{(+)}(\lambda,\theta_q)\equiv \sin\theta_q \int_{\xi_\lambda-\kappa^{(+)}}^{\xi_\lambda} d\theta_1 f_\lambda(\theta_1) \tilde{J}_{\lambda\tilde{k}}^{(+)}(\lambda,\theta_q,\theta_1) ,
\label{calJ^(+)-2c3d2}
\end{align}
in terms of $\tilde{J}_{\lambda\tilde{k}}^{(+)}$ given by Eq.\ (\ref{tJ-w}) with Eq.\ (\ref{w-2b4f5}).
Let us substitute Eq.\ (\ref{tW^(2c3d2)-2}) into Eq.\ (\ref{eta-def2}).
Terms with derivatives of $\theta_{{\rm c}1}$ cancel out, and we obtain
\begin{align}
&\, \eta^{(2{\rm c}3{\rm d}2)}
\notag \\
=&\,  -g_*^2 \int_0^1 \frac{d\lambda}{\lambda}\Biggl\{ \int_0^{\frac{\pi}{2}} d\theta_q 
\frac{\partial^2 f_{\tilde{k}}(\theta_q)}{\partial\tilde{k}^2} \Biggr|_{\tilde{k}=0}\,
\notag \\
&\,\times \Biggl[ \int_0^{\xi_{\lambda}}d\theta_1  f_{\lambda}(\theta_1)-(\lambda\rightarrow 0)\Biggr]
\notag \\
&\,
+\int_{\xi_\lambda-\frac{\pi}{2}}^{\frac{\pi}{2}} d\theta_q 
\Biggl[\bar{f}_{0}(\theta_q) {\cal J}^{(+2)}_{0}(\lambda,\theta_q)+2\frac{\partial\bar{f}_{\tilde{k}}(\theta_q)}{\partial\tilde{k}}\Biggr|_{\tilde{k}=0} {\cal J}^{(+1)}_{0}(\lambda,\theta_q)
\Biggr]
\notag \\
&\,+\bar{f}_{0}({\textstyle\frac{\pi}{2}}){\cal J}^{(+1)}_0(\lambda,{\textstyle\frac{\pi}{2}})
-(\lambda\rightarrow 0) \Biggr\},
\label{eta^(2c3d2)-0}
\end{align}
where ${\cal J}_0^{(+1)}$ and ${\cal J}_0^{(+2)}$ are given by Eqs.\ (\ref{calJ_0^(pm1)-2c3d2}) and (\ref{calJ_0^(pm2)-2c3d2}), respectively,
and we have used that ${\partial f_{\tilde{k}}(\theta_q)}/{\partial\tilde{k}} \bigr|_{\tilde{k}=0}={\partial f_0(\theta_q)}/{\partial\theta_q}=0$ holds at $\theta_q=\frac{\pi}{2}$
for Eq.\ (\ref{f_k}).
One can show that the first integral over $\theta_q$ in the curly brackets of Eq.\ (\ref{eta^(2c3d2)-0}) vanishes.
The other integrals can be evaluated both numerically and analytically. We obtain
\begin{align}
\eta^{(2{\rm c}3{\rm d}2)}=-\left(\frac{5}{288}+\frac{\sqrt{3}}{24\pi}
-\frac{5}{64\pi^2}\right)g_*^2 
= -0.0324g_*^2 .
\label{eta^(2c3d2)}
\end{align}

\begin{figure}[b]
\begin{center}
\includegraphics[width=0.9\linewidth]{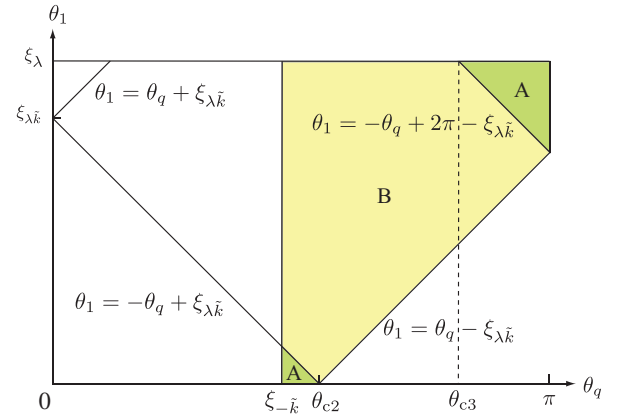}
\end{center}
\caption{Distinct regions of the double integral over $(\theta_q,\theta_1)$ for $(\lambda,\tilde{k})=(0.9,0.25)$.
The ranges of integration over $s_1$ for regions A and B are
$s_1\in [-1,1]$ and $s_1\in [s_{{\rm c}1},1]$, respectively, with $s_{{\rm c}1}$ given by Eq.\ (\ref{s_c1}).
\label{Fig8}}
\end{figure}

Third, we focus on the third term in the square brackets of Eq.\ (\ref{W^(2c3d)}) and set $\tilde{\bf q}\rightarrow-\tilde{\bf q}$ in the integrand for convenience.
The corresponding contribution can be written in the coordinate system of Eq.\ (\ref{SpheC}) as
\begin{align}
&\,\delta\tilde{W}_{\infty}^{(2{\rm c}3{\rm d}3)}(\tilde{k}) 
\notag \\
=&\, -2g_*^2 \int_0^1 \frac{d\lambda}{\lambda}\Biggl[\int_{\xi_{-\tilde{k}}}^\pi d\theta_q f_0(\theta_q)\int_0^{\xi_\lambda} d\theta_1 f_\lambda(\theta_1)
\notag \\
&\,\times 
\frac{1}{2}\int_{-1}^1ds_1 \frac{\varTheta(\cos\theta_{k1}-\cos\xi_{\lambda\tilde{k}})}{1+\lambda^2\tilde{k}^2+2\lambda\tilde{k}\cos\theta_{k1}}
-(\lambda\rightarrow 0)\Biggr] ,
\label{tW^(2c3d3)-01}
\end{align}
where $f_\lambda$, $\xi_\lambda$, and $\theta_{k1}$ are given by Eqs.\ (\ref{f_k}), (\ref{xi_k-def}), and (\ref{theta_k1}), respectively.
We can draw Fig.\ \ref{Fig8} that divides the $(\theta_q,\theta_1)$ plane into two regions
according to the range of integration over $s_1$: region A with $s_1\in [-1,1]$ and region B with $s_1\in [s_{{\rm c}1},1]$, 
where $s_{{\rm c}1}$ is given by Eq.\ (\ref{s_c1}).
The integral of $s_1$ over B and A defines ${J}_{\lambda\tilde{k}}^{(-)}$ and ${J}_{\lambda\tilde{k}}^{}$
given by Eqs.\ (\ref{J_x^(pm)-2b4f7}) and (\ref{J_x-2b4f7}), respectively.
We express ${J}_{\lambda\tilde{k}}^{}={J}_{\lambda\tilde{k}}^{(-)}-{J}_{\lambda\tilde{k}}^{(+)}$ for region A,
use the symmetry ${J}_{\lambda\tilde{k}}^{(+)}(\theta_q,\theta_1)=-{J}_{\lambda\tilde{k}}^{(-)}(\theta_q,-\theta_1)$ 
on the lower left corner, and set $\theta_1\rightarrow -\theta_1$ there subsequently. 
We can thereby transform Eq.\ (\ref{tW^(2c3d3)-01}) into
\begin{align}
&\,\delta\tilde{W}_{\infty}^{(2{\rm c}3{\rm d}3)}(\tilde{k}) 
\notag \\
=&\, -2g_*^2 \int_0^1 \frac{d\lambda}{\lambda}\Biggl[\int_{\xi_{-\tilde{k}}}^\pi d\theta_q\int_{\theta_q-\xi_{\lambda\tilde{k}}}^{\xi_\lambda} d\theta_1
f_0(\theta_q)f_\lambda(\theta_1)J_{\lambda\tilde{k}}^{(-)}(\theta_q,\theta_1)
\notag \\
&\,
-\int_{\theta_{{\rm c}3}}^\pi \! d\theta_q\int_{2\pi-\xi_{\lambda\tilde{k}}-\theta_q}^{\xi_\lambda} \! d\theta_1
f_0(\theta_q)f_\lambda(\theta_1)J_{\lambda\tilde{k}}^{(+)}(\theta_q,\theta_1)
-(\lambda\rightarrow 0)\Biggr] .
\label{tW^(2c3d3)-02}
\end{align}
Let us substitute Eq.\ (\ref{tW^(2c3d3)-02}) into Eq.\ (\ref{eta-def2}).
We then obtain
\begin{align}
&\, \eta^{(2{\rm c}3{\rm d}3)}
\notag \\
=&\,  -g_*^2 \int_0^1 \frac{d\lambda}{\lambda}\Biggl\{\int_{\frac{\pi}{2}}^\pi\! d\theta_q \int_{\theta_q-\frac{\pi}{2}}^{\xi_\lambda} \! d\theta_1
f_0(\theta_q)   f_{\lambda}(\theta_1) \lambda^2 J_0^{(-2)}(\theta_q,\theta_1)
\notag \\
&\,-\int_{\frac{3\pi}{2}-\xi_\lambda}^\pi \! d\theta_q \int_{\frac{3\pi}{2}-\theta_q}^{\xi_\lambda}\! d\theta_1
f_0(\theta_q)   f_{\lambda}(\theta_1)\lambda^2 J_0^{(+2)}(\theta_q,\theta_1)
\notag \\
&\,+f_0({\textstyle\frac{\pi}{2}})\int_0^{\xi_\lambda}d\theta_1 f_{\lambda}(\theta_1)\left[\lambda J_0^{(-1)}({\textstyle\frac{\pi}{2}},\theta_1)
-\frac{1}{4}\frac{\partial J_0^{(-)}(\theta_q,\theta_1)}{\partial\theta_q}\Biggr|_{\theta_q=\frac{\pi}{2}}\,\right]
\notag \\
&\,
+\lambda^2 \int_{\frac{\pi}{2}}^\pi d\theta_q f_0(\theta_q)f_{\lambda}(\theta_q-{\textstyle\frac{\pi}{2}})\Biggl[ J_0^{(-1)}(\theta_q,\theta_1)
\notag \\
&\,
-\frac{1}{4}\frac{\partial J_0^{(-)}(\theta_q,\theta_1)}{\partial\theta_1} \Biggr]_{\theta_1=\theta_q-\frac{\pi}{2}}
-\lambda^2 \int_{\frac{3\pi}{2}-\xi_\lambda}^\pi d\theta_q f_0(\theta_q)f_{\lambda}({\textstyle\frac{3\pi}{2}}-\theta_q)
\notag \\
&\,\times \Biggl[J_0^{(+1)}(\theta_q,\theta_1)
-\frac{1}{4}\frac{\partial J_0^{(+)}(\theta_q,\theta_1)}{\partial\theta_1}\Biggr]_{\theta_1=\frac{3\pi}{2}-\theta_q}
-(\lambda\rightarrow 0) \Biggr\},
\label{eta^(2c3d3)-0}
\end{align}
where we have used $J_0^{(-)}(\theta_q,\theta_q\!-\!{\textstyle\frac{\pi}{2}})=J_0^{(+)}(\theta_q,{\textstyle\frac{3\pi}{2}}\!-\!\theta_q)=0$ 
for Eq.\ (\ref{J_x^(pm)-2b4f7}) and ${\partial f_0(\theta_q)}/{\partial\theta_q}\bigr|_{\theta_q=\frac{\pi}{2}}=0$ for Eq.\ (\ref{f_k}).
We obtain $J_0^{(\pm n)}(\theta_q,\theta_1)\equiv \partial^n J_x^{(\pm)}(\theta_q,\theta_1)/\partial x^2\bigr|_{x=0}$ for $n=0,1,2$ 
from Eq.\ (\ref{J_x^(pm)-2b4f7}) as
\begin{align}
J_0^{(\pm n)}(\theta_q,\theta_1)=(-1)^n \frac{\cos[(n+1)(\theta_q\pm\theta_1)]}{2(n+1)\sin\theta_q\sin\theta_1} .
\end{align}
The fourth and fifth terms in the curly brackets of Eq.\ (\ref{eta^(2c3d3)-0}) can be combined, through $\theta_q= \theta_q'+\frac{\pi}{2}$ and 
 $\theta_q= \frac{3\pi}{2}-\theta_q'$, respectively,  into a single integral over $\theta_q'\in[0,\xi_\lambda]$.
Equation (\ref{eta^(2c3d3)-0}) can be evaluated both  numerically and analytically, including the triple one 
by the order $\theta_q \rightarrow \lambda\rightarrow \theta_1$.
We eventually obtain 
\begin{align}
\eta^{(2{\rm c}3{\rm d}3)}=\left(  \frac{1}{48}+\frac{17}{32\pi^2}\right) g_*^2
= 0.0747g_*^2 .
\label{eta^(2c3d3)}
\end{align}

Fourth, we focus on the fourth term in the square brackets of Eq.\ (\ref{W^(2c3d)}) and set $\tilde{\bf q}\rightarrow-\tilde{\bf q}$ in the integrand for convenience.
We can transform the contribution
in the same way as the third term above.
We thereby obtain
\begin{align}
&\,\delta\tilde{W}_{\infty}^{(2{\rm c}3{\rm d}4)}(\tilde{k}) 
\notag \\
=&\, -2g_*^2 \int_0^1 \frac{d\lambda}{\lambda}\Biggl\{\int_{\xi_{-\tilde{k}}}^\pi d\theta_q\int_{\theta_q-\xi_{\lambda\tilde{k}}}^{\xi_\lambda} d\theta_1
f_{-\tilde{k}}(\theta_q)f_\lambda(\theta_1)J_{\lambda\tilde{k}}^{(-)}(\theta_q,\theta_1)
\notag \\
&\,
-\int_{\theta_{{\rm c}3}}^\pi  \! d\theta_q\int_{2\pi-\xi_{\lambda\tilde{k}}-\theta_q}^{\xi_\lambda}\! d\theta_1
f_{-\tilde{k}}(\theta_q)f_\lambda(\theta_1)J_{\lambda\tilde{k}}^{(+)}(\theta_q,\theta_1)
-(\lambda\rightarrow 0)\Biggr\},
\label{tW^(2c3d4)-02}
\end{align}
where $J_{\lambda\tilde{k}}^{(\pm)}(\theta_q,\theta_1)$ is now defined in terms of Eq.\ (\ref{s_c1}) by
\begin{align}
J_{\lambda\tilde{k}}^{(\pm)}(\theta_q,\theta_1)\equiv  \frac{1}{2} \int_{s_{{\rm c}1}}^{\mp1}ds_1
=\frac{2\cos(\theta_q\pm \theta_1)+\lambda\tilde{k}}{4\sin\theta_q\sin\theta_1},
\label{J_k^(pm)-2c3d4}
\end{align}
satisfying $J_{\lambda\tilde{k}}^{(-)}-J_{\lambda\tilde{k}}^{(+)}=1$ and $J_{\lambda\tilde{k}}^{(+)}(\theta_q,\theta_1)=
-J_{\lambda\tilde{k}}^{(-)}(\theta_q,-\theta_1)$.
Substitution of Eq.\ (\ref{tW^(2c3d4)-02}) into Eq.\ (\ref{eta-def2})
yields $\eta^{(2{\rm c}3{\rm d}4)}$, which turns out to have the same expression as Eq.\ (\ref{eta^(2c3d3)-0}) except for the replacement of
$f_0(\theta_q)\lambda^2J_0^{(\pm2)}(\theta_q,\theta_1)$ by
\begin{align*}
\frac{\partial^2 f_{-\tilde{k}}(\theta_q)}{\partial\tilde{k}^2}\Biggr|_{\tilde{k}=0}
J_0^{(\pm)}(\theta_q,\theta_1)+2\frac{\partial f_{-\tilde{k}}(\theta_q)}{\partial\tilde{k}}\Biggr|_{\tilde{k}=0} \lambda J_0^{(\pm1)}(\theta_q,\theta_1) .
\end{align*} 
The coefficients $J_0^{(\pm n)}(\theta_q,\theta_1)\equiv \partial^n J_x^{(\pm)}(\theta_q,\theta_1)/\partial x^2\bigr|_{x=0}$ for $n=0,1$
are obtained easily from Eq.\ (\ref{J_k^(pm)-2c3d4}). 
Substituting them, we can evaluate $\eta^{(2{\rm c}3{\rm d}4)}$ both numerically and analytically,
including the triple one by the order $\theta_q \rightarrow \lambda\rightarrow \theta_1$.
We eventually obtain 
\begin{align}
\eta^{(2{\rm c}3{\rm d}4)}=\left(\frac{1}{48}-\frac{5}{64\pi^2}\right) g_*^2
= 0.0129g_*^2 .
\label{eta^(2c3d4)}
\end{align}

Fifth,  we focus on the fifth term in the square brackets of Eq.\ (\ref{W^(2c3d)}).
We can transform the contribution
in the same way as the fourth term in the curly brackets of Eq.\ (\ref{tW^(2b4e)}), i.e., from Eq.\ (\ref{tW^(2b4e4)-01})
through Eq.\ (\ref{tW^(2b4e4)}). 
The key difference lies in the additional factor $\varTheta(|\tilde{\bf k}+\tilde{\bf q}|-1)$,
which introduces $\xi_{\tilde{k}}$ defined by Eq.\ (\ref{xi_k-def}) as the upper limit of the $\theta_q$ integral.
Also noting $\xi_{\tilde{k}}\gtrsim\theta_{{\rm c}2}\gtrsim\frac{\pi}{2}$ from Eqs.\ (\ref{xi_k-def}) and (\ref{theta_c2}), 
we obtain the expression of $\delta\tilde{W}_{\infty}^{(2{\rm c}3{\rm d}5)}$
in place of Eq.\ (\ref{tW^(2b4e4)}) as
\begin{align}
&\,\delta\tilde{W}_{\infty}^{(2{\rm c}3{\rm d}5)}(\tilde{k}) 
\notag \\
=&\, -2g_*^2 \int_0^1 \frac{d\lambda}{\lambda}\Biggl\{\int_0^{\theta_{{\rm c}2}} \!d\theta_q \int_0^{\xi_{\lambda\tilde{k}}-\theta_q}\!d\theta_1
f_0(\theta_q) f_0(\theta_1) J_{\lambda\tilde{k}}(\lambda,\theta_q,\theta_1)
\notag \\
&\, +\Biggl(\int_0^{\theta_{{\rm c}1}} d\theta_q \int_{\xi_{\lambda\tilde{k}}-\theta_q}^{\xi_{\lambda\tilde{k}}+\theta_q} d\theta_1 
+\int_{\theta_{{\rm c}1}}^{\theta_{{\rm c}2}} d\theta_q \int_{\xi_{\lambda\tilde{k}}-\theta_q}^{\xi_\lambda-\tilde{\kappa}_\lambda} d\theta_1 
\notag \\
&\,
+\int_{\theta_{{\rm c}2}}^{\xi_{\tilde{k}}} d\theta_q \int_{\theta_q-\xi_{\lambda\tilde{k}}}^{\xi_\lambda-\tilde{\kappa}_\lambda} d\theta_1
\Biggr) 
f_0(\theta_q) f_0(\theta_1) J_{\lambda\tilde{k}}^{(-)}(\lambda,\theta_q,\theta_1)
\notag \\
&\, +\int_{\theta_{{\rm c}1}}^{\xi_{\tilde{k}}} d\theta_q  \bar{f}_0(\theta_q)\Bigl[\tilde{\cal J}_{\tilde{k}}(\lambda,\theta_q)
-{\cal J}_{\tilde{k}}^{(-)}(\lambda,\theta_q)\Bigr]
-(\lambda\rightarrow 0)\Biggr\},
\label{tW^(2c3d5)-2}
\end{align}
where $J_{\lambda\tilde{k}}$, $J_{\lambda\tilde{k}}^{(-)}$, $\tilde{\cal J}_{\tilde{k}}$, and ${\cal J}_{\tilde{k}}^{(-)}$ are the same as those 
for the 2b-4e contribution given by Eq.\ (\ref{J_x-2b4e}), (\ref{J_x^(pm)-2b4e}), (\ref{tcalJ}), and (\ref{calJ^(pm)}), respectively.

Substitution of Eq.\ (\ref{tW^(2c3d5)-2}) into Eq.\ (\ref{eta-def2}) yields
\begin{align}
&\,\eta^{(2{\rm c}3{\rm d}5)}
\notag \\
=&\, -g_*^2 \int_0^1 \frac{d\lambda}{\lambda}\Biggl\{\int_0^{\frac{\pi}{2}} \!d\theta_q \int_0^{\frac{\pi}{2}-\theta_q}\!d\theta_1
f_0(\theta_q) f_0(\theta_1)\lambda^2 J_0^{(2)}(\lambda,\theta_q,\theta_1)
\notag \\
&\,
+\left(\int_0^{\xi_\lambda-\frac{\pi}{2}} \! d\theta_q \int_{\frac{\pi}{2}-\theta_q}^{\frac{\pi}{2}+\theta_q} \! d\theta_1 
+\int_{\xi_\lambda-\frac{\pi}{2}}^{\frac{\pi}{2}} \! d\theta_q 
\int_{\frac{\pi}{2}-\theta_q}^{\xi_\lambda} \! d\theta_1 \right)f_0(\theta_q) f_0(\theta_1)
\notag \\
&\, \times \lambda^2 J_0^{(-2)}(\lambda,\theta_q,\theta_1)
+\int_{\xi_\lambda-\frac{\pi}{2}}^{\frac{\pi}{2}} d\theta_q  \bar{f}_0(\theta_q)\Bigl[\tilde{\cal J}_0^{(2)}(\lambda,\theta_q)
\notag \\
&\, - {\cal J}_0^{(-2)}(\lambda,\theta_q)\Bigr]
+\lambda^2 \int_0^{\xi_\lambda-\frac{\pi}{2}} d\theta_q  f_0(\theta_q) 
 f_0({\textstyle\frac{\pi}{2}}+\theta_q) 
\notag \\
&\,\times \Biggl[ J_0^{(-1)}(\lambda,\theta_q,\theta_1)
+\frac{1}{4}
\frac{\partial J_0^{(-)}(\lambda,\theta_q,\theta_1)}{\partial\theta_1}\Biggr]_{\theta_1=\frac{\pi}{2}+\theta_q}
-\lambda^2 \int_0^{\frac{\pi}{2}} d\theta_q  
\notag \\
&\,\times  f_0(\theta_q) 
 f_0({\textstyle\frac{\pi}{2}}\!-\!\theta_q) 
 \Biggl[ J_0^{(+1)}(\lambda,\theta_q,\theta_1)
+\!\frac{1}{4}
\frac{\partial J_0^{(+)}(\lambda,\theta_q,\theta_1)}{\partial\theta_1}\Biggr]_{\theta_1=\frac{\pi}{2}-\theta_q}
\notag \\
&\, -f_0(\xi_\lambda) \int_{\xi_\lambda-\frac{\pi}{2}}^{\frac{\pi}{2}} d\theta_q
 f_0(\theta_q) \tilde{\kappa}_\lambda^{(1)}
\Biggl[
2\lambda J_0^{(-1)}(\lambda,\theta_q,\theta_1)-\tilde{\kappa}_\lambda^{(1)}
\notag \\
&\,
\times  \frac{\partial J_0^{(-)}(\lambda,\theta_q,\theta_1)}{\partial\theta_1}
- \tilde{\kappa}_\lambda^{(1)}\frac{d \ln f_0(\theta_1)}{d\theta_1}J_0^{(-)}(\lambda,\theta_q,\theta_1)
\notag \\
&\,+\frac{\tilde{\kappa}_\lambda^{(2)}}{\tilde{\kappa}_\lambda^{(1)}} 
J_0^{(-)}(\lambda,\theta_q,\theta_1)
\Biggr]_{\theta_1=\xi_\lambda}
+ f_0({\textstyle\frac{\pi}{2}})\int_0^{\xi_\lambda} d\theta_1
 f_0(\theta_1)
 \notag \\
&\, \times \Biggl[ 
\lambda J_0^{(-1)}(\lambda,\theta_q,\theta_1)
+\frac{1}{4} \frac{\partial J_0^{(-)}(\lambda,\theta_q,\theta_1)}{\partial\theta_q} \Biggr]_{\theta_q=\frac{\pi}{2}}
\notag \\
&\,+\bar{f}_0({\textstyle\frac{\pi}{2}}) \Bigl[\tilde{\cal J}_0^{(1)}(\lambda,{\textstyle\frac{\pi}{2}})
- {\cal J}_0^{(-1)}(\lambda,{\textstyle\frac{\pi}{2}})\Bigr]-(\lambda\rightarrow 0)\Biggr\} ,
\label{eta^(2c3d5)-0}
\end{align}
where we have used $J_0^{(-)}(\lambda,\theta_q,\theta_q+{\textstyle\frac{\pi}{2}})=J_0^{(+)}(\lambda,\theta_q,{\textstyle\frac{\pi}{2}}-\theta_q)=0$ for Eq.\ (\ref{J_x^(pm)-2b4e}) and ${\partial f_0(\theta_q)}/{\partial\theta_q}\bigr|_{\theta_q=\frac{\pi}{2}}=0$ for Eq.\ (\ref{f_k}).
The quantities $(J_0^{(2)},J_0^{(-2)},\tilde{\kappa}_\lambda^{(n)},{\cal J}_0^{(-1)},{\cal J}_0^{(-2)},\tilde{\cal J}_0^{(n)})$ $(n=1,2)$ are given
by Eqs.\  (\ref{J_0^(2)}), (\ref{J_0^-2}), (\ref{tkappa^(12)}), (\ref{calJ^(pm1)-2b4e}), (\ref{calJ^(pm2)-2b4e}), and (\ref{tcalJ^(12)-2b4e}), respectively,
while $J_0^{(\pm n)}(\lambda,\theta_q,\theta_1)\!\equiv\!\partial^n J_x^{(\pm)}(\lambda,\theta_q,\theta_1)/\partial x^n$ for $n=1,2$ 
can be obtained easily from Eq.\ (\ref{J_x^(pm)-2b4e}). 
Substituting them, we can evaluate all the integrals of Eq.\ (\ref{eta^(2c3d5)-0}) both numerically and analytically, including the triple one by the order $\theta_q \rightarrow \lambda\rightarrow \theta_1$.
We eventually obtain 
\begin{align}
\eta^{(2{\rm c}3{\rm d}5)}=\frac{5}{288} g_*^2
= 0.0174g_*^2 .
\label{eta^(2c3d5)}
\end{align}

Sixth, we focus on the sixth term in the square brackets of Eq.\ (\ref{W^(2c3d)}).
We can transform the contribution
in the same way as the fifth term above.
We thereby obtain an expression of $\delta\tilde{W}_{\infty}^{(2{\rm c}3{\rm d}6)}$, which is apparently identical with
Eq.\ (\ref{tW^(2c3d5)-2}) except for the replacement of $f_0(\theta_q)$ by $f_{\tilde{k}}(\theta_q)$.
However, the basic functions $(J_{\lambda\tilde{k}}^{(\pm)},\tilde{J}_{\lambda\tilde{k}}^{(\pm)})$ are now defined by
\begin{subequations}
\label{J^(pm)-tJ-2c3d6}
\begin{align}
J_{\lambda\tilde{k}}^{(\pm)}(\lambda,\theta_q,\theta_q)\equiv &\,\frac{1}{2} \int_{s_{{\rm c}1}}^{\mp 1}ds_1 \frac{1}{1+\lambda^2\tilde{k}^2+2\lambda\tilde{k}\cos\theta_{k1}} 
\notag \\
&\,\times \frac{1}{1+\lambda^2\tilde{q}^{\prime 2}+2\lambda\tilde{q}'\cos\theta_{q'1}} ,
\label{J^(pm)-2c3d6}
\\
\tilde{J}_{\lambda\tilde{k}}^{(\pm)}(\lambda,\theta_q,\theta_q)\equiv &\,\frac{1}{2} \int_{s_{{\rm c}2}}^{\mp 1}ds_1 \frac{1}{1+\lambda^2\tilde{k}^2+2\lambda\tilde{k}\cos\theta_{k1}} 
\notag \\
&\,\times \frac{1}{1+\lambda^2\tilde{q}^{\prime 2}+2\lambda\tilde{q}'\cos\theta_{q'1}} ,
\label{tJ-2c3d6}
\end{align}
\end{subequations}
and $({\cal J}_{\tilde{k}}^{(\pm)},\tilde{\cal J}_{\tilde{k}})$ are given in terms of Eq.\ (\ref{tJ-2c3d6}) as Eq.\ (\ref{calJ^(pm)}) and (\ref{tcalJ}).

Substitution of $\delta\tilde{W}_{\infty}^{(2{\rm c}3{\rm d}6)}$ into Eq.\ (\ref{eta-def2})
yields $\eta^{(2{\rm c}3{\rm d}6)}$, whose expression can also be obtained from  Eq.\ (\ref{eta^(2c3d5)-0}) by replacing (i)
$f_0(\theta_q)\lambda^2J_0^{(2)}(\lambda,\theta_q,\theta_1)$ by
\begin{subequations}
\begin{align}
&\,f_0(\theta_q)\lambda^2J_0^{(2)}(\lambda,\theta_q,\theta_1)+2\frac{\partial f_{\tilde{k}}(\theta_q)}{\partial\tilde{k}}\Biggr|_{\tilde{k}=0} \lambda J_0^{(1)}(\lambda,\theta_q,\theta_1) 
\notag \\
&\,+\frac{\partial^2 f_{\tilde{k}}(\theta_q)}{\partial\tilde{k}^2}\Biggr|_{\tilde{k}=0}J_0(\lambda,\theta_q,\theta_1),
\end{align} 
(ii) $f_0(\theta_q)\lambda^2J_0^{(-2)}(\lambda,\theta_q,\theta_1)$ by
\begin{align}
&\,f_0(\theta_q)\lambda^2J_0^{(- 2)}(\lambda,\theta_q,\theta_1)+2\frac{\partial f_{\tilde{k}}(\theta_q)}{\partial\tilde{k}}\Biggr|_{\tilde{k}=0} \lambda J_0^{(-1)}(\lambda,\theta_q,\theta_1) 
\notag \\
&\,+\frac{\partial^2 f_{\tilde{k}}(\theta_q)}{\partial\tilde{k}^2}\Biggr|_{\tilde{k}=0}J_0^{(-)}(\lambda,\theta_q,\theta_1),
\end{align} 
(iii) $\bar{f}_0(\theta_q)[\tilde{\cal J}_0^{(2)}(\lambda,\theta_q)- {\cal J}_0^{(-2)}(\lambda,\theta_q)]$ by
\begin{align}
&\,\bar{f}_0(\theta_q)\left[\tilde{\cal J}_0^{(2)}(\lambda,\theta_q)- {\cal J}_0^{(-2)}(\lambda,\theta_q)\right]
\notag \\
&\,+2\frac{\partial \bar{f}_{\tilde{k}}(\theta_q)}{\partial\tilde{k}}\Biggr|_{\tilde{k}=0}\left[\tilde{\cal J}_0^{(1)}(\lambda,\theta_q)- {\cal J}_0^{(-1)}(\lambda,\theta_q)\right]  ,
\end{align}
(iv) $- \tilde{\kappa}_\lambda^{(1)}[{d \ln f_0(\theta_1)}/{d\theta_1}]J_0^{(-)}(\lambda,\theta_q,\theta_1)$ by
\begin{align}
\left[2\frac{\partial\ln f_{\tilde{k}}(\theta_1)}{\partial\tilde{k}}\Biggr|_{\tilde{k}=0}- \tilde{\kappa}_\lambda^{(1)}\frac{d \ln f_0(\theta_1)}{d\theta_1}\right]
J_0^{(-)}(\lambda,\theta_q,\theta_1).
\end{align}
\end{subequations}
The quantities $J_0^{(\pm n)}(\lambda,\theta_q,\theta_1)$,
${\cal J}_0^{(\pm n)}(\lambda,\theta_q)$,
and $\tilde{\cal J}_0^{(n)}(\lambda,\theta_q)$
are obtained as Eqs.\ (\ref{J_0^(pm n)-2c3d6}), (\ref{calJ^(pm1,2)-2c3d6}), and (\ref{tcalJ^(1,2)-2c3d6}), respectively.
Substituting them, we can evaluate all the integrals of $\eta^{(2{\rm c}3{\rm d}6)}$ both numerically and analytically, including the triple one by the order 
$\theta_q \rightarrow \lambda\rightarrow \theta_1$ by using $J_0^{(2)}=J_0^{(- 2)}-J_0^{(+2)}$ and 
$J_0^{(+2)}(\lambda,\theta_q,\theta_1)=-J_0^{(-2)}(\lambda,\pi-\theta_q,\theta_1)$.
We eventually obtain 
\begin{align}
\eta^{(2{\rm c}3{\rm d}6)}=-\left(\frac{17}{288}-\frac{\sqrt{3}}{48\pi}+\frac{9}{64\pi^2}\right) g_*^2
= -0.0618g_*^2 .
\label{eta^(2c3d6)}
\end{align}

Adding Eqs.\ (\ref{eta^(2c3d1)}), (\ref{eta^(2c3d2)}), (\ref{eta^(2c3d3)}), (\ref{eta^(2c3d4)}), (\ref{eta^(2c3d5)}), and (\ref{eta^(2c3d6)})
yields the 2c-3d contribution to $\eta$ as
\begin{align}
\eta^{(2{\rm c}3{\rm d})}=\left(-\frac{\sqrt{3}}{48\pi}+\frac{3}{16\pi^2}\right) g_*^2 =0.0075g_*^2.
\label{eta^(2c3d)}
\end{align}

\subsection{Sum of various 2c contributions}

The net 2c contribution is obtained by adding Eqs.\ (\ref{eta^(2c0)}), (\ref{eta^(2c3c)}), and (\ref{eta^(2c3d)}) as
\begin{align}
\eta^{(2{\rm c})}=\left(-\frac{3}{16}-\frac{\sqrt{3}}{48\pi}+\frac{3}{16\pi^2}\right)g_*^2= -0.1800g_*^2 .
\label{eta^(2c)}
\end{align}

\section{Summary}

Collecting Eqs.\ (\ref{eta^(2a)}), (\ref{eta^(2b)}), and (\ref{eta^(2c)}),
we obtain the coherence exponent for single-component Bose-Einstein condensates at $d\lesssim 4$ as
\begin{align}
\eta=\left(\frac{1}{16}-\frac{\sqrt{3}}{32\pi}\right)g_*^2 .
\end{align}
Substituting Eq.\ (\ref{epsilon-def}), we arrive at Eq.\ (\ref{eta-total}).

\section*{Acknowledgment}
This work is supported by Yamada Science Foundation.

\appendix

\section{Expansions of ${\cal J}_{\tilde{k}}^{(\pm)}(\lambda,\theta_q)$ and $\tilde{\cal J}_{\tilde{k}}(\lambda,\theta_q)$}
\label{App-calJ-exp}

\subsection{The 2b-4e contribution}
\label{App: A1}

We expand Eqs.\ (\ref{calJ^(pm)}) and (\ref{tcalJ}) up to the second order in $\tilde{k}$.
For this purpose, we substitute Eq.\ (\ref{tJ_x^(pm)-2b4e}) to express them as
\begin{subequations}
\label{calJ^(pm)-tcalJ-App}
\begin{align}
{\cal J}_{\tilde{k}}^{(\pm)}(\lambda,\theta_q)=&\, \frac{1}{4\lambda\tilde{k}}
\int_{\xi_\lambda-\kappa_\lambda^{(\pm)}}^{\xi_\lambda}d\theta_1 \bar{f}_0(\theta_1)w_{\lambda\tilde{k}}^{(\pm)}(\lambda,\theta_q,\theta_1),
\label{calJ^(pm)-App}
\end{align}
\begin{align}
\tilde{\cal J}_{\tilde{k}}(\lambda,\theta_q)=&\, \frac{1}{4\lambda\tilde{k}}
\int_{\xi_\lambda-\tilde{\kappa}_\lambda}^{\xi_\lambda}d\theta_1 \bar{f}_0(\theta_1)w_{\lambda\tilde{k}}^{(-)}(\lambda,\theta_q,\theta_1),
\label{tcalJ-App}
\end{align}
\end{subequations}
where $\bar{f}_0$ is given in Eq.\ (\ref{barf_k}), and $w_x^{(\pm)}$ are defined by
\begin{align}
w_x^{(\pm)}(\lambda,\theta_q,\theta_1)\equiv&\, \ln \Bigl\{1+\lambda^2+2\lambda\cos\theta_1
\notag \\
&\,+2x\bigl[\lambda\cos\theta_q
+\cos(\theta_q\pm \theta_1) \bigr]+x^2\Bigr\} ,
\label{w^(pm)}
\end{align}
satisfying $w_0^{(\pm)}(\lambda,\theta_q,\xi_\lambda)=0$, as shown by using Eq.\ (\ref{xi_k-def}).
Next, we write $({\cal J}_{\tilde{k}}^{(\pm)},\tilde{\cal J}_{\tilde{k}},\kappa_\lambda^{(\pm)},\tilde{\kappa}_\lambda, 
w_x^{(\pm)})$ in series of $\tilde{k}$ as
\begin{subequations}
\label{AppA-expansions}
\begin{align}
{\cal J}_{\tilde{k}}^{(\pm)}(\lambda,\theta_q)=&\, \sum_{n=1}^\infty\frac{{\cal J}_0^{(\pm n)}(\lambda,\theta_q)}{n!}\tilde{k}^n ,
\label{calJ^(pm)-exp}
\\
\tilde{\cal J}_{\tilde{k}}(\lambda,\theta_q)=&\, \sum_{n=1}^\infty\frac{\tilde{\cal J}_0^{(n)}(\lambda,\theta_q)}{n!}\tilde{k}^n ,
\label{tcalJ-exp}
\\
\kappa_\lambda^{(\pm)}(\theta_q,\tilde{k})=&\,\sum_{n=1}^\infty\frac{\kappa_\lambda^{(\pm n)}(\theta_q)}{n!}\tilde{k}^n,
\label{kappa^(pm)-exp}
\\
\tilde{\kappa}_\lambda(\theta_q,\tilde{k})=&\,\sum_{n=1}^\infty\frac{\tilde{\kappa}_\lambda^{(n)}(\theta_q)}{n!}\tilde{k}^n,
\label{tkappa-exp}
\\
w_x^{(\pm)}(\lambda,\theta_q,\theta_1)=&\,\sum_{n=0}^\infty\frac{w_0^{(\pm n)}(\lambda,\theta_q,\theta_1)}{n!}x^n .
\label{w^(pm)-exp}
\end{align}
\end{subequations}
The fact that $n=0$ terms are absent except Eq.\ (\ref{w^(pm)-exp})
will be confirmed shortly.
Let us expand the integral of Eq.\ (\ref{calJ^(pm)-App}) up to the second order in $\kappa_\lambda^{(\pm)}$ and substitute
Eqs.\ (\ref{calJ^(pm)-exp}), (\ref{kappa^(pm)-exp}), and (\ref{w^(pm)-exp}) into the resulting expression.
Comparing the coefficients of $\tilde{k}^n$, 
we can express ${\cal J}_0^{(\pm n)}$ for $n=1,2$ in terms of 
$\kappa_\lambda^{(\pm n')}$ and $w_0^{(\pm n')}$ with $n'\leq n$ as
\begin{subequations}
\label{calJ^(pm1,2)}
\begin{align}
&\,{\cal J}_0^{(\pm 1)}(\lambda,\theta_q)
\notag \\
=&\,
 \frac{1}{4}\bar{f}_0(\xi_\lambda)\kappa_\lambda^{(\pm 1)}\left[
 w_0^{(\pm1)}(\lambda,\theta_q,\xi_\lambda)
-\frac{\kappa_\lambda^{(\pm 1)}}{2\lambda}\frac{\partial w_0^{(\pm0)}(\lambda,\theta_q,\theta_1)}{\partial\theta_1}\Biggr|_{\theta_1=\xi_\lambda}\right],
\label{calJ^(pm1)}
\end{align}
\begin{align}
&\,{\cal J}_0^{(\pm 2)}(\lambda,\theta_q)
\notag \\
=&\,\frac{1}{4}\bar{f}_0(\xi_\lambda)\Biggl\{\lambda \kappa_\lambda^{(\pm 1)} w_0^{(\pm2)}(\lambda,\theta_q,\xi_\lambda) 
+\kappa_\lambda^{(\pm 2)}w_0^{(\pm1)}(\lambda,\theta_q,\xi_\lambda)
\notag \\
&\,
-\bigl(\kappa_\lambda^{(\pm 1)}\bigr)^2\Biggl[\frac{\bar{f}_0'(\xi_\lambda)}{\bar{f}_0(\xi_\lambda)}w_0^{(\pm1)}(\lambda,\theta_q,\xi_\lambda)
+\frac{\partial w_0^{(\pm1)}(\lambda,\theta_q,\theta_1)}{\partial\theta_1}\Biggr|_{\theta_1=\xi_\lambda}\Biggr]
\notag \\
&\, 
-\frac{\kappa_\lambda^{(\pm 1)}\kappa_\lambda^{(\pm 2)}}{\lambda}\frac{\partial w_0^{(\pm0)}(\lambda,\theta_q,\theta_1)}{\partial\theta_1}\Biggr|_{\theta_1=\xi_\lambda}
+\frac{\bigl(\kappa_\lambda^{(\pm 1)}\bigr)^3}{3\lambda}
\Biggl[2\frac{\bar{f}_0'(\xi_\lambda)}{\bar{f}_0(\xi_\lambda)}
\notag \\
&\,
\times \frac{\partial w_0^{(\pm0)}(\lambda,\theta_q,\theta_1)}{\partial\theta_1}\Biggr|_{\theta_1=\xi_\lambda}
+\frac{\partial^2 w_0^{(\pm0)}(\lambda,\theta_q,\theta_1)}{\partial\theta_1^2}\Biggr|_{\theta_1=\xi_\lambda}\,\Biggr]\Biggr\} .
\label{calJ^(pm2)}
\end{align}
\end{subequations}
The coefficients $\tilde{\cal J}_0^{(n)}$ in Eq.\ (\ref{tcalJ-exp}) for $n=1,2$ are obtained from ${\cal J}_0^{(\pm n)}$
above by the replacement $(\kappa_\lambda^{(\pm n')},w_0^{(\pm n')})\rightarrow (\tilde{\kappa}_\lambda^{(n')},w_0^{(- n')})$.

The expansion coefficients of Eqs.\ (\ref{kappa^(pm)-exp}) and (\ref{tkappa-exp}) for $n\leq 2$ are obtained from Eqs.\ (\ref{kappa^(pm)-1}) 
and (\ref{tkappa}) as
\begin{subequations}
\label{kappa^(pm12)}
\begin{align}
\kappa_\lambda^{(\pm 1)}=&\, \frac{\cos(\theta_q\mp\xi_\lambda)}{\sin\xi_{\lambda}} ,
\label{kappa^(pm1)}
\\
\kappa_\lambda^{(\pm 2)}=&\, \cot\xi_\lambda
\sin^2\theta_q+\cot^3\xi_\lambda\cos^2\theta_q\mp\sin2\theta_q ,
\label{kappa^(pm2)}
\end{align}
\end{subequations}
\begin{align}
\tilde{\kappa}_\lambda^{(1)}=&\, 2\cot\xi_\lambda\cos\theta_q ,
\hspace{5mm}\tilde{\kappa}_\lambda^{(2)}= 4\cot^3\xi_\lambda\cos^2\theta_q .
\label{tkappa^(12)}
\end{align}
Moreover, the coefficients concerning $w_0^{(\pm n)}$  in Eq.\ (\ref{calJ^(pm1,2)})
can be calculated elementarily from Eq.\ (\ref{w^(pm)}) as
\begin{subequations}
\label{w^(pm)-derives}
\begin{align}
w_0^{(\pm 1)}(\lambda,\theta_q,\xi_\lambda)
=&\, -2\cos(\theta_q\mp\xi_\lambda),
\label{w^(pm)-derives1}
\\
w_0^{(\pm 2)}(\lambda,\theta_q,\xi_\lambda)
=&\, -2\cos2(\theta_q\mp\xi_\lambda),
\\
\frac{\partial w_0^{(\pm 0)}(\lambda,\theta_q,\theta_1)}{\partial\theta_1}\Biggr|_{\theta_1=\xi_\lambda}
=&\,
-2\lambda\sin\xi_\lambda,
\\
\frac{\partial w_0^{(\pm 1)}(\lambda,\theta_q,\theta_1)}{\partial\theta_1}\Biggr|_{\theta_1=\xi_\lambda}
=&\,2\sin(3\xi_\lambda\mp\theta_q) ,
\\
\frac{\partial^2 w_0^{(\pm 0)}(\lambda,\theta_q,\theta_1)}{\partial\theta_1^2}\Biggr|_{\theta_1=\xi_\lambda}
=&\,\lambda^2(1-4\sin^2\xi_\lambda),
\label{w^(pm)-derives5}
\end{align}
\end{subequations}
where we have used Eq.\ (\ref{xi_k-def}). 
Substituting Eqs.\ (\ref{barf_k}), (\ref{kappa^(pm12)}), and (\ref{w^(pm)-derives}) into Eq.\ (\ref{calJ^(pm1,2)}),
we obtain 
\begin{align}
{\cal J}_0^{(\pm 1)}(\lambda,\theta_q)=-\frac{\cos^2(\theta_q\mp\xi_\lambda)}{2\pi} ,
\label{calJ^(pm1)-2b4e}
\end{align}
and Eq.\ (\ref{calJ^(pm2)-2b4e}).
The same calculation with $(\tilde{\kappa}^{(n')},w_0^{(-n')})$ in place of $({\kappa}^{(\pm n')},w_0^{(\pm n')})$
yields the following expressions for the first two expansion coefficients of Eq.\ (\ref{tcalJ-exp}):
\begin{subequations}
\label{tcalJ^(12)-2b4e}
\begin{align}
\tilde{\cal J}_0^{(1)}(\lambda,\theta_q)
=&\,\frac{2}{\pi}\sin\xi_\lambda\cos\xi_\lambda\sin\theta_q\cos\theta_q,
\label{tcalJ^(1)-2b4e}
\end{align}
\begin{align}
\tilde{\cal J}_0^{(2)}(\lambda,\theta_q)
=&\,\frac{4}{\pi}\cos\xi_\lambda\cos\theta_q\Biggl[ \cos\xi_\lambda\cos2(\theta_q+\xi_\lambda)
\notag \\
&\,-\cot\xi_\lambda\cos\theta_q
\sin(\theta_q\!+\!3\xi_\lambda)+\frac{8\cos^3\xi_\lambda\cos^2\theta_q}{3}\Biggr] .
\label{tcalJ^(2)-2b4e}
\end{align}
\end{subequations}

\subsection{The 2b-4f5 contribution}

Integration of Eq.\ (\ref{tJ_k-J_k1-2b4f5}) can be performed elementarily.
The result is expressible by using Eqs.\ (\ref{s_c2}), (\ref{xi_k-def}), and (\ref{SpheC}) as
\begin{align}
\tilde{J}_{\lambda\tilde{k}}^{(\pm)}(\lambda,\theta_q,\theta_1)=\frac{w_{\lambda\tilde{k}}^{(\pm)}(\lambda,\theta_q,\theta_1)}{4\lambda\tilde{k}\sin\theta_q\sin\theta_1},
\label{tJ-w}
\end{align}
with
\begin{align}
w_x^{(\pm)}(\lambda,\theta_q,\theta_1)=&\,\lambda^2+2\lambda\cos\theta_1+2x\bigl[\lambda\cos\theta_q+\cos(\theta_q\pm\theta_1)\bigr] 
\notag \\
&\, +x^2 .
\label{w-2b4f5}
\end{align}
The latter functions satisfy $w_0^{(\pm)}(\lambda,\theta_q,\xi_\lambda)=0$ once again,
and the right-hand sides of Eqs.\ (\ref{w^(pm)-derives1})-(\ref{w^(pm)-derives5}) for Eq.\ (\ref{w-2b4f5}) are replaced by
$-2\cos(\theta_q\mp\xi_\lambda)$, $2$,
$-2\lambda\sin\xi_\lambda$, $-2\sin(\xi_\lambda\pm\theta_q)$, 
and $\lambda^2$, respectively.
Let us substitute these coefficients and Eq.\ (\ref{kappa^(pm12)}) into Eq.\ (\ref{calJ^(pm2)}),
where we should also replace $\bar{f}_0$ by
\begin{align}
\bar{\tilde{f}}_\lambda(\theta_1)\equiv \frac{\tilde{f}_\lambda(\theta_1)}{\sin\theta_1}=\frac{2}{\pi}\frac{\sin\theta_1}{(1+\lambda^2+2\lambda\cos\theta_1)^2},
\label{barf_k-2b4f5}
\end{align}
as seen by comparing Eq.\ (\ref{calJ^(pm)-2b4f5}) with Eq.\ (\ref{calJ^(pm)}).
Simplifying the resulting expression, we obtain ${\cal J}^{(\pm 2)}_0$ for the 2b-4f5 contribution as
\begin{align}
{\cal J}^{(\pm 2)}_0(\lambda,\theta_q)
=&\, \frac{1}{\pi}\Biggl[\lambda\cos(\theta_q\mp\xi_\lambda)
+\frac{\cos^2(\theta_q\mp\xi_\lambda)}{\sin\xi_\lambda}\sin(\xi_\lambda\pm \theta_q)
\notag \\
&\,
+\frac{4\lambda\cos^3(\theta_q\mp\xi_\lambda)}{3}
\Biggr] .
\label{calJ_0^(pm2)-2b4f5}
\end{align}

\subsection{The 2b-4f6 contribution}

Integration of Eq.\ (\ref{tJ_x^(pm)-2b4f6}) can be performed elementarily.
The result is expressible by using 
Eqs.\ (\ref{xi_k-def}),  (\ref{SpheC}), (\ref{q'}), and (\ref{s_c2}) as
Eq.\ (\ref{tJ-w}), where $w_x^{(\pm)}$ are given by
\begin{align}
w_x^{(\pm)}(\lambda,\theta_q,\theta_1)=&\,
\frac{1}{1-\lambda^2-2\lambda\cos\theta_1-2\lambda x\cos\theta_q}
\notag \\
&\,-\frac{1}{1+2x\cos(\theta_q\pm\theta_1)+x^2} .
\label{w-2b4f6}
\end{align}
They also satisfy $w_0^{(\pm)}(\lambda,\theta_q,\xi_\lambda)=0$,
and the right-hand sides of Eqs.\ (\ref{w^(pm)-derives1})-(\ref{w^(pm)-derives5}) for Eq.\ (\ref{w-2b4f6}) are given explicitly by
$-2\cos(\theta_q\mp\xi_\lambda)$, $2\bigl[1-4\cos^2(\theta_q\pm\xi_\lambda)+4\lambda^2\cos^2\theta_q\bigr]$,
$-2\lambda\sin\xi_\lambda$, $-2\sin(\xi_\lambda\pm\theta_q)-8\lambda^2\sin\xi_\lambda\cos\theta_q$, 
and $\lambda^2(1+8\sin^2\xi_\lambda)$, respectively.
Substitution of these coefficients and Eq.\ (\ref{kappa^(pm12)}) into Eq.\ (\ref{calJ^(pm2)}) yields
\begin{align}
&\,{\cal J}^{(\pm 2)}_0(\lambda,\theta_q)
\notag \\
=&\, \frac{\cos(\theta_q\mp\xi_\lambda)}{\pi}\Biggl\{ \lambda
\Bigl[1-4\cos^2(\theta_q\pm\xi_\lambda)+4\lambda^2\cos^2\theta_q\Bigr]
\notag \\
&\,
+\frac{\cos(\theta_q\mp\xi_\lambda)}{\sin\xi_\lambda}
\Bigl[
\sin(\xi_\lambda\pm\theta_q)+4\lambda^2\sin\xi_\lambda\cos\theta_q\Bigr]
\notag \\
&\, +\frac{4\lambda}{3}\cos^2(\theta_q\mp\xi_\lambda)\Biggr\}.
\label{calJ_0^(pm2)-2b4f6}
\end{align}
Repeating the calculation with Eq.\ (\ref{tkappa^(12)}) and $w_0^{(-n)}$ in place of Eq.\ (\ref{kappa^(pm12)}) and $w_0^{(\pm n)}$, respectively,
we obtain $\tilde{\cal J}^{(2)}_0$ for the 2b-4f6 contribution as
\begin{align}
&\,\tilde{\cal J}^{(2)}_0(\lambda,\theta_q)
\notag \\
=&\,\frac{2}{\pi}\cos\xi_\lambda\cos\theta_q\Biggl\{ \lambda\Bigl[1-4\cos^2(\theta_q-\xi_\lambda)\Bigr]
\notag \\
&\,
+2\cot\xi_\lambda\cos\theta_q\sin(\xi_\lambda-\theta_q)
+\frac{4\lambda^3\cos^2\theta_q}{3}\Biggl\}.
\label{tcalJ_0^(2)-2b4f6}
\end{align}

\subsection{The 2b-4f7 contribution}

Integration of Eq.\ (\ref{tJ_x^(pm)-2b4f7}) can be performed elementarily.
The result is expressible by using 
Eqs.\ (\ref{xi_k-def}),  (\ref{SpheC}), (\ref{q'}), and (\ref{s_c2}) as
Eq.\ (\ref{tJ-w}), where $w_x^{(\pm)}$ are now given by
\begin{align}
w_x^{(\pm)}(\lambda,\theta_q,\theta_1)=&\,\ln \frac{1+2x\cos(\theta_q\pm \theta_1)+x^2}{
1-\lambda^2-2\lambda\cos\theta_1-2\lambda x\cos\theta_q} .
\label{w-2b4f7}
\end{align}
They also satisfy $w_0^{(\pm)}(\lambda,\theta_q,\xi_\lambda)=0$,
and the right-hand sides of Eqs.\ (\ref{w^(pm)-derives1})-(\ref{w^(pm)-derives5}) for Eq.\ (\ref{w-2b4f7}) are given explicitly by
$-2\cos(\theta_q\mp\xi_\lambda)$, $2\bigl[1-2\cos^2(\theta_q\pm\xi_\lambda)+2\lambda^2\cos^2\theta_q\bigr]$,
$-2\lambda\sin\xi_\lambda$, $-2\sin(\xi_\lambda\pm\theta_q)-4\lambda^2\sin\xi_\lambda\cos\theta_q$, 
and $\lambda^2(1+4\sin^2\xi_\lambda)$, respectively.
Let us substitute these coefficients and Eq.\ (\ref{kappa^(pm12)}) into Eq.\ (\ref{calJ^(pm2)}),
where we should also use $\bar{f}_\lambda$ in place of $\bar{f}_0$, 
as seen by comparing Eq.\ (\ref{calJ^(pm)-2b4f7}) with Eq.\ (\ref{calJ^(pm)}), where $\bar{f}_\lambda$ is defined by Eq.\ (\ref{barf_k}).
Simplifying the resulting expression, we obtain ${\cal J}^{(\pm 2)}_0$ for the 2b-4f7 contribution as
\begin{subequations}
\begin{align}
&\,{\cal J}^{(\pm 2)}_0(\lambda,\theta_q)
\notag \\
=&\,
\frac{\cos(\theta_q\mp\xi_\lambda)}{\pi}\Biggl\{ \lambda\Bigl[1-2\cos^2(\theta_q\pm\xi_\lambda)+2\lambda^2\cos^2\theta_q\Bigr]
\notag \\
&\,
+\frac{\cos(\theta_q\mp\xi_\lambda)}{\sin\xi_\lambda}
\Bigl[\sin(\xi_\lambda\pm\theta_q)+2\lambda^2\sin\xi_\lambda\cos\theta_q\Bigr]
\notag \\
&\,
+\frac{4\lambda}{3}\cos^2(\theta_q\mp\xi_\lambda)
\Biggr\}.
\label{calJ_0^(pm2)-2b4f7}
\end{align}
The same calculation of using $(\tilde{\kappa}^{(n')},w_0^{(-n')})$ in place of $({\kappa}^{(\pm n')},w_0^{(\pm n')})$
yields 
\begin{align}
&\,\tilde{\cal J}_0^{(2)}(\lambda,\theta_q)
\notag \\
=&\,\frac{2\cos\xi_\lambda\cos\theta_q}{\pi}\Biggl\{ \lambda\Bigl[1-2\cos^2(\theta_q-\xi_\lambda)\Bigr]-\frac{8\lambda\cos^2\xi_\lambda\cos^2\theta_q}{3}
\notag \\
&\,
+2\cot\xi_\lambda\cos\theta_q\bigl[2\lambda\sin\xi_\lambda\cos(\theta_q+\xi_\lambda)
+\sin(\xi_\lambda-\theta_q)\bigr]\Biggl\} .
\label{tcalJ_0^(2)-2b4f7}
\end{align}
\end{subequations}

\subsection{The 2c-3d2 contribution}

Equation (\ref{calJ^(+)-2c3d2}) differs from Eq.\ (\ref{calJ^(pm)-App}) in that
$(\bar{f}_\lambda,w_x)$ given by Eqs.\ (\ref{barf_k}) and Eq.\ (\ref{w-2b4f5}), respectively, 
are used in place of $(\bar{f}_0,w_x)$.
With these modifications in Eq.\ (\ref{calJ^(pm1,2)}), we obtain
\begin{subequations}
\label{calJ_0^(pm12)-2c3d2}
\begin{align}
{\cal J}^{(+1)}_0(\lambda,\theta_q)=-\frac{\cos^2(\theta_q\mp\xi_\lambda)}{2\pi} ,
\label{calJ_0^(pm1)-2c3d2}
\end{align}
\begin{align}
{\cal J}^{(+2)}_0(\lambda,\theta_q)
=&\, \frac{1}{\pi}\Biggl[\lambda\cos(\theta_q-\xi_\lambda)
+\frac{\cos^2(\theta_q-\xi_\lambda)}{\sin\xi_\lambda}\sin(\xi_\lambda+ \theta_q)
\notag \\
&\,+\frac{2\lambda\cos^3(\theta_q-\xi_\lambda)}{3}
\Biggr] .
\label{calJ_0^(pm2)-2c3d2}
\end{align}
\end{subequations}
The difference of Eq.\ (\ref{calJ_0^(pm2)-2c3d2}) from Eq.\ (\ref{calJ_0^(pm2)-2b4f5}) is caused
by the replacement $\bar{\tilde{f}}_\lambda\rightarrow \bar{f}_\lambda$.

\subsection{The 2c-3d6 contribution}

We derive $J_0^{(\pm n)}(\lambda,\theta_q,\theta_1)\!\equiv\!\partial^n J_x^{(\pm)}(\lambda,\theta_q,\theta_1)/\partial x^n$
for $n=0,1,2$ besides $({\cal J}_0^{(\pm n)},\tilde{\cal J}_0^{(n)})$ for $n=1,2$
defined with Eq.\ (\ref{J^(pm)-tJ-2c3d6}).
Using Eqs.\ (\ref{SpheC}), (\ref{s_c2}), and (\ref{s_c1}), we can calculate the integral of Eq.\ (\ref{J^(pm)-2c3d6}) analytically 
to obtain
\begin{align}
&\,J_x^{(\pm)}(\lambda,\theta_q,\theta_1)
\notag \\
=&\, \frac{1}{4x\sin\theta_q\sin\theta_1(\lambda^2+2\lambda\cos\theta_1+2\lambda x \cos\theta_q)}
\notag \\
&\,\times
\Biggl\{\ln\frac{1+2x\cos(\theta_q\pm \theta_1)+x^2}{1+\lambda^2+2\lambda\cos\theta_1+2x\bigl[\lambda\cos\theta_q+\cos(\theta_q\pm\theta_1)\bigr]+x^2}
\notag \\
&\, +\ln \Bigl(1+\lambda^2+2\lambda\cos\theta_1+2\lambda x\cos\theta_q\Bigr)\Biggr\}.
\label{J_x^(pm)-2c3d6}
\end{align}
Equation (\ref{tJ-2c3d6}) can be integrated similarly, which is expressible as Eq.\ (\ref{tJ-w}) with 
\begin{align}
&\,w_x^{(\pm)}(\lambda,\theta_q,\theta_1)
\notag \\
=&\, \frac{1}{\lambda^2+2\lambda\cos\theta_1+2\lambda x\cos\theta_q}
\notag \\
&\,\times
\Biggl\{\ln\frac{1+2x\cos(\theta_q\pm \theta_1)+x^2}{1+\lambda^2+2\lambda\cos\theta_1+2x\bigl[\lambda\cos\theta_q+\cos(\theta_q\pm\theta_1)\bigr]+x^2}
\notag \\
&\, -\ln \Bigl(1-\lambda^2-2\lambda\cos\theta_1-2\lambda x\cos\theta_q\Bigr)\Biggr\}.
\label{w^(pm)-2c3d6}
\end{align}
They also satisfy $w_0^{(\pm)}(\lambda,\theta_q,\xi_\lambda)=0$,
and the right-hand sides of Eqs.\ (\ref{w^(pm)-derives1})-(\ref{w^(pm)-derives5}) for Eq.\ (\ref{w^(pm)-2c3d6}) are given explicitly by
$-2\cos(\theta_q\mp\xi_\lambda)$, $2(1+4\cos^2\xi_\lambda\cos^2\theta_q-4\sin^2\xi_\lambda\sin^2\theta_q)$,
$-2\lambda\sin\xi_\lambda$, $2[-\sin(\xi_\lambda\pm\theta_q)+2\lambda\sin\xi_\lambda\cos(\theta_q\pm\xi_\lambda)]$, and $\lambda^2$, respectively.
Substituting these coefficients and Eq.\ (\ref{kappa^(pm12)}) into Eq.\ (\ref{calJ^(pm1,2)}),
we obtain ${\cal J}^{(\pm 1)}_0$ and  ${\cal J}^{(\pm 2)}_0$ for the 2c-3d6 contribution as
\begin{subequations}
\label{calJ^(pm1,2)-2c3d6}
\begin{align}
{\cal J}^{(\pm 1)}_{0}(\lambda,\theta_q)
=-\frac{\cos^2(\theta_q\mp\xi_\lambda)}{2\pi},
\label{calJ^(pm1)-2c3d6}
\end{align}
\begin{align}
&\,{\cal J}_0^{(\pm 2)}(\lambda,\theta_q)
\notag \\
=&\, \frac{1}{\pi}\cos(\theta_q\mp\xi_\lambda)\Biggl[\lambda 
\left(\frac{1}{2}+2\cos^2\xi_\lambda\cos^2\theta_q-2\sin^2\xi_\lambda\sin^2\theta_q\right)
\notag \\
&\,
\pm\frac{\cos\theta_q\sin\theta_q}{\sin\xi_\lambda}\Biggr].
\label{calJ^(pm2)-2c3d6}
\end{align}
\end{subequations}
The same calculation of using $(\tilde{\kappa}^{(n')},w_0^{(-n')})$ in place of $({\kappa}^{(\pm n')},w_0^{(\pm n')})$
yields 
\begin{subequations}
\label{tcalJ^(1,2)-2c3d6}
\begin{align}
\tilde{\cal J}_0^{(1)}(\lambda,\theta_q)=\frac{2}{\pi} \sin\xi_\lambda\cos\xi_\lambda\sin\theta_q\cos\theta_q,
\label{tcalJ^(1)-2c3d6}
\end{align}
\begin{align}
&\,\tilde{\cal J}_0^{(2)}(\lambda,\theta_q)
\notag \\
=&\, -\frac{\lambda^2}{\pi}(1-4\sin^2\xi_\lambda)\cos\theta_q\Bigl(
\sin^2\theta_q
+\cot\xi_\lambda\sin\theta_q\cos\theta_q\Bigr).
\label{tcalJ^(2)-2c3d6}
\end{align}
\end{subequations}

The coefficients $J_0^{(\pm n)}(\lambda,\theta_q,\theta_1)\equiv \partial^n J_x^{(\pm)}(\lambda,\theta_q,\theta_1)/\partial x^n\bigr|_{x=0}$
for $n=0,1,2$ are obtained from Eq.\ (\ref{J_x^(pm)-2c3d6}) as
\begin{subequations}
\label{J_0^(pm n)-2c3d6}
\begin{align}
J_0^{(\pm)}(\lambda,\theta_q,\theta_1)=&\,\frac{\cos(\theta_q\pm\theta_1)}{2\sin\theta_q\sin\theta_1(1+\lambda^2+2\lambda\cos\theta_1)},
\end{align}
\begin{align}
&\,J_0^{(\pm 1)}(\lambda,\theta_q,\theta_1)
\notag \\
=&\,\frac{1}{4\sin\theta_q\sin\theta_1(1+\lambda^2+2\lambda\cos\theta_1)^2}\Bigl[
1+\lambda^2+2\lambda \cos\theta_1
\notag \\
&\,
-4\cos^2(\theta_q\pm\theta_1)-2(\lambda^2+2\lambda \cos\theta_1)\cos^2(\theta_q\pm\theta_1)
\notag \\
&\,
-4\lambda\cos\theta_q\cos(\theta_q\pm\theta_1)\Bigr],
\end{align}
\begin{align}
&\,J_0^{(\pm 2)}(\lambda,\theta_q,\theta_1)
\notag \\
=&\,\frac{\Bigl[2\cos(\theta_q\pm\theta_1)+\lambda\cos\theta_q(3\!+\!\lambda^2\!+\!2\lambda\cos\theta_1)\Bigr]
\cos2(\theta_q\pm\theta_1)}{\sin\theta_q\sin\theta_1(1+\lambda^2+2\lambda\cos\theta_1)^3}
\notag \\
&\,
+\frac{(\lambda^2+2\lambda \cos\theta_1)(3+\lambda^2+2\lambda \cos\theta_1)\cos3(\theta_q\pm\theta_1)}{3\sin\theta_q\sin\theta_1(1+\lambda^2+2\lambda\cos\theta_1)^3}
\notag \\
&\,
+\frac{2\lambda\cos\theta_q\Bigl[1+2\lambda\cos\theta_q\cos(\theta_q\pm\theta_1)]}{\sin\theta_q\sin\theta_1(1+\lambda^2+2\lambda\cos\theta_1)^3}.
\end{align}
\end{subequations}

\section{Derivation of $\delta\tilde{W}^{(2{\rm b}4{\rm f})}_\infty$}
\label{App-2b4f}

\begin{figure}[b]
\begin{center}
\includegraphics[width=0.95\linewidth]{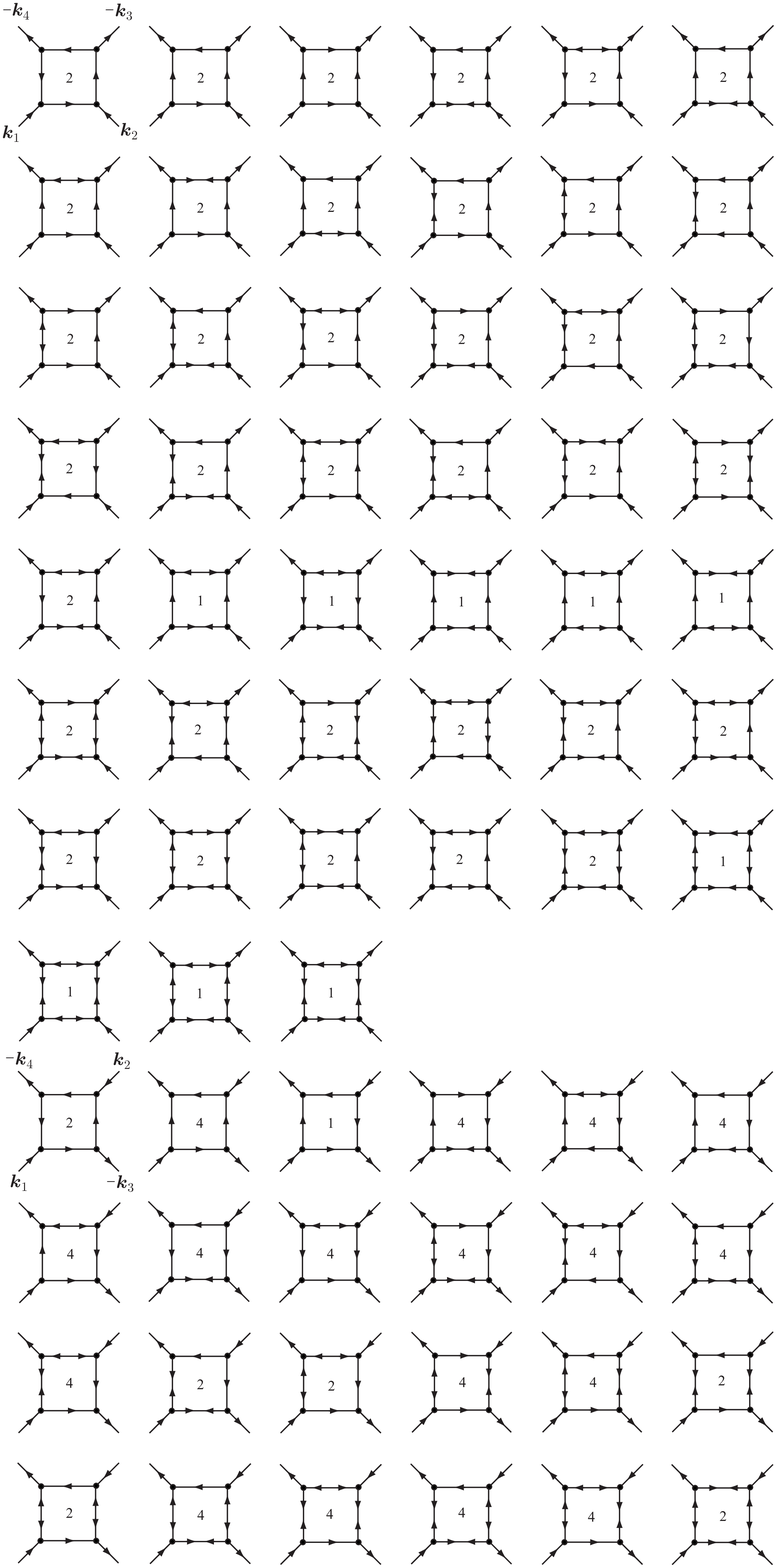}
\end{center}
\caption{Diagrammatic expressions of $W^{(4{\rm f})}_{\Lambda,1122}$.
The number inside each loop  indicates its weight.
The total number of diagrams is $3^4+3^4=162$.
\label{Fig9a}}
\end{figure}
\begin{figure}[t]
\begin{center}
\includegraphics[width=0.95\linewidth]{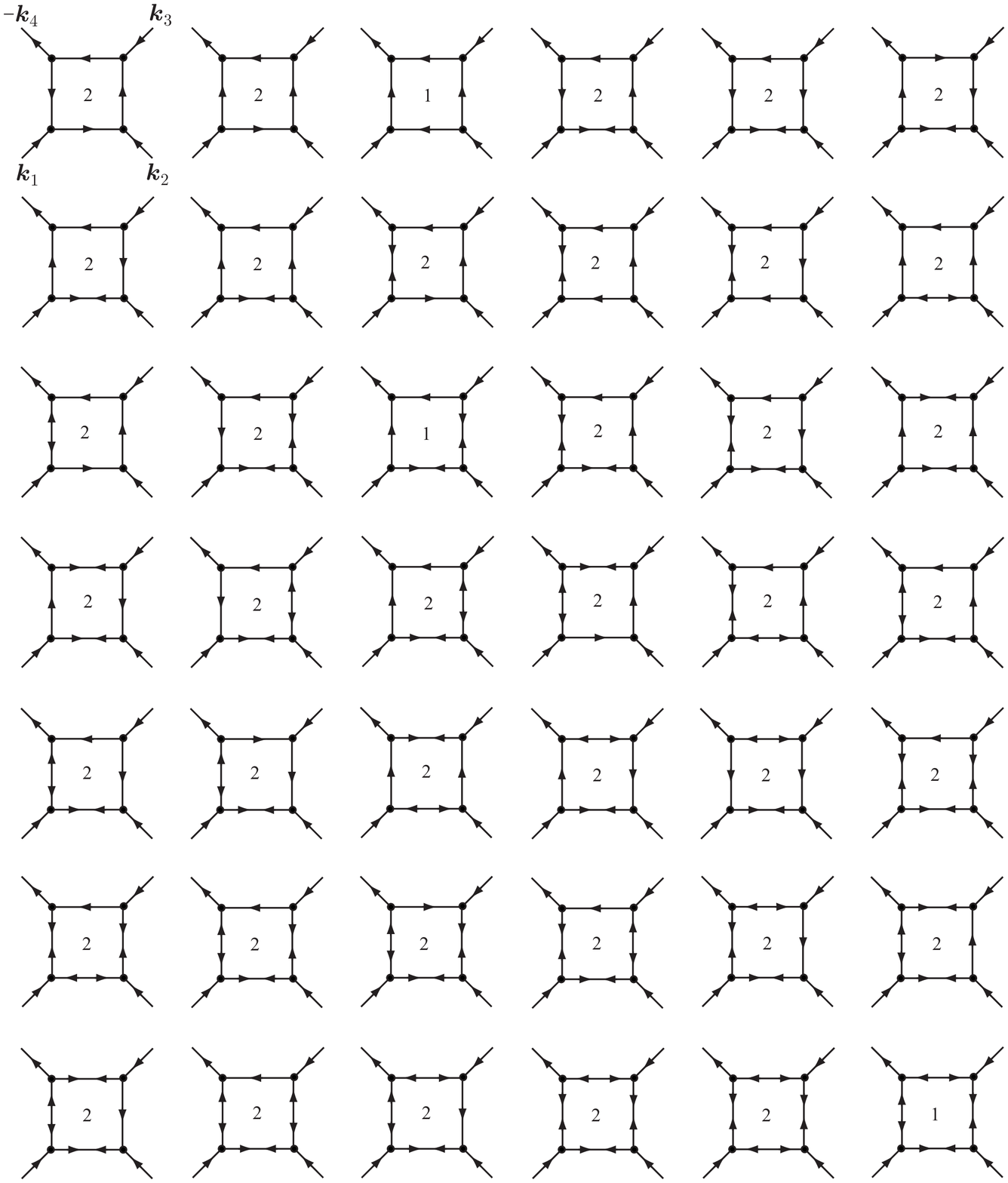}
\end{center}
\caption{Diagrammatic expressions of $W^{(4{\rm f})}_{\Lambda,1112}$.
The number inside each loop  indicates its weight.
The total number of diagrams is $3^4=81$.
\label{Fig9b}}
\end{figure}
\begin{figure}[t]
\begin{center}
\includegraphics[width=0.95\linewidth]{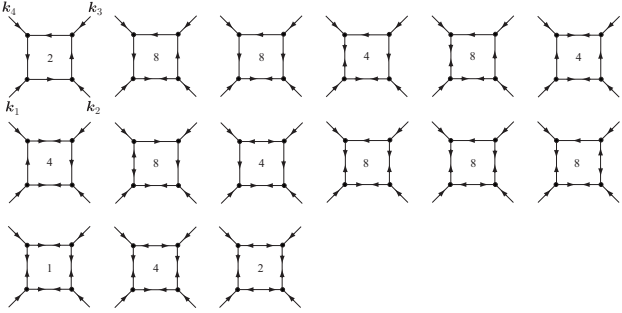}
\end{center}
\caption{Diagrammatic expressions of $W^{(4{\rm f})}_{\Lambda,1111}$.
The number inside each loop  indicates its weight.
The total number of diagrams is $3^4=81$.
\label{Fig9c}}
\end{figure}

The contribution of Fig.\ \ref{Fig1} (4f) to Eq.\ (\ref{tW^(4)}) is  expressible diagrammatically as Figs.\ \ref{Fig9a}-\ref{Fig9c}
by using the relevant vertices of Eq.\ (\ref{Gamma^(3)}) 
and adding an incoming (outgoing) arrow for $j=1$ ($j=2$) around each vertex.
The corresponding analytic expressions of $W^{(4{\rm f})}_{\Lambda,1122}$, $W^{(4{\rm e})}_{\Lambda,1112}$, and $W^{(4{\rm e})}_{\Lambda,1111}$ can be 
obtained from Eq.\ (46e) of I. To present them concisely, we introduce the function
\begin{align}
&\, \chi_{ABCD}^{1,2,3}
\notag \\
\equiv &\,  \frac{\delta_{{\bf k}_1+{\bf k}_2+{\bf k}_3+{\bf k}_4}}{\beta}\int\frac{d^dq}{(2\pi)^d}
\notag \\
&\,\times \Bigl[\dot{A}(q){B}(|{\bf k}_1+{\bf q}|){C}(|{\bf k}_1+{\bf k}_2+{\bf q}|)
{D}(|{\bf k}_1+{\bf k}_2+{\bf k}_3+{\bf q}|)
\notag \\
&\,+{A}(q)\dot{B}(|{\bf k}_1+{\bf q}|){C}(|{\bf k}_1+{\bf k}_2+{\bf q}|){D}(|{\bf k}_1+{\bf k}_2+{\bf k}_3+{\bf q}|)
\notag \\
&\,+{A}(q){B}(|{\bf k}_1+{\bf q}|)\dot{C}(|{\bf k}_1+{\bf k}_2+{\bf q}|){D}(|{\bf k}_1+{\bf k}_2+{\bf k}_3+{\bf q}|)
\notag \\
&\,+{A}(q){B}(|{\bf k}_1+{\bf q}|){C}(|{\bf k}_1+{\bf k}_2+{\bf q}|)\dot{D}(|{\bf k}_1+{\bf k}_2+{\bf k}_3+{\bf q}|)\Bigr] .
\label{chi_ABCD}
\end{align}
Using it, we can express the two sets of diagrams in Figs.\ \ref{Fig9a} analytically as
\begin{subequations}
\begin{align}
\chi^{1,2,3}_{{\rm a}1}=&\,6\chi_{GGGG}^{1,2,3}-6\chi_{GFGG}^{1,2,3}-6\chi_{GGGF}^{1,2,3}-4\chi_{GGFG}^{1,2,3}
-4\chi_{FGGG}^{1,2,3}
\notag \\
&\,+5\chi_{FFGG}^{1,2,3}+5\chi_{GFFG}^{1,2,3}+5\chi_{GGFF}^{1,2,3}+5\chi_{FGGF}^{1,2,3}+2\chi_{FGFG}^{1,2,3}
\notag \\
&\,+7\chi_{GFGF}^{1,2,3}-4\chi_{FFFG}^{1,2,3}-4\chi_{FGFF}^{1,2,3}-6\chi_{GFFF}^{1,2,3}-6\chi_{FFGF}^{1,2,3}
\notag \\
&\,+6\chi_{FFFF}^{1,2,3} ,
\end{align}
\begin{align}
\chi^{1,2,3}_{{\rm a}2}\equiv &\, 7\chi_{GGGG}^{1,3,2}-6\chi_{FGGG}^{1,3,2}-6\chi_{GFGG}^{1,3,2}-6\chi_{GGFG}^{1,3,2}-6\chi_{GGGF}^{1,3,2}
\notag \\
&\, +5\chi_{FFGG}^{1,3,2}+5\chi_{GGFF}^{1,3,2}+5\chi_{FGGF}^{1,3,2}+5\chi_{GFFG}^{1,3,2}+6\chi_{GFGF}^{1,3,2}
\notag \\
&\, +6\chi_{FGFG}^{1,3,2}-4\chi_{GFFF}^{1,3,2}-4\chi_{FGFF}^{1,3,2}-4\chi_{FFGF}^{1,3,2}-4\chi_{FFFG}^{1,3,2}
\notag \\
&\, + 2\chi_{FFFF}^{1,3,2}.
\end{align}
Similarly, Figs.\ \ref{Fig9b} and \ref{Fig9c} can be written as
\begin{align}
\chi^{1,2,3}_{{\rm b}}\equiv &\, 5\chi_{GGGG}^{1,2,3}-6\chi_{GFGG}^{1,2,3}-6\chi_{GGFG}^{1,2,3}
-4\chi_{FGGG}^{1,2,3}-4\chi_{GGGF}^{1,2,3}
\notag \\
&\,+6\chi_{FFGG}^{1,2,3}+6\chi_{GGFF}^{1,2,3}+7\chi_{GFFG}^{1,2,3}+2\chi_{FGGF}^{1,2,3}+5\chi_{FGFG}^{1,2,3}
\notag \\
&\,+5\chi_{GFGF}^{1,2,3}-6\chi_{FFFG}^{1,2,3}-6\chi_{GFFF}^{1,2,3}-4\chi_{FGFF}^{1,2,3}-4\chi_{FFGF}^{1,2,3}
\notag \\
&\,+5\chi_{FFFF}^{1,2,3},
\end{align}
\begin{align}
\chi^{1,2,3}_{{\rm c}}\equiv &\, 2\chi_{GGGG}^{1,2,3}-4\chi_{GFGG}^{1,2,3}-4\chi_{GGFG}^{1,2,3}
-4\chi_{FGGG}^{1,2,3}-4\chi_{GGGF}^{1,2,3}
\notag \\
&\,+5\chi_{FFGG}^{1,2,3}+5\chi_{GGFF}^{1,2,3}+5\chi_{GFFG}^{1,2,3}+5\chi_{FGGF}^{1,2,3}+6\chi_{FGFG}^{1,2,3}
\notag \\
&\,+6\chi_{GFGF}^{1,2,3}-6\chi_{FFFG}^{1,2,3}-6\chi_{GFFF}^{1,2,3}-6\chi_{FGFF}^{1,2,3}-6\chi_{FFGF}^{1,2,3}
\notag \\
&\,+7\chi_{FFFF}^{1,2,3},
\end{align}
\end{subequations}
respectively.
Functions $W^{(4{\rm f})}_{\Lambda,1122}$, $W^{(4{\rm e})}_{\Lambda,1112}$, and $W^{(4{\rm e})}_{\Lambda,1111}$ are
given in terms of them by symmetrizing the arguments as
\begin{subequations}
\label{W^(4f)}
\begin{align}
W^{(4{\rm f})}_{\Lambda,1122}({\bf k}_1,{\bf k}_2,{\bf k}_3,{\bf k}_4)
=&\, \frac{(2\Psi_\Lambda g_\Lambda)^4}{2}(\chi_{{\rm a}1}^{1,2,3}+\chi_{{\rm a}1}^{2,1,3}+\chi_{{\rm a}1}^{1,2,4}
\notag \\
&\,+\chi_{{\rm a}1}^{2,1,4}+\chi_{{\rm a}2}^{1,3,2}+\chi_{{\rm a}2}^{2,3,1}\Bigr),
\end{align}
\begin{align}
W^{(4{\rm f})}_{\Lambda,1112}({\bf k}_1,{\bf k}_2,{\bf k}_3,{\bf k}_4)
=&\, \frac{(2\Psi_\Lambda g_\Lambda)^4}{2}\Bigl(\chi_{{\rm b}}^{1,2,3}+\chi_{{\rm b}}^{1,3,2}+\chi_{{\rm b}}^{2,3,1}
\notag \\
&\,+\chi_{{\rm b}}^{2,1,3}+\chi_{{\rm b}}^{3,1,2}+\chi_{{\rm b}}^{3,2,1}\Bigr),
\end{align}
\begin{align}
W^{(4{\rm f})}_{\Lambda,1111}({\bf k}_1,{\bf k}_2,{\bf k}_3,{\bf k}_4)
=&\, \frac{(2\Psi_\Lambda g_\Lambda)^4}{2}\Bigl(\chi_{{\rm c}}^{1,2,3}+\chi_{{\rm c}}^{1,3,2}+\chi_{{\rm c}}^{2,3,1}
\notag \\
&\,+\chi_{{\rm c}}^{2,1,3}+\chi_{{\rm c}}^{3,1,2}+\chi_{{\rm c}}^{3,2,1}\Bigr).
\end{align}
\end{subequations}
Using Eqs.\ (\ref{tW^(4)}) and (\ref{W^(4f)}),
we obtain the key function in Eq.\ (\ref{tW^(2b)}) as
\begin{align}
&\, 2\delta \tilde{W}^{(4{\rm f})}_x(\tilde{\bf k}_1,\tilde{\bf k}_2;-\tilde{\bf k}_1,-\tilde{\bf k}_2)
+\delta \tilde{W}^{(4{\rm f})}_x(\tilde{\bf k}_1,-\tilde{\bf k}_1;\tilde{\bf k}_2,-\tilde{\bf k}_2)
\notag \\
=&\, \frac{z_{\Lambda,-}^2}{\Lambda^{3-d}} \frac{(2g_\Lambda\Psi_\Lambda)^4}{2}\delta_{{\bf k}_1+{\bf k}_2+{\bf k}_3+{\bf k}_4,{\bf 0}}
\notag \\
&\,\times
\Bigl[2\Bigl(\chi_{\rm a}^{1,2,-1}+\chi_{\rm a}^{2,1,-1}+\chi_{\rm a}^{1,2,-2}+\chi_{\rm a}^{2,1,-2}
+\chi_{\rm b}^{1,-1,2}+\chi_{\rm b}^{2,-1,1}\Bigr)
\notag \\
&\, +\Bigl(\chi_{\rm a}^{1,-1,2}+\chi_{\rm a}^{-1,1,2}+\chi_{\rm a}^{1,-1,-2}+\chi_{\rm a}^{-1,1,-2}
+\chi_{\rm b}^{1,2,-1}+\chi_{\rm b}^{-1,2,1}\Bigr)
\notag \\
&\,-\Bigl(\chi_{\rm c}^{1,2,-1}+\chi_{\rm c}^{1,-1,2}+\chi_{\rm c}^{2,-1,1}+\chi_{\rm c}^{2,1,-1}
+\chi_{\rm c}^{-1,1,2}+\chi_{\rm c}^{-1,2,1}\Bigr)
\notag \\
&\,-\Bigl(\chi_{\rm c}^{-2,1,2}+\chi_{\rm c}^{-2,2,1}+\chi_{\rm c}^{1,2,-2}+\chi_{\rm c}^{1,-2,2}
+\chi_{\rm c}^{2,-2,1}+\chi_{\rm c}^{2,1,-2}\Bigr)
\notag \\
&\,-\Bigl(\chi_{\rm c}^{-1,-2,1}+\chi_{\rm c}^{-1,1,-2}+\chi_{\rm c}^{-2,1,-1}+\chi_{\rm c}^{-2,-1,1}
+\chi_{\rm c}^{1,-1,-2}
\notag \\
&\,
+\chi_{\rm c}^{1,-2,-1}\Bigr)
-\Bigl(\chi_{\rm c}^{2,-1,-2}+\chi_{\rm c}^{2,-2,-1}+\chi_{\rm c}^{-1,-2,2}+\chi_{\rm c}^{-1,2,-2}
\notag \\
&\,
+\chi_{\rm c}^{-2,2,-1}+\chi_{\rm c}^{-2,-1,2}\Bigr)
+\Bigl(\chi_{\rm d}^{1,2,-1}+\chi_{\rm d}^{1,-1,2}+\chi_{\rm d}^{2,-1,1}+\chi_{\rm d}^{2,1,-1}
\notag \\
&\, +\chi_{\rm d}^{-1,1,2}+\chi_{\rm d}^{-1,2,1}\Bigr)\Bigr] .
\label{tW^4f)-0}
\end{align}
Equation (\ref{chi_ABCD}) satisfies $\chi_{ABCD}^{1,2,3}=\chi_{ABCD}^{-1,-2,-3}$ as shown with ${\bf q}\rightarrow -{\bf q}$ in the integrand, 
and also
$\chi_{ABCD}^{1,2,-1}= \chi_{BCDA}^{2,-1,-2}=\chi_{CDAB}^{-1,-2,1}=\chi_{DABC}^{-2,1,2}$ and
$\chi_{ABCD}^{1,2,-2}=\chi_{BCDA}^{2,-2,-1}= \chi_{CDAB}^{-2,-1,1}= \chi_{DABC}^{-1,1,2}$
owing to $\delta_{{\bf k}_1+{\bf k}_2+{\bf k}_3+{\bf k}_4,{\bf 0}}$.
These equalities enable us to classify the terms in the square brackets of Eq.\ (\ref{tW^4f)-0}) into the four categories of superscripts: 
$(1,2,-1)$, $(1,-2,-1)$, $(1,-1,2)$, $(1,-1,-2)$.
We can thereby transform Eq.\ (\ref{tW^4f)-0}) into
\begin{align}
&\, 2\delta \tilde{W}^{(4{\rm f})}_x(\tilde{\bf k}_1,\tilde{\bf k}_2;-\tilde{\bf k}_1,-\tilde{\bf k}_2)
+\delta \tilde{W}^{(4{\rm f})}_x(\tilde{\bf k}_1,-\tilde{\bf k}_1;\tilde{\bf k}_2,-\tilde{\bf k}_2)
\notag \\
=&\, \frac{z_{\Lambda,-}^2}{\Lambda^{3-d}} \frac{(2g_\Lambda\Psi_\Lambda)^4}{2}\delta_{{\bf k}_1+{\bf k}_2+{\bf k}_3+{\bf k}_4,{\bf 0}}
 \Bigl(
\chi_{GGGG}^{1,2,-1}+2\chi_{FGGG}^{1,2,-1}-2\chi_{GFGG}^{1,2,-1}
\notag \\
&\,+2\chi_{GGFG}^{1,2,-1}-2\chi_{GGGF}^{1,2,-1}
-\chi_{FFGG}^{1,2,-1}-\chi_{GFFG}^{1,2,-1}
-\chi_{GGFF}^{1,2,-1}-\chi_{FGGF}^{1,2,-1}
\notag \\
&\, -4\chi_{FGFG}^{1,2,-1}+6\chi_{GFGF}^{1,2,-1}+2\chi_{FFFG}^{1,2,-1}-2\chi_{GFFF}^{1,2,-1}+2\chi_{FGFF}^{1,2,-1}
\notag \\
&\, -2\chi_{FFGF}^{1,2,-1}+\chi_{FFFF}^{1,2,-1}
+\chi_{GGGG}^{1,-2,-1}-2\chi_{FGGG}^{1,-2,-1}+2\chi_{GFGG}^{1,-2,-1}
\notag \\
&\,-2\chi_{GGFG}^{1,-2,-1}+2\chi_{GGGF}^{1,-2,-1}
-\chi_{FFGG}^{1,-2,-1}-\chi_{GFFG}^{1,-2,-1}-\chi_{GGFF}^{1,-2,-1}
\notag \\
&\,-\chi_{FGGF}^{1,-2,-1}
+6\chi_{FGFG}^{1,-2,-1}
-4\chi_{GFGF}^{1,-2,-1}-2\chi_{FFFG}^{1,-2,-1}+2\chi_{GFFF}^{1,-2,-1}
\notag \\
&\, -2\chi_{FGFF}^{1,-2,-1}+2\chi_{FFGF}^{1,-2,-1}+\chi_{FFFF}^{1,-2,-1}
+ 4 \chi_{GGGG}^{1,-1,2}-4\chi_{GFGG}^{1,-1,2}
\notag \\
&\,-4\chi_{GGGF}^{1,-1,2}-2\chi_{FFGG}^{1,-1,2}-2\chi_{GFFG}^{1,-1,2}-2\chi_{GGFF}^{1,-1,2}
-2\chi_{FGGF}^{1,-1,2}
\notag \\
&\,+10\chi_{GFGF}^{1,-1,2}+4\chi_{FFFG}^{1,-1,2}+4\chi_{FGFF}^{1,-1,2}
-6\chi_{FFFF}^{1,-1,2}
+4\chi_{GFGG}^{1,-1,-2}
\notag \\
&\,+4\chi_{GGGF}^{1,-1,-2}-2\chi_{FFGG}^{1,-1,-2}-2\chi_{GFFG}^{1,-1,-2}-2\chi_{GGFF}^{1,-1,-2}
-2\chi_{FGGF}^{1,-1,-2}
\notag \\
&\,+4\chi_{FGFG}^{1,-1,-2}-6\chi_{GFGF}^{1,-1,-2}-4\chi_{FFFG}^{1,-1,-2}
-4\chi_{FGFF}^{1,-1,-2}+10\chi_{FFFF}^{1,-1,-2}\Bigr).
\label{tW^4f)-1}
\end{align}
Subsequently, we write $G=F+\delta G$ and expand the resulting expression in terms of $\delta G$ given by Eq.\ (\ref{dG})
up to the second order.
Terms of the zeroth order cancel out in $\delta \tilde{W}^{(4{\rm f})}_x$, and we obtain
\begin{align}
&\, 2\delta \tilde{W}^{(4{\rm f})}_x(\tilde{\bf k}_1,\tilde{\bf k}_2;-\tilde{\bf k}_1,-\tilde{\bf k}_2)
+\delta \tilde{W}^{(4{\rm f})}_x(\tilde{\bf k}_1,-\tilde{\bf k}_1;\tilde{\bf k}_2,-\tilde{\bf k}_2)
\notag \\
=&\, \frac{z_{\Lambda,-}^2}{\Lambda^{3-d}} \frac{(2g_\Lambda\Psi_\Lambda)^4}{2}\delta_{{\bf k}_1+{\bf k}_2+{\bf k}_3+{\bf k}_4,{\bf 0}}
\Bigl(\chi_{\delta G FFF}^{1,2,-1}-\chi_{F\delta G FF}^{1,2,-1}
\notag \\
&\, +\chi_{FF\delta G F}^{1,2,-1} -\chi_{FFF\delta G }^{1,2,-1}
+3 \chi_{\delta G F \delta G F}^{1,2,-1}+ \chi_{ F \delta G F\delta G}^{1,2,-1}
-\chi_{\delta G FFF}^{1,-2,-1}
\notag \\
&\, +\chi_{F\delta G FF}^{1,-2,-1}-\chi_{FF\delta G F}^{1,-2,-1}+\chi_{FF F\delta G}^{1,-2,-1}
+ \chi_{\delta G F \delta G F}^{1,-2,-1}+ 3\chi_{ F \delta G F\delta G}^{1,-2,-1}
\notag \\
&\, +2\chi_{\delta G FFF}^{1,-1,2}
+2\chi_{FF\delta G F}^{1,-1,2}
-2 \chi_{\delta G \delta G F F}^{1,-1,2}-2 \chi_{F \delta G \delta G F}^{1,-1,2}
\notag \\
&\,-2 \chi_{FF \delta G \delta G }^{1,-1,2}-2 \chi_{ \delta G FF\delta G }^{1,-1,2}+ 6\chi_{ \delta G F\delta GF }^{1,-1,2}+4\chi_{ F\delta GF \delta G }^{1,-1,2}
\notag \\
&\, -2\chi_{\delta G FFF}^{1,-1,-2}
-2\chi_{FF\delta G F}^{1,-1,-2}
+2 \chi_{\delta G \delta G F F}^{1,-1,-2}+2 \chi_{F \delta G \delta G F}^{1,-1,-2}
\notag \\
&\,+2 \chi_{FF \delta G \delta G }^{1,-1,-2}+2 \chi_{ \delta G FF\delta G }^{1,-1,-2}+ 2\chi_{ \delta G F\delta GF }^{1,-1,-2}
+4\chi_{ F\delta GF \delta G }^{1,-1,-2} \bigr),
\label{tW^(4f)}
\end{align}
within ${\rm O}\bigl((\delta G)^2\bigr)$.

Let us substitute Eqs.\ (\ref{g_Lambda-asymp}) and (\ref{tW^(4f)}) into Eq.\ (\ref{tW^(2b)}).
We then find
through ${\bf q}\rightarrow -{\bf q}$ in the integrand 
that terms of ${\rm O}\bigl((\delta G)^1\bigr)$ also cancel out in $\delta \tilde{W}^{(2{\rm b}4{\rm f})}_x$.
Subsequently, we substitute the leading-order expressions of Eqs.\ (\ref{F}) and (\ref{dG}), 
transform wave vectors into dimensionless forms by Eq.\ (\ref{tk-k}),
make changes of variables such as $\lambda\tilde{\bf q}+\tilde{\bf q}_1= \tilde{\bf q}_1'$ and $\tilde{\bf q}=-\tilde{\bf q}'$
for $\varTheta(\tilde{q}_1-1)\varTheta(|\lambda\tilde{\bf k}+\tilde{\bf q}_1|-1)\varTheta(|\lambda\tilde{\bf k}+\lambda\tilde{\bf q}+\tilde{\bf q}_1|-1)\delta(|\lambda\tilde{\bf q}+\tilde{\bf q}_1|-1)$ to yield a common factor $\delta(\tilde{q}_1-1)$,
and approximate $d\approx 4$ in the integrand as justified
for $\epsilon\ll 1$.
We can thereby express $\delta\tilde{W}_{\infty}^{(2{\rm b}4{\rm f})}$ in terms of Eqs.\ (\ref{phi^(3d)}) and (\ref{phi^(4f)}) as Eq.\ (\ref{tW^(2b4f)}).

\end{document}